\documentclass[aps,prb,preprint,showpacs,superscriptaddress,floatfix]{revtex4-2}

\usepackage{graphicx}
\usepackage{amsmath,amssymb,amsfonts}

\begin{document}


\title{Non-centrosymmetric, transverse structural modulation
in SrAl$_4$, and elucidation of its
origin in the BaAl$_4$ family of compounds}

\author{Sitaram Ramakrishnan}
\email{niranj002@gmail.com}
\affiliation{Department of Quantum Matter, AdSE,
Hiroshima University, Higashi-Hiroshima 739-8530, Japan}

\author{Surya Rohith Kotla}
\affiliation{Laboratory of Crystallography,
University of Bayreuth, 95447 Bayreuth, Germany}

\author{Hanqi Pi}
\affiliation{Beijing National Laboratory for Condensed Matter Physics, Institute of Physics, Chinese Academy of Sciences, Beijing 100190, China}
\affiliation{School of Physics, University of Chinese Academy of Sciences, Beijing 100049, China}

\author{Bishal Baran Maity}
\affiliation{Department of Condensed Matter Physics
and Materials Science,
Tata Institute of Fundamental Research, Mumbai 400005, India}

\author{Jia Chen}
\email{chenjia@zhejianglab.com}
\affiliation{Zhejiang Laboratory, Hangzhou 311121, China}

\author{Jin-Ke Bao}
\affiliation{Department of Physics, Materials Genome Institute and International Center
for Quantum and Molecular Structures, Shanghai University,
 Shanghai 200444, People’s Republic of China}

\author{Zhaopeng Guo}
\affiliation{Beijing National Laboratory for Condensed Matter Physics, Institute of Physics, Chinese Academy of Sciences, Beijing 100190, China}
\affiliation{School of Physics, University of Chinese Academy of Sciences, Beijing 100049, China}

\author{Masaki Kado}
\affiliation{Department of Quantum Matter, AdSE,
Hiroshima University, Higashi-Hiroshima 739-8530, Japan}

\author{Harshit Agarwal}
\affiliation{Laboratory of Crystallography,
University of Bayreuth, 95447 Bayreuth, Germany}

\author{Claudio Eisele}
\affiliation{Laboratory of Crystallography,
University of Bayreuth, 95447 Bayreuth, Germany}

\author{Minoru Nohara}
\affiliation{Department of Quantum Matter, AdSE,
Hiroshima University, Higashi-Hiroshima 739-8530, Japan}

\author{Leila Noohinejad}
\affiliation{P24, PETRA III, Deutsches Elektronen-Synchrotron DESY,
Notkestrasse 85, 22607 Hamburg, Germany}

\author{Hongming Weng}
\email{hmweng@iphy.ac.cn}
\affiliation{Beijing National Laboratory for Condensed Matter Physics, Institute of Physics, Chinese Academy of Sciences, Beijing 100190, China}
\affiliation{Songshan Lake Materials Laboratory, Dongguan, Guangdong 523808, China}
\affiliation{School of Physics, University of Chinese Academy of Sciences, Beijing 100049, China}

\author{Srinivasan Ramakrishnan}
\affiliation{Department of Physics, Indian Institute of Science
Education and Research, Pune, 411008, India}

\author{Arumugam Thamizhavel}
\affiliation{Department of Condensed Matter Physics
and Materials Science,
Tata Institute of Fundamental Research, Mumbai 400005, India}

\author{Sander van Smaalen}
\email{smash@uni-bayreuth.de}
\affiliation{Laboratory of Crystallography,
University of Bayreuth, 95447 Bayreuth, Germany}

\date{\today}

\begin{abstract}
At ambient conditions SrAl$_4$ adopts the BaAl$_4$
structure type with space group $I4/mmm$.
It undergoes a charge-density-wave (CDW) transition
at $T_{CDW}$ = 243 K, followed by a structural
transition at $T_{S}$ = 87 K.
Temperature-dependent single-crystal X-ray
diffraction (SXRD) leads to the observation
of incommensurate superlattice reflections at
$\mathbf{q} = \sigma\,\mathbf{c}^{*}$ with
$\sigma = 0.1116$ at 200 K.
The CDW has orthorhombic symmetry with the
noncentrosymmetric superspace group
$F222(0\,0\,\sigma)00s$,
where $F222$ is a subgroup of $Fmmm$ as well as
of $I4/mmm$.
Atomic displacements mainly represent a transverse wave,
with displacements that are 90 deg out of phase
between the two diagonal directions
of the $I$-centered unit cell,
resulting in a helical wave.
Small longitudinal displacements are provided
by the second harmonic modulation.
The orthorhombic phase realized in SrAl$_4$ is
similar to that found in EuAl$_4$,
except that no second harmonic could be determined
for the latter compound.
Electronic structure calculations and phonon
calculations by density functional theory (DFT)
have failed to reveal the mechanism of CDW formation.
No clear Fermi surface nesting nor electron-phonon
coupling nor the involvement of Dirac points
could be established.
However, DFT reveals that Al atoms dominate the
density of states near the Fermi level,
thus, corroborating the SXRD measurements.
SrAl$_4$ remains incommensurately modulated
at the structural transition, where the
symmetry lowers from orthorhombic to $\mathbf{b}$-unique
monoclinic.
The present work draws a comparison on the modulated
structures of non-magnetic SrAl$_4$ and magnetic EuAl$_4$
elucidating their similarities and differences, and firmly
establishing that although substitution of Eu to Sr
plays little to no role in the structure, the
transition temperatures are affected by the atomic sizes.
We have identified a simple criterion, that correlates
the presence of a phase transition with the interatomic
distances.
Only those compounds
$X$Al$_{4-x}$Ga$_x$ ($X$ = Ba, Eu, Sr, Ca; $0 < x <4$)
undergo phase transitions, for which the ratio $c/a$
falls within the narrow range $2.51 < c/a < 2.54$.
\end{abstract}

\maketitle

\clearpage

\section{\label{sec:sral4_introduction}Introduction}

The manifestation of charge density waves (CDWs)
was initially found to occur in quasi-one-dimensional (1D)
electronic systems, like NbSe$_3$ and
K$_{0.3}$MoO$_3$ \cite{gruner1988a,gruener1994a,monceau2012a}.
These compounds possess Fermi surfaces with co-planar
sections that allow so-called Fermi surface nesting (FSN),
thus explaining the stabilization of CDWs.
The nesting vector of the periodic structure becomes
the wave vector of the CDW in the valence bands,
as well as the wave vector of the
modulation of the atomic positions (periodic
lattice distortion).
Recent research has expanded
the criteria for the occurrence of CDWs \cite{zhu2017a}.
Materials need not support 1D or quasi-two-dimensional (2D)
electron bands, but they can involve complex
three-dimensional (3D) electronic system,
if certain conditions are fulfilled.
The mechanism of stabilization of
CDW in these systems is provided by
$\mathbf{q}$-dependent electron-phonon coupling (EPC).
Several examples exist of 3D compounds with CDWs,
which include
CuV$_2$S$_4$ \cite{flemingrm1981a,kawaguchi2012a,okadah2004a,ramakrishnan2019a},
$R$$_3$Co$_4$Sn$_{13}$ ($R$ = La, Ce) \cite{slebarski2013a,otomo2016a,welsch2019a},
$R$$_5$Ir$_4$Si$_{10}$ ($R$ = Dy, Ho, Er, Yb, Lu) \cite{ramakrishnan2017a},
Sm$_2$Ru$_3$Ge$_5$ \cite{bugaris2017a,kuo2020a},
TmNiC$_2$ \cite{kolincio2020a,roman2023a},
$R_2$Ir$_3$Si$_5$ ($R$ = Lu, Er, Ho)
\cite{ramakrishnan2021a,ramakrishnan2020a,ramakrishnan2023a},
CuIr$_{2-x}$Cr$_x$Te$_4$ \cite{zeng2022a},
and kagome materials like $A$V$_3$Sb$_5$ ($A$ = K, Rb, Cs)
\cite{guguchia2023a,kautzsch2023a,xiao2023a,zhou2023a,wang2023b,yang2023a}
and FeGe \cite{teng2022a,teng2023a,zhou2023b}.

Compounds $X$Al$_4$\,/\,$X$Ga$_4$ ($X$ = Eu, Ca, Sr, Ba)
crystallize in the tetragonal BaAl$_4$ structure type
with space group $I4/mmm$.
These systems have attracted a
lot of attention in the recent years,
because of their properties as topological
quantum materials
\cite{haussermann2002a,stavinoha2018a,wang2021a,shang2021a,mori2022a,gen2023a,wang2023a}.
EuAl$_4$, containing divalent Eu, is a 3D CDW compound
that also undergoes four successive
antiferromagnetic (AFM)
transitions below 20 K
\cite{nakamura2015a,shimomura2019a,kaneko2021a,ramakrishnan2022a,meier2022a}.
It has been reported to have a chiral spin structure
and skyrmions \cite{meier2022a,takagir2022a},
similar to skyrmions reported in other
divalent Eu-based systems, such as EuPtSi \cite{kaneko2019a,onuki2020a}.
A skyrmion state of magnetic order, and the non-trivial
band topology were furthermore established by
measurements of the topological Hall resistivity,
the muon-spin rotation and relaxation ($\mu$SR),
and the magnetostriction \cite{shang2021a,zhu2022a,gen2023a}.
Our recent investigation into the modulated structure
of EuAl$_4$ reveals a breaking of the fourfold
rotational symmetry, resulting in an orthorhombic $Fmmm$
symmetry of the CDW state below 145 K.
There is no evidence of a lattice distortion
away from tetragonal symmetry.
The reduction towards orthorhombic symmetry is
brought about solely by
the transverse CDW modulation in EuAl$_4$.
More recent work proposes the loss of inversion symmetry
\cite{nih2024a,yangr2024a}.
Alternatively, centrosymmetric $Immm$ symmetry was
proposed for the
CDW state of EuAl$_4$ on the basis of inelastic x-ray
scattering experiments \cite{korshunovan2024a}.

Replacement of the divalent rare-earth element Eu
by divalent alkaline earth metals Ba, Sr and Ca
results in isostructural compounds.
SrAl$_4$ and CaAl$_4$ undergo CDW transitions
or other structural transitions (STs).
However, BaAl$_4$ does not undergo any phase transition
\cite{wang2021a,mori2022a}.
Out of all the materials in the $X$Al$_4$ series,
the compound that is most similar to EuAl$_4$ would
be SrAl$_4$.
SrAl$_4$ exhibits a CDW transition at $T_{CDW} = 243$ K,
which is at a much higher temperature than the CDW
transition of EuAl$_4$ \cite{nakamura2016a,niki2020a}.
Isostructural CaAl$_4$ undergoes a ST at 443 K,
which is at an even higher temperature than
for SrAl$_4$ \cite{miller1993a}.
SrAl$_4$ undergoes a second transition at
$T_S$ = 87 K \cite{nakamura2016a}.
It was suggested that the symmetry of the
low-temperature phase is monoclinic
\cite{nakamura2016a}.
The mechanism of CDW formation in either SrAl$_4$
or EuAl$_4$ presently is not understood, as both
compounds possess a complex 3D electronic structure
\cite{nakamura2015a,kobata2016a,nakamura2016a,wangll2023a}.

Here, we present the results of temperature-dependent
single-crystal x-ray diffraction (SXRD) experiments
on SrAl$_4$.
We find that the CDW transition is accompanied
by a reduction in symmetry from tetragonal to
$F$-centered orthorhombic, analogous to EuAl$_4$
\cite{ramakrishnan2022a}.
However, second-order satellite reflections
in the SXRD data set of SrAl$_4$ point towards
a loss of inversion symmetry, resulting in the
superspace group $F222(0\,0\,\sigma)00s$ for
the CDW phase.
Reconsideration of our SXRD data on EuAl$_4$
shows a marginally better fit for $F222(0\,0\,\sigma)00s$
than for originally published $Fmmm(0\,0\,\sigma)s00$
\cite{ramakrishnan2022a}.
The lack of resolving power of those data
probably is the result of the lack of second-order
satellite reflections in the SXRD data on EuAl$_4$
\cite{ramakrishnan2022a}.
Nevertheless, details of both structure models
for EuAl$_4$ are provided in
the Supplemental Material \cite{sral4suppmat2023a}.

Either model for EuAl$_4$ and SrAl$_4$ leads
to qualitatively similar variations of
interatomic distances along the incommensurate
coordinate of the CDW in these compound.
This confirms the earlier conclusion, that
the network of Al atoms governs the CDW in
SrAl$_4$ as well as in EuAl$_4$ \cite{ramakrishnan2022a}.
This finding is supported by solid solution
samples of isostructural
SrAl$_{4-x}$Si$_x$ and SrAl$_{4-x}$Ge$_x$,
since the CDW is suppressed by the disorder
at the Al site, and induces superconductivity
in the case of Si-doping \cite{zhang2013a,zevalkink2017a}.
From SXRD and physical properties measurements,
we confirm that below 100 K SrAl$_4$
undergoes a second phase transition
that is characterized by a monoclinic lattice
distortion, but across which the CDW modulation
remains incommensurate.
In this paper, we present the modulated crystal
structure of SrAl$_4$, and we discuss the similarities
and subtle differences in the CDWs of SrAl$_4$ and EuAl$_4$.
Furthermore, we establish that the value of $c/a$
must be within the narrow range $2.51 < c/a < 2.54$,
in order for a compound of this family to undergo a phase
transition.
This allows us to predict whether new materials
developed in this series could or could not
undergo a CDW or structural transition.

\section{\label{sec:sral4_experimental}%
Experimental and computational details}

\subsection{\label{sec:sral4_crystal_growth}%
Crystal growth}

Single crystals of SrAl$_4$ were grown by the
Al self-flux method according to \cite{nakamura2016a}.
Crystals were grown at three places, using similar
methods.

At the Lab. of Crystallography in Bayreuth,
the elements strontium (Alfa Aesar, 99.95\% purity)
and aluminium (Alfa Aesar, 99.9995\%)
were filled into an alumina crucible in the ratio 1:19.
The crucible was sealed in an evacuated
quartz-glass ampoule.
It was heated to a temperature of 1173 K
and held at this temperature for 2 hours.
The crucible was then cooled to 823 K
with a rate of 0.5 K/hour,
at which point the crystals were separated
from the molten metal by centrifugation.
Small crystals were selected and annealed
in vacuum for 72 hours at 723 K.
The 1:4 stoichiometry of the product was confirmed
by structure refinement against SXRD data.
The resulting crystals A were used for the
SXRD experiments.

At the Tata Institute in Mumbai,
the elements were filled into an alumina crucible
with the ratio 1:23.
Crystal growth was at 1323 K for 24 hours,
after which it was cooled to 973 K
with a rate of 1 K/hour, followed by centrifugation.
The longer growth time resulted in
crystal B that is larger than crystal A.
These crystals were employed without annealing
for measurement of the specific heat.
At the Department of Quantum Matter in Hiroshima,
the same procedure was
followed, resulting in crystal C that was used
for the measurement of the electrical resistivity.

\subsection{\label{sec:sral4_sxrd_data_collection}%
Single-crystal x-ray diffraction data collection}

SXRD experiments were performed at Beamline P24 of
PETRA III at DESY in Hamburg, employing a four-circle
Huber diffractometer with Euler geometry,
and radiation of wavelength 0.50000 \AA{}.
The temperature of the crystal was controlled by a
CRYOCOOL open-flow helium cryostat.
Complete data sets were measured at temperatures
of 293, 200, 120, 100, 75 and 20 K, covering all phases.
Each run of data collection comprises 3640 frames,
corresponding to a rotation of the crystal over
364 deg, which was repeated 10 times.
These data were binned to a data set of 364 frames
of 1 deg rotation and 10 s exposure time
\cite{paulmann2020a}.
Further details are provided in the
Supplemental Material \cite{sral4suppmat2023a}.

\subsection{\label{sec:sral4_sxrd_processing}%
Single-crystal x-ray diffraction data processing}

The EVAL15 software suite \cite{schreursamm2010a}
was used for processing the SXRD data.
Each temperature comprises two runs, one with and
another without 2$\theta$ offset of the detector.
The two runs were integrated separately, and
subsequently merged in the module ANY of EVAL15.
SADABS \cite{sheldrick2008} was used for scaling
and absorption correction with Laue symmetry $4/mmm$
for the data measured at 293 K (periodic phase).
Different Laue symmetries were employed for the
runs collected at 200, 120 and 100 K (CDW phase),
depending on the symmetry of the structure model
being tested (compare Tables S2--S4 in the
Supplemental Material \cite{sral4suppmat2023a}).
Laue symmetry $2/m$ ($\mathbf{b}$-unique) was used
for the SXRD data at 20 K (low-temperature phase).
The resulting reflection files were imported
into the software
JANA2006 \cite{petricekv2014a, petricekv2016a}
for structure refinements.
Tables \ref{tab:sral4_cdw_crystalinfo} and
\ref{tab:sral4_f222_ampl}, and
Tables S2--S6 in the Supplemental Material
\cite{sral4suppmat2023a} give details of
the analysis, crystallographic information,
atomic coordinates and modulation amplitudes.
\begin{table}[p]
\caption{\label{tab:sral4_cdw_crystalinfo}%
Crystallographic data of crystal A of
SrAl$_4$ at 293 K (periodic phase) and 200 K (CDW phase).
Details are given for two different structure models
for the CDW phase.}
\scriptsize
\centering
\begin{minipage}{95mm}
\begin{ruledtabular}
\begin{tabular}{cccc}
Temperature (K) & 293       & 200 & 200 \\
Model          & periodic   & C   & D \\
Crystal system & Tetragonal & Orthorhombic & Orthorhombic \\
Space/Superspace group & $I4/mmm$  &
$Fmmm(0\,0\,\sigma)s00$ & $F222(0\,0\,\sigma)00s$ \\
 No. \cite{stokesht2011a} & 139 & {69.1.17.2} & {22.1.17.2}  \\
$a$ (\AA{}) &4.4893(2) & \multicolumn{2}{c}{6.3326(4)}  \\
$b$ (\AA{}) &4.4893    & \multicolumn{2}{c}{6.3331(5)}  \\
$c$ (\AA{}) &11.2764(5)  & \multicolumn{2}{c}{11.2541(5)}  \\
Volume (\AA{}$^3$) & 227.26(3) & \multicolumn{2}{c}{451.35(4)}  \\
Wavevector \textbf{q} & --
& \multicolumn{2}{c}{0.1116(2)$\mathbf{c}^{*}$}  \\
$Z$ & 2 & \multicolumn{2}{c}{4} \\
Wavelength (\AA{}) & 0.50000 & \multicolumn{2}{c}{0.50000}  \\
Detector distance (mm) &260 & \multicolumn{2}{c}{260}  \\
$2\theta$-offset (deg) &0, 25 & \multicolumn{2}{c}{0, 25}  \\
$\chi$-offset (deg) &-60 & \multicolumn{2}{c}{-60}  \\
Rotation per image (deg) & 1 & \multicolumn{2}{c}{1}  \\
$(\sin(\theta)/\lambda)_{max}$ (\AA{}$^{-1}$) &0.746821& \multicolumn{2}{c}{0.745874} \\
Absorption, $\mu$ (mm$^{-1}$) & 4.844 & \multicolumn{2}{c}{4.878}  \\
T$_{min}$, T$_{max}$ & 0.3118, 0.3522 & \multicolumn{2}{c}{0.3121, 0.3515}  \\
Criterion of observability & $I>3\sigma(I)$ & \multicolumn{2}{c}{$I>3\sigma(I)$} \\
No. of reflections measured, \\
$(m = 0)$  &  630  & \multicolumn{2}{c}{493}  \\
$(m = 1)$  & -     & \multicolumn{2}{c}{969} \\
$(m = 2)$  & -     & \multicolumn{2}{c}{971} \\
No. of unique reflections,   \\
$(m = 0)$ (obs/all) & 137/139 & \multicolumn{2}{c}{197/211}  \\
$(m = 1)$ (obs/all) & -- & \multicolumn{2}{c}{323/392}  \\
$(m = 2)$ (obs/all) & -- & \multicolumn{2}{c}{94/403} \\
$R_{int}$ $(m = 0)$ (obs/all) &0.0241/0.0241 & \multicolumn{2}{c}{0.0229/0.0229}  \\
$R_{int}$ $(m = 1)$ (obs/all) &-- & \multicolumn{2}{c}{0.0979/0.0978} \\
$R_{int}$ $(m = 2)$ (obs/all) &-- & \multicolumn{2}{c}{0.0845/0.1010} \\
No. of parameters &9 & 25 & 30 \\
$R_{F }$ $(m = 0)$  (obs) &0.0144 &0.0277 & 0.0269 \\
$R_{F }$ $(m = 1)$ (obs) &- &0.0502 & 0.0493 \\
$R_{F }$ $(m = 2)$ (obs) &- &0.1865 & 0.0646 \\
$wR_{F }$ $(m = 0)$ (all) &0.0186 &0.0311 & 0.0303 \\
$wR_{F }$ $(m = 1)$ (all) &- &0.0658 & 0.0645 \\
$wR_{F }$ $(m = 2)$ (all) &- &0.2480 & 0.0934 \\
$wR_{F }$ (all) (all) &0.0186 &0.0560 & 0.0472 \\
GoF (obs/all) &1.28/1.27 &1.88/1.53 & 1.64/1.29 \\
$\Delta\rho_{min}$, $\Delta\rho_{max}$(e \AA$^{-3}$) &
 -0.29, 0.47 & -4.01, 4.87 &  -1.89, 2.99 \\
\end{tabular}
\end{ruledtabular}
\end{minipage}
\end{table}
\begin{table}[ht]
\begin{center}
\small
\caption{\label{tab:sral4_f222_ampl}%
Amplitudes of the modulation functions of crystal A at
200 K for superspace group $F222(0\,0\,\sigma)00s$.
The relative coordinates, $x$, $y$ and $z$ of the basic
position are also specified.
Similar information for $Fmmm(0\,0\,\sigma)s00$
is provided in Tables S5 and S6.
Values of modulation amplitudes have been multiplied
by the corresponding lattice parameter,
in order to obtain values in \AA{}.
}
\begin{ruledtabular}
\begin{tabular}{ccccc}
Atom                  & Sr & Al1a &  Al1b & Al2 \\
$x$                   & 0  & 0.25 & -0.25 & 0  \\
$y$                   & 0  & 0.25 & -0.25 & 0  \\
$z$                   & 0  & 0.25 & -0.25 & 0.38359(12)  \\
$A_{1,x}\, a$ (\AA{}) & 0.1939(40) & 0.1942(126) & 0.1616(115) & 0.1945(68) \\
$A_{1,y}\, b$ (\AA{}) & 0          & 0           & 0           & 0.0413(92) \\
$A_{1,z}\, c$ (\AA{}) & 0          & 0           & 0           & 0 \\
$B_{1,x}\, a$ (\AA{}) & 0          & 0           & 0           & 0.0536(52) \\
$B_{1,y}\, b$ (\AA{}) &-0.1301(62) &-0.0713(127) &-0.2162(114) &-0.1325(102) \\
$B_{1,z}\, c$ (\AA{}) & 0          & 0           & 0           & 0 \\
$A_{2,x}\, a$ (\AA{}) & 0          & 0           & 0           & 0 \\
$A_{2,y}\, b$ (\AA{}) & 0          & 0           & 0           & 0 \\
$A_{2,z}\, c$ (\AA{}) & 0.0127(26) & 0.0028(85)  & 0.0069(74)  & 0.0004(60) \\
$B_{2,x}\, a$ (\AA{}) & 0          & 0           & 0           & 0 \\
$B_{2,y}\, b$ (\AA{}) & 0          & 0           & 0           & 0 \\
$B_{2,z}\, c$ (\AA{}) & 0          & 0           & 0           &-0.0004(59)  \\
\end{tabular}
\end{ruledtabular}
\end{center}
\end{table}

\subsection{\label{sec:sral4_dft_calculations}%
Density Functional Theory (DFT) Calculations}

Density functional theory (DFT) calculations were
performed for the tetragonal crystal structure of SrAl$_4$
with space group of $I4/mmm$,
employing the Vienna \textit{ab initio} simulation
package (VASP) \cite{kresse1996efficient}.
The Projector-augmented-wave (PAW) method
\cite{blochl1994projector,PAW1999}
and generalized gradient approximations (GGA)
\cite{perdew1996generalized} with Perdew-Burke-Ernzerhorf (PBE)
type pseudopotentials were chosen to deal with exchange-correlations.

The conventional unit cell was fully relaxed with a
$12 \times 12\times 4$ $k$-mesh sampling until the
energy convergence tolerance fell below $10^{-7}$ eV
and the force was less than $10^{-3}$ eV/\AA{}.
The cutoff energy of the plane-wave basis was chosen as $520$ eV.
The DFT optimized results are close to the experimental values
(Table \ref{tab:sral4_cdw_crystalinfo}).
An $18 \times 18\times 18$ $k$-mesh was implemented for
the Brillouin zone (BZ) integral sampling of the
primitive unit cell.
The Fermi surface was calculated by using the WannierTools
package \cite{WU2018} with tight-binding Hamiltonian
constructed by the WANNIER90 code \cite{wan2008,wan90},
which is based on the maximally localized Wannier function
(MLWF) method \cite{MLWF2012}.
The bare charge susceptibility is carried out using a
$k$-mesh of $200 \times 200 \times 200$.
The phonon spectrum calculations were performed within
the framework of density functional perturbation theory
(DFPT) \cite{DFPT1991,DFPT1997}
and the finite displacement method,
implemented by the PHONOPY \cite{togo2015first} package
combined with VASP,
and by Quantum ESPRESSO
\cite{Giannozzi_2009,Giannozzi_2017}.

\subsection{\label{sec:sral4_physical_properties}%
Physical properties}

The dc electrical resistivity was measured between
6 and 296 K, employing a standard four-probe method.
The small size of the specimen prevented the
identification of lattice directions, so that the
single experimental run has resulted in $\rho(T)$
along an unspecified, general direction.
Basically, the temperature dependence of the
electrical resistivity confirms the data
reported in Ref. \cite{nakamura2016a}.
The present resistivity data are given in Section S6
of the Supplemental Material \cite{sral4suppmat2023a}.
A commercial physical property measurement system
(PPMS, Quantum Design, USA) was used for measuring
the specific heat data in a heating run.

\section{\label{sec:sral4_results_discussion}%
Results and discussion}

\subsection{\label{sec:sral4_orthocdw}%
The orthorhombic CDW phase}

SrAl$_4$ has been reported to undergo two phase
transitions: one at $T_{CDW} = 243$ K and
another below $90$ K \cite{nakamura2016a}.
Cooling the crystal from 293 to 200 K revealed
superlattice reflections with
$\mathbf{q} =  0.1116(2)\,\mathbf{c}^{*}$ at 200 K,
similar to the diffraction of isostructural
EuAl$_4$ below $145$ K \cite{shimomura2019a,ramakrishnan2022a}.
The data measured at $200$ K (incommensurate CDW phase)
could be refined successfully
(Table \ref{tab:sral4_cdw_crystalinfo}).
Like for EuAl$_4$ \cite{shimomura2019a,ramakrishnan2022a},
we do not observe any lattice distortion within the
CDW phase, thus preserving the tetragonal lattice
symmetry for the basic structure.

In order to elucidate the modulated CDW structure
at $200$ K, we have tested different
superspace groups as symmetry of the crystal structure
(See Tables S2--S4 in the
Supplemental Material \cite{sral4suppmat2023a}).
%
The candidate models are based on symmetries
$I4/mmm$ (model A) and its centrosymmetric (Table S2)
and noncentrosymmetric (Table S3) subgroups.
The best fit to the SXRD data is obtained for
model D with the noncentrosymmetric superspace group
$F222(0\,0\,\sigma)00s$.
This superspace group is a proper subgroup of
centrosymmetric $Fmmm(0\,0\,\sigma)s00$,
the latter which was previously found for the
CDW phase of EuAl$_4$ \cite{ramakrishnan2022a}.
For SrAl$_4$, model C with superspace group
$Fmmm(0\,0\,\sigma)s00$ also leads to a good fit
to the SXRD data, except for the second-order
satellite reflections, the latter which have
been observed by SXRD, but which are not present in the
available data sets for EuAl$_4$
\cite{ramakrishnan2022a,moyajm2022a,korshunovan2024a}.
Based on their significantly worse fit to the
SXRD data, $I$-centered orthorhombic and monoclinic
superspace groups can be excluded as symmetry for
SrAl$4_4$.

The analysis thus gave model D with superspace
symmetry $F222(0\,0\,\sigma)00s$ as structure model
for the incommensurately modulated CDW state
of SrAl$_4$.
Although with less strong evidence for either one
(due to the missing
second-order satellite reflections), both
centrosymmetric $Fmmm(0\,0\,\sigma)s00$ and
noncentrosymmetric $F222(0\,0\,\sigma)00s$
remain candidates for the symmetry of the CDW
of EuAl$_4$.

\begin{figure}
\includegraphics[width=80mm,height=80mm,keepaspectratio]{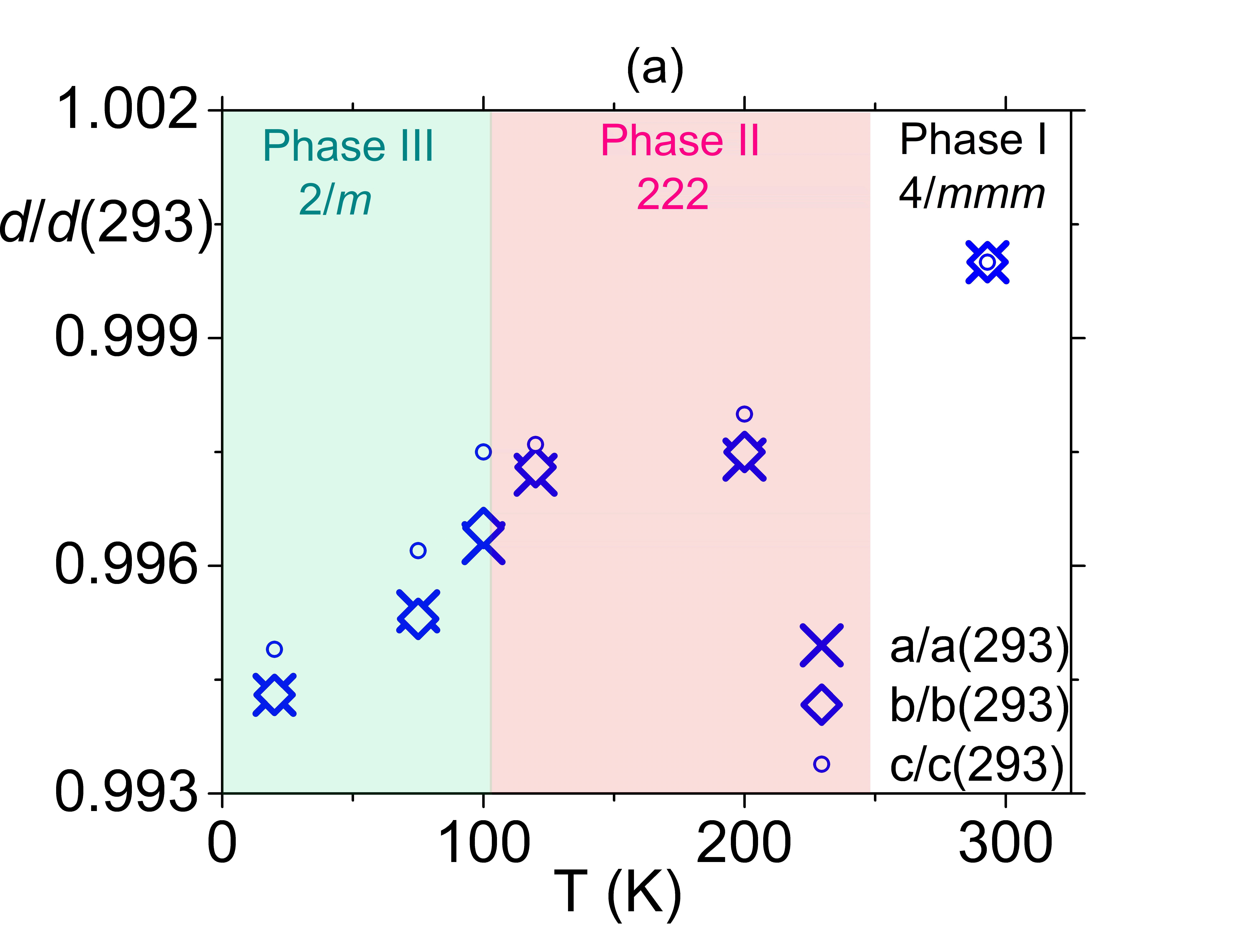}
\includegraphics[width=80mm,height=80mm,keepaspectratio]{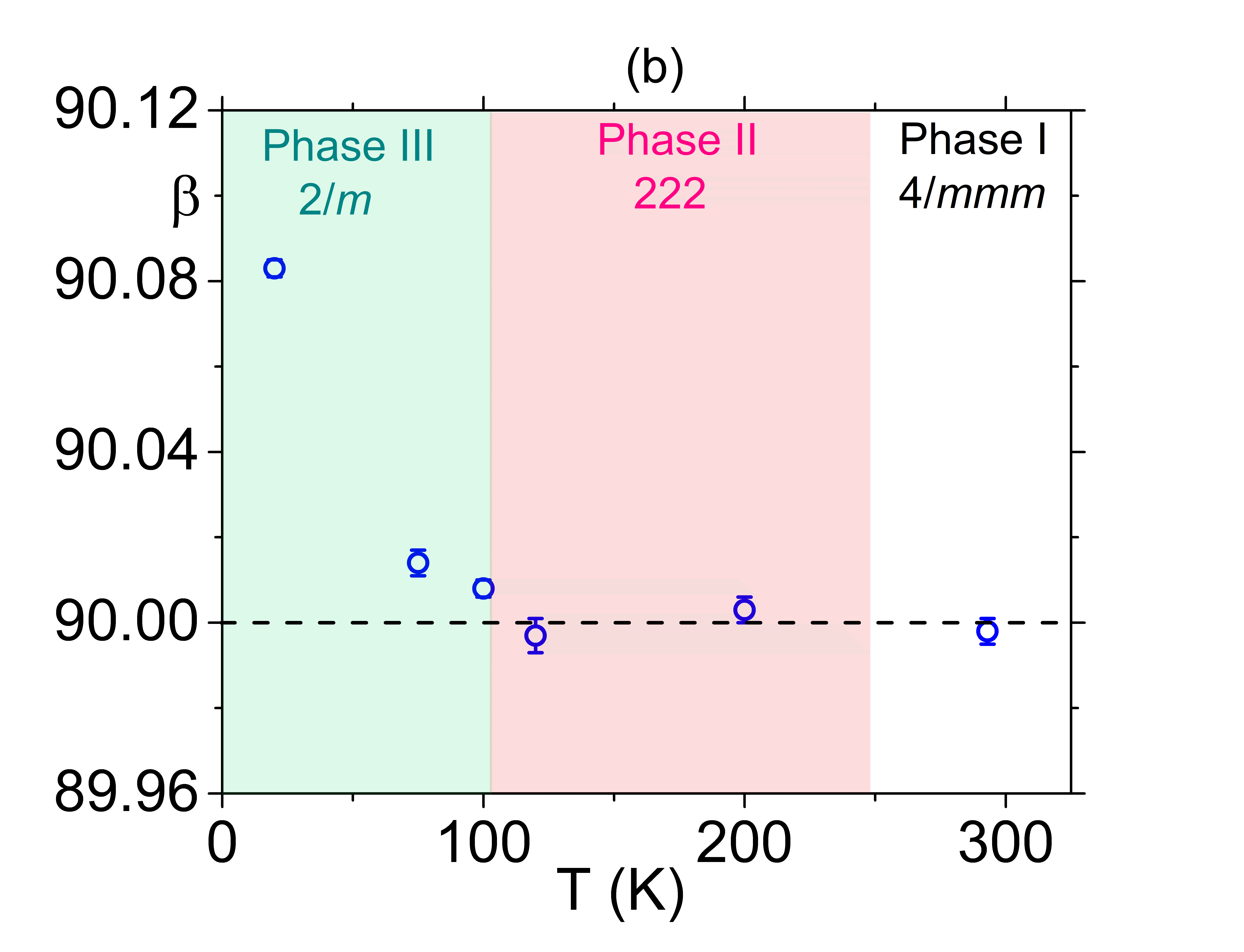}
\\
\includegraphics[width=80mm,height=80mm,keepaspectratio]{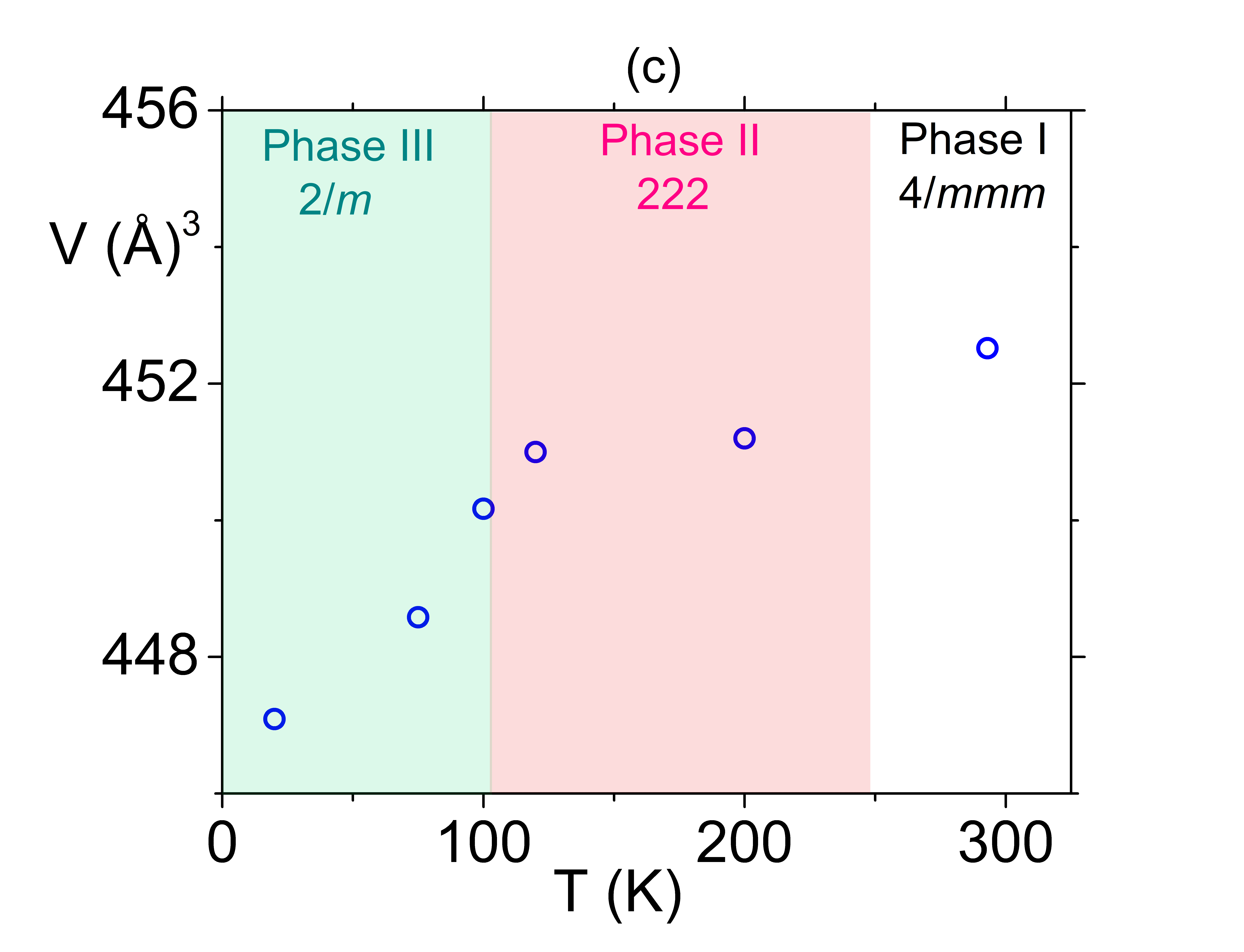}
\includegraphics[width=80mm,height=80mm,keepaspectratio]{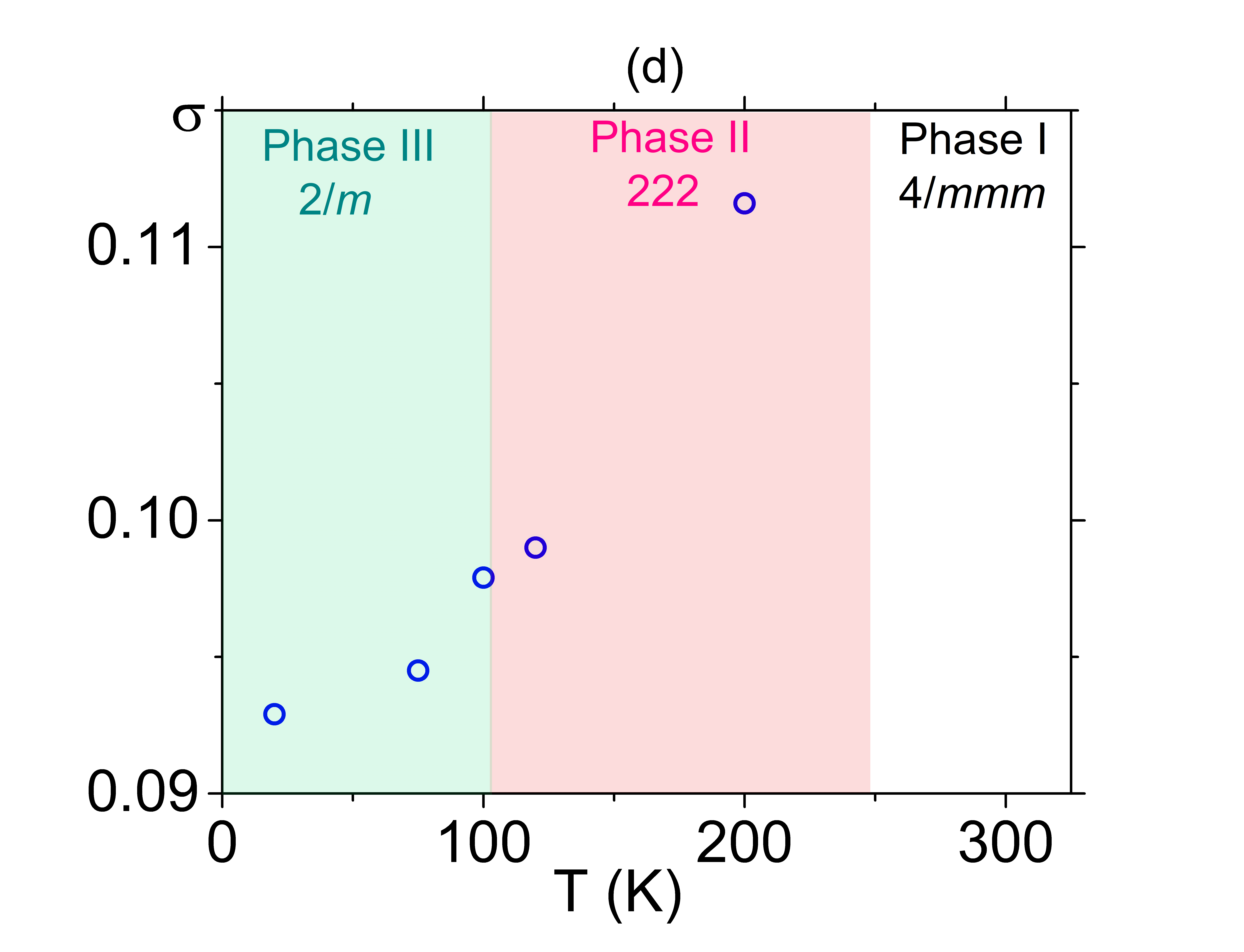}
\caption{\label{fig:sral4_lattice_qvec}%
Lattice parameters and modulation wave vector
within the temperature range 20--293 K.
(a) Lattice parameters ($F$-centered setting)
relative to their values at $T = 293$ K, with
$a(293) = 6.3488(3)$, $b(293) = 6.3488$ and
$c(293) = 11.2764(3)4$ \AA{}.
(b) The lattice parameter $\beta$ in deg.
(c) Volume of the unit cell.
(d) Component $\sigma$ of $\mathbf{q} = (0,\,0,\,\sigma)$,
which remains incommensurate down to 20 K.}
\end{figure}

\begin{figure}
\includegraphics[width=80mm]{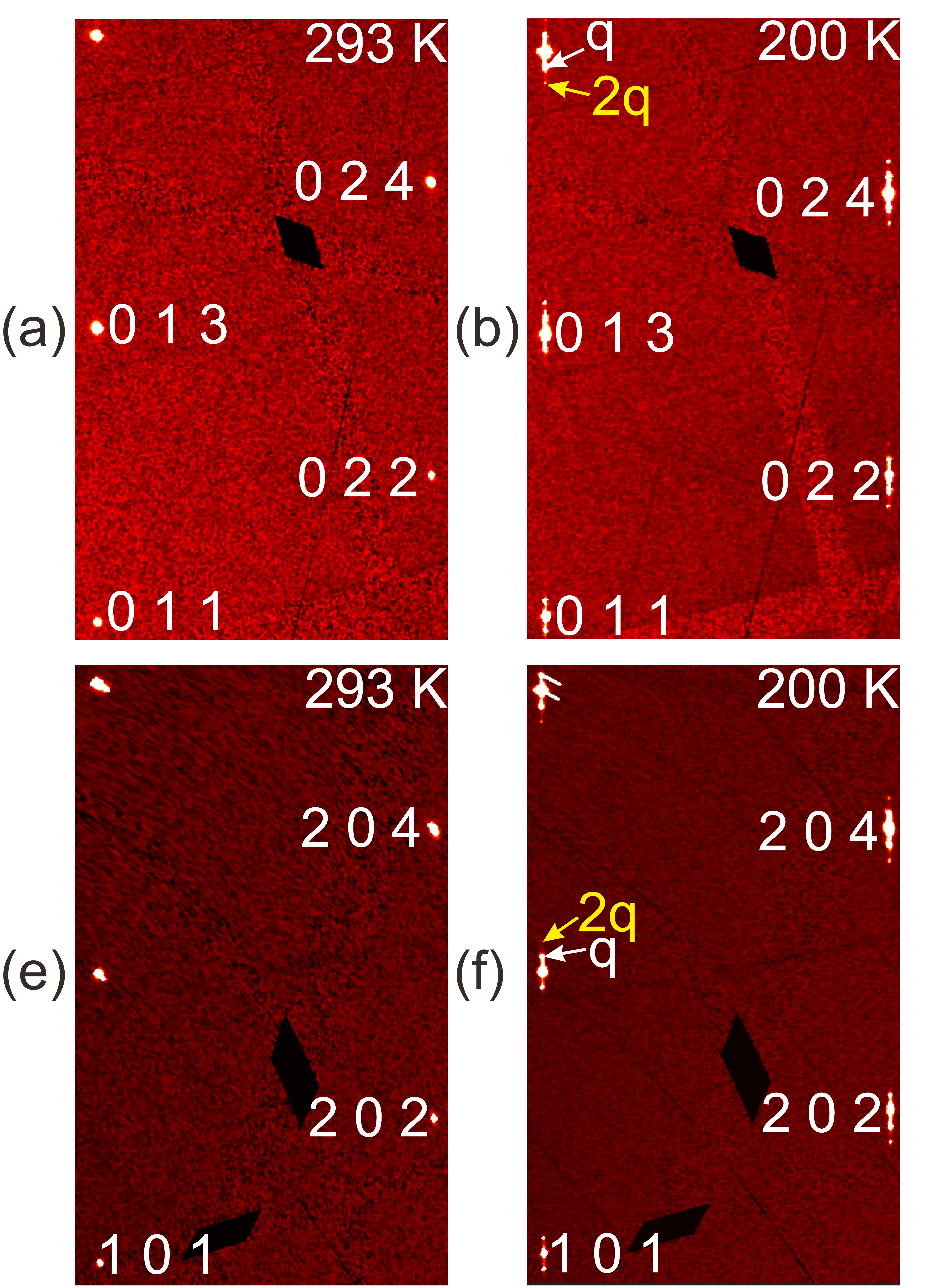}
\includegraphics[width=80mm]{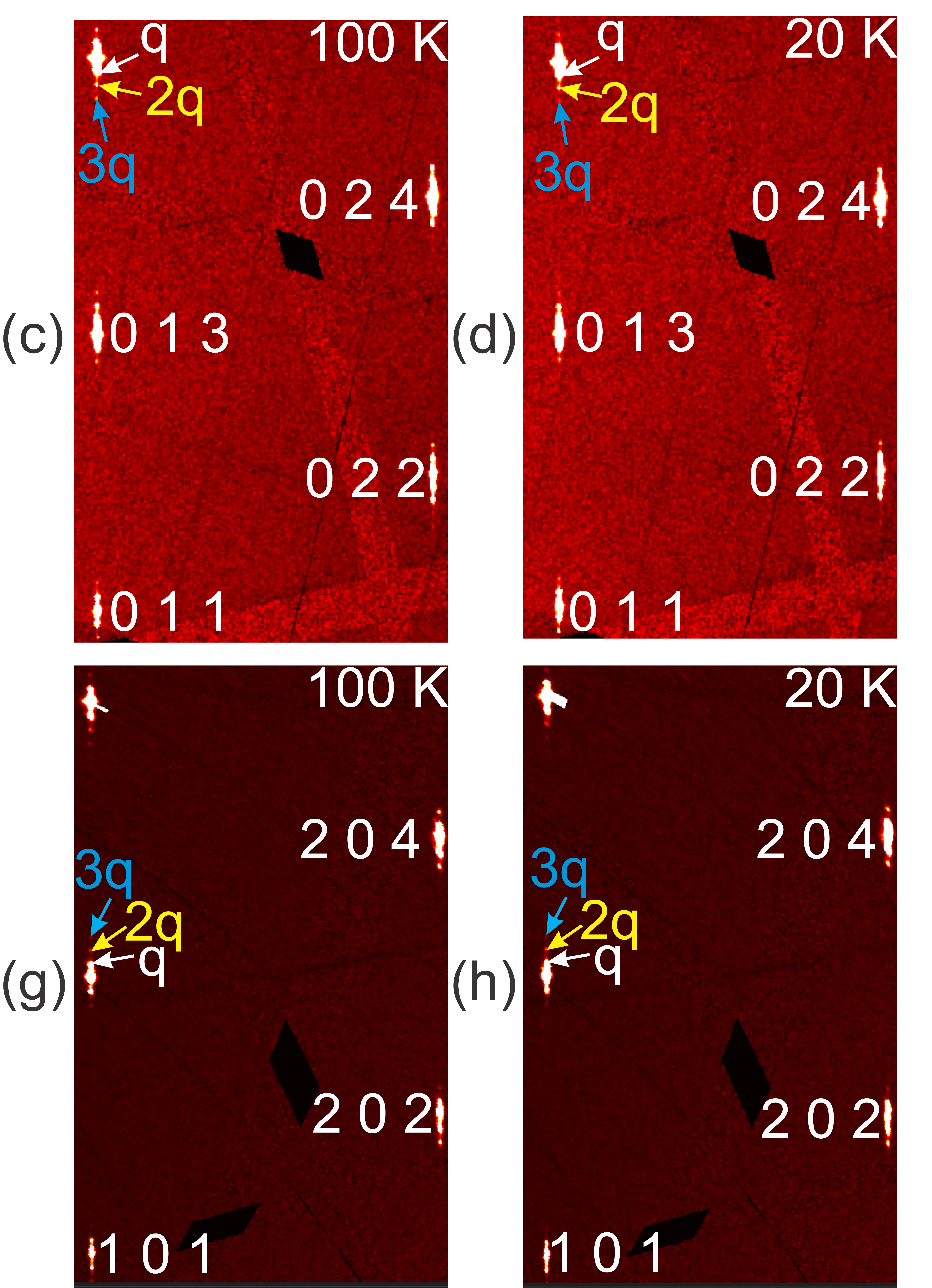}
\caption{\label{fig:sral4_unwarp}%
Excerpts of the reciprocal layers of diffraction for
(a)--(d) $(0\,k\,l)$ and (e)--(f) $(h\,0\,l)$,
as reconstructed from the measured SXRD data
\cite{crysalis}.
(a)--(d) and (e)--(f) display scattering for
temperatures of 293, 200, 100 and 20 K,
as indicated.
Indices are given for selected main reflections.
Selected satellite reflections are indicated by arrows.
Up to $3^{rd}$-order satellites are visible
at 20 and 100 K.
Dark gaps are due to insensitive pixels between
the active modules of the PILATUS3X CdTe 1M detector.
}
\end{figure}

Upon further cooling below 200 K, the length
of the modulation wave vector gradually shrinks
[Fig. \ref{fig:sral4_lattice_qvec}(d)].
As a consequence, main reflections and nearby
satellite reflections are not always resolved
at $120$ K and below (compare the streaky maxima
in Fig. \ref{fig:sral4_unwarp}).
A reliable extraction was not possible of the
values for the integrated intensities of
individual Bragg reflections for $T < 120$ K.

\subsection{\label{sec:sral4_location_CDW}%
Location of the CDW}

\begin{figure}
\includegraphics[height=80mm]{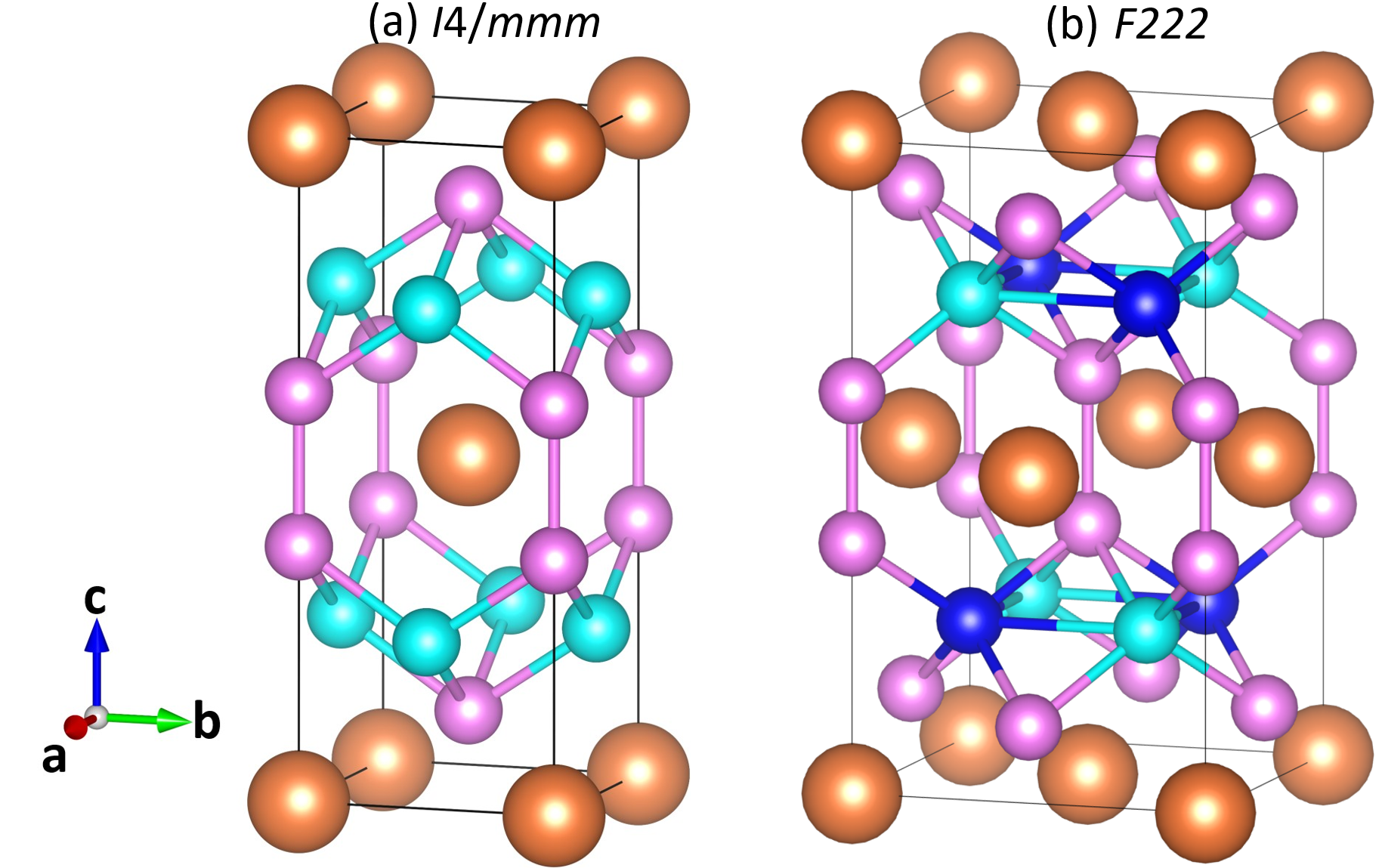}
\caption{\label{fig:sral4_basic_unit_cell}%
(a) Crystal structure of SrAl$_4$ with
space group $I4/mmm$ in the periodic phase at 293 K.
Depicted is the $I$-centered unit cell with basis vectors
$\mathbf{a}_I$, $\mathbf{b}_I$ and $\mathbf{c}_I$.
Orange spheres correspond to the Sr atoms;
green-bluish spheres represent Al1 atoms;
and pink spheres stand for Al2 atoms.
Shortest interatomic distances are:
$d$[Sr--Sr] = 4.4893(4) \AA{},
$d$[Al1--Al1] = 3.1774(2) \AA{},
$d$[Al2--Sr] = 4.3768(11) \AA{},
$d$[Al2--Al1] = 2.7034(6) \AA{} and
$d$[Al2--Al2] = 2.6250(15) \AA{}.
(b) Average crystal structure of SrAl$_4$ with
space group $F222$ in the CDW phase at 200 K.
Depicted is the $F$-centered unit cell with basis vectors
$\mathbf{a}_F$, $\mathbf{b}_F$ and $\mathbf{c}_F$.
In the CDW phase, the Al1 atom splits into the Al1a
and Al1b atoms, where the Al1b atoms are represented
by dark blue spheres.
Shortest interatomic distances are:
$d$[Sr--Sr] = 4.4780(5) \AA{},
$d$[Al1a--Al1a] = 4.4780(5) \AA{},
$d$[Al1a--Al1b] = 3.1663(4) \AA{},
$d$[Al1b--Al1b] = 4.4780(5) \AA{},
$d$[Al2--Sr] = 4.3165(11) \AA{},
$d$[Al2--Al1a] = 2.6967(6) \AA{},
$d$[Al2--Al1b] =  2.6967(6) \AA{} and
$d$[Al2--Al2] = 2.6211(14) \AA{}.
The crystal structure was drawn using VESTA \cite{momma2008vesta}.}
\end{figure}

SrAl$_4$ is isostructural to EuAl$_4$ with
a tetragonal crystal structure at room
temperature \cite{parthe1983a},
that is preserved as basic structure at lower
temperatures (Fig. \ref{fig:sral4_basic_unit_cell}).

The modulation functions exhibit the following
features.
Firstly, for the centrosymmetric model C,
the largest modulation is along $\mathbf{a}_F$.
It has about twice the amplitude in SrAl$_4$,
and values are nearly equal for all three
independent atoms---Sr, Al1 and Al2---like
it is the case for EuAl$_4$
(Table S6 and \cite{ramakrishnan2022a}).
A smaller amplitude is found along $\mathbf{b}_F$
exclusively for the Al1 atom.
Major difference with respect to EuAl$_4$
is the presence of second-order harmonics with
non-zero amplitudes exclusively along $\mathbf{c}_F$.
Although these values are much smaller than
the principal modulation, they might be responsible
for the displacements along $\mathbf{c}$ suggested
by Korshunov \textit{et al.} on the basis of
inelastic x-ray scattering experiments \cite{korshunovan2024a}.
The present results show that the alternative
symmetry of $Immm(0\,0\,\sigma)s00$ \cite{korshunovan2024a}
is not required
for achieving $z$-displacements.

The situation is slightly different for the
noncentrosymmetric model D.
The independent atom Al1 splits into two sites,
Al1a and Al1b, in such a way that Al1a is
bonded to four Al1b and vice versa
[Fig. \ref{fig:sral4_basic_unit_cell}(b)].
Modulations of similar magnitudes are again
present along $\mathbf{a}_F$ for all four
independent atoms.
However, all atoms now possess modulations
along $\mathbf{b}_F$, which are of different
magnitudes, especially Al1a and Al1b
(Table \ref{tab:sral4_f222_ampl}).
The second-order harmonic amplitudes remain
small and represent the $z$ displacement.

\begin{figure}
\includegraphics[width=80mm]{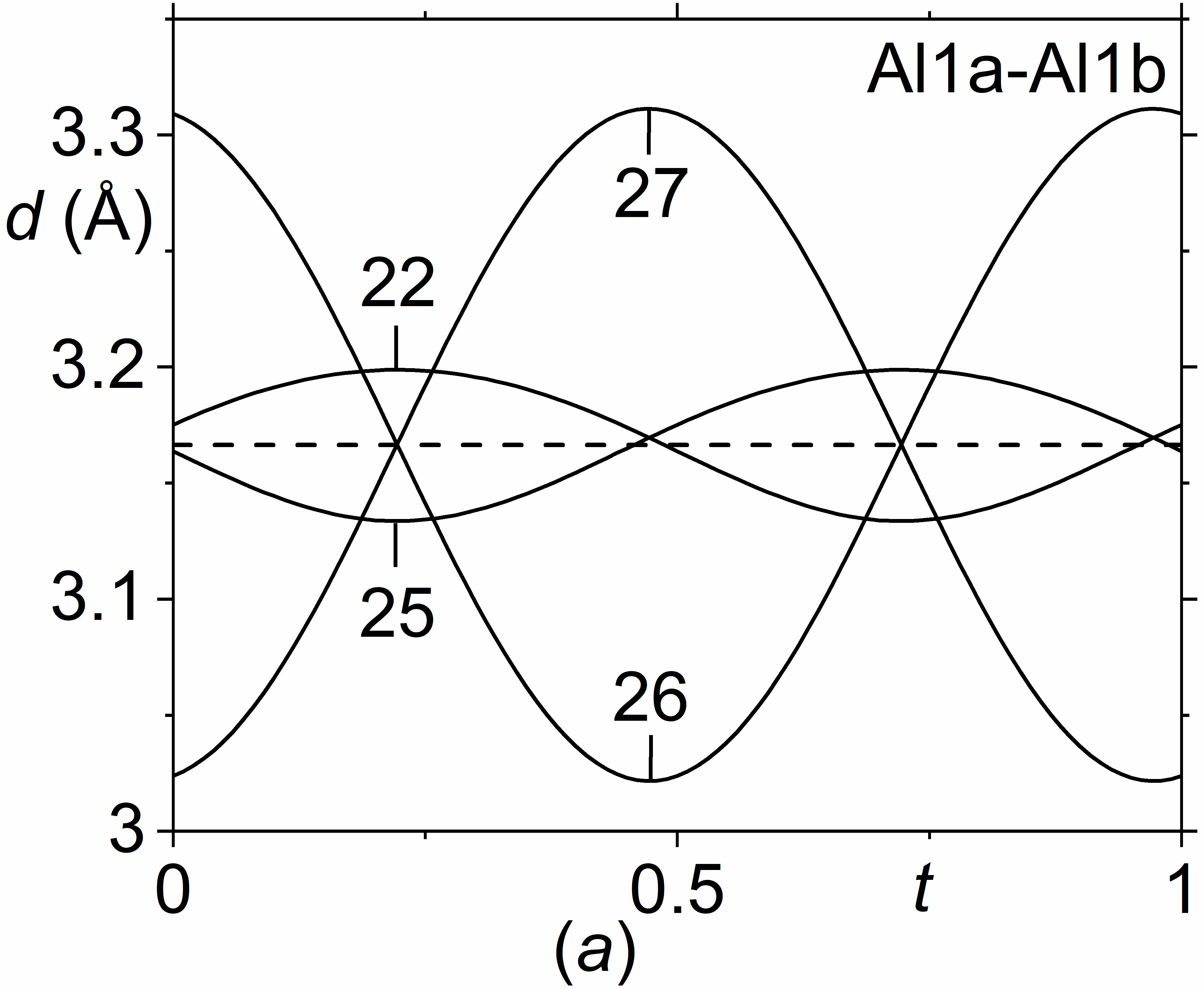}
\hfill
\includegraphics[width=80mm]{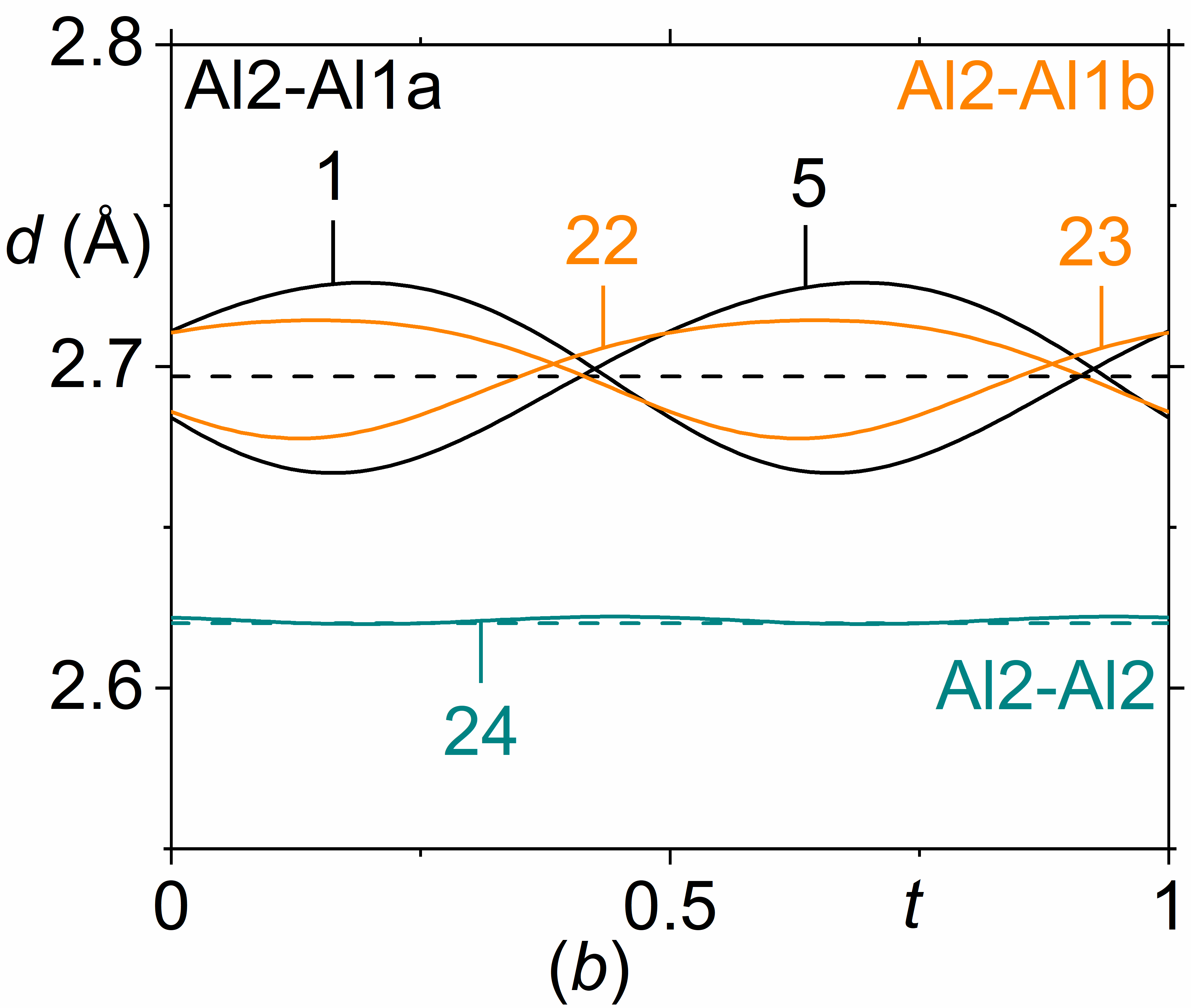}
\caption{\label{fig:sral4_t_plot_aluminium}%
$t$-Plot of interatomic distances (\AA{})
in the CDW phase of SrAl$_4$ at 200 K.
(a) $d$[Al1a--Al1b], and
(b) $d$[Al2--Al1a], $d$[Al2--Al1b] and $d$[Al2--Al2],
where the first atom is the central atom.
$t$-Plots display interatomic
distances as a function of the phase $t$ of the
modulation wave \cite{vansmaalen2007a}.
In case of multiple curves, each value of $t$
gives the distances from a central atom
towards its neighboring atoms.
The number on each curve is the number of the symmetry
operator that is applied to the second atom of the
bond pair.
Symmetry operators are listed in Table S13
in the supporting information.}
\end{figure}

Surprisingly, the different models C and D
as well as the corresponding models for EuAl$_4$
provide a qualitatively similar picture for
the modulations of the interatomic distances.
Largest variation of the bonding contacts is
between Al1a and Al1b, with Al2-Al1a and
Al2-Al1b of secondary importance
(Fig. \ref{fig:sral4_t_plot_aluminium}).
$t$-Plots are given in the Supplementary Material
\cite{sral4suppmat2023a} for other interatomic
distances as well as for the comparison of models
C and D, and for a similar comparison for EuAl$_4$.
We conclude that the Al1 atoms govern the
formation of the CDW in both SrAl$_4$ and EuAl$_4$.
The CDW modulation principally is a transverse
wave, a feature that remains valid for both
centrosymmetric model C and noncentrosymmetric
model D \cite{ramakrishnan2022a}.

\subsection{\label{sec:sral4_criteria_CDW_baal4}%
Criteria for the formation of CDW
in BaAl$_4$ type compounds}

Only a few of the isostructural compounds
$X$Al$_{4-x}$Ga$_{x}$ ($X$ = Ba, Eu, Sr, Ca; $0 < x < 4$)
undergo phase transitions.
First, it is noticed that chemical disorder tends
to suppress a CDW transition, if the disorder is
at sites carrying the CDW.
The transition temperature is lowered upon increasing
doping, while a few percent of doping can be sufficient
to completely suppress the CDW transition
\cite{liuy2010a,yand2019a}.
The sensitivity of the CDW for chemical disorder
explains the absence of CDW transitions in
EuAl$_{4-x}$Ga$_x$ ($0 < x <4$, except
$x=0$ and $x=2$) \cite{stavinoha2018a},
and the absence of CDW transitions in
SrAl$_{4-x}$Si$_x$ and SrAl$_{4-x}$Ge$_x$
\cite{zhang2013a,zevalkink2017a}.

\begin{figure}
\includegraphics[width=80mm]{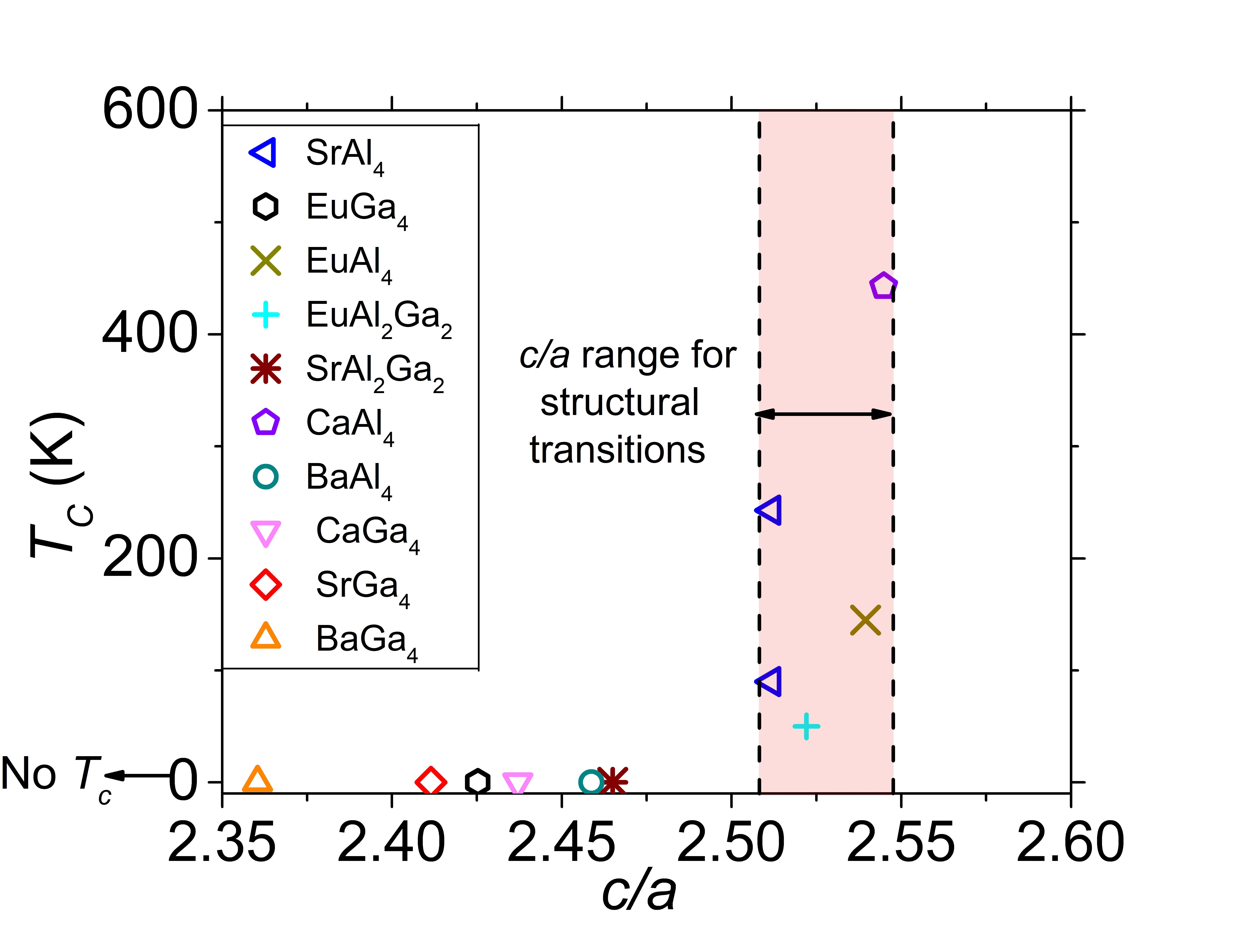}
\caption{\label{fig:sral4_criteria}%
Value of $T_{c}$ = $T_{S}$ or $T_{CDW}$ as a function
of the ratio $c/a$ for eleven compounds
$X$Al$_{4-x}$Ga$_{x}$.
Only four compounds undergo phase transitions.}
\end{figure}

Thus, restricting the analysis to compounds
$X$Al$_{4-x}$Ga$_{x}$ with $x = 0,\:2$ or $4$,
we make the observation that structural and CDW phase
transitions take place in compounds that have
the ratio $c/a$ within a narrow range:
$2.51 < c/a < 2.54$ (Fig. \ref{fig:sral4_criteria}).
EuGa$_4$ has a CDW phase transition for pressures
exceeding 2 GPa \cite{nakamura2015a}.
One might then speculate, that the ratio $c/a$ of
EuGa$_4$ at high pressures will be within the critical range
(Fig. \ref{fig:sral4_criteria}).
This criterion is useful to predict CDWs in new
compounds of this family.
At present there is no information regarding
whether SrAl$_2$Ga$_2$ exhibits CDW or other
structural transitions.

\subsection{\label{sec:sral4_monoclinic}%
Evidence of monoclinic distortion below 100 K}

The orthorhombic phase is realized through
the orthorhombic symmetry of the CDW modulation
displacements (Table \ref{tab:sral4_f222_ampl}),
while the lattice and basic structure remain tetragonal
(Fig. \ref{fig:sral4_lattice_qvec} and Table S4).
The structural transition at $T_{S} = 87$ K is
accompanied by a lattice distortion.
In particular, we find the angle $\beta$ to be
different from 90 deg within the low-temperature
phase [Fig. \ref{fig:sral4_lattice_qvec}(b)].
These values indicate that the low-temperature
phase is $\mathbf{b}$-unique monoclinic,
in agreement with Nakamura \textit{et al.} \cite{nakamura2016a}.

The structural phase transition leads to a
twinned crystal, such that the monoclinic lattice
distortion is visible as split reflections in $q$-scans
along the direction of $\mathbf{c}^{*}$.
For example, for the reflection $(0\,2\,4)$,
a secondary peak appears below $100$ K between the
main reflection and first-order satellite
(Fig. S1 in the Supplemental Material \cite{sral4suppmat2023a}).
Furthermore, these $q$-scans show that the
incommensurate CDW satellite reflections persist
into the low-temperature phase.
Across $T_{S}$, they continue to grow in intensity and
the length of $\mathbf{q}$ continues to decrease
[Fig. \ref{fig:sral4_lattice_qvec}(d)].

Due to overlap between main and satellite reflections,
as it is the result of the short modulation wave
vector [Fig. \ref{fig:sral4_lattice_qvec}(d)],
structure refinements failed.
However, measured intensities are of sufficient
quality to distinguish between tetragonal,
orthorhombic and monoclinic symmetries on the
basis of $R_{int}$ values for averaging equivalent
reflections.
These quantities clearly favor monoclinic, $\mathbf{b}$-unique
symmetry for the low-temperature phase
(Table \ref{tab:sral4 compare rint}).
\begin{table}[ht]
\small
\caption{Comparison of models on basis of SSG and $R_{int}$ at 20 K.
Criterion of observability: $I>3\sigma(I)$}
\label{tab:sral4 compare rint}%
\centering
\begin{ruledtabular}
\begin{tabular}{ccccc}
SSG  &$I4/mmm(0\,0\,\sigma)0000$ & $Fmmm(0\,0\,\sigma)s00$ & $F$2/$m(\sigma{_1}0\sigma{_2})00$ & \\
$R_{int}$ ($m = 0$)(obs/all)$\%$  & 5.07/5.07  & 3.78/3.79 & 1.62/1.62  \\
$R_{int}$ ($m = 1$)(obs/all)$\%$  & 38.47/38.47  & 30.86/30.86  & 5.33/5.35   \\
$R_{int}$ ($m = 2$)(obs/all)$\%$  & 64.42/64.43  & 38.41/38.41   & 6.23/9.16  \\
Unique ($m = 0$) (obs/all)   & 46/50   & 138/202   & 255/344  \\
Unique ($m = 1$) (obs/all)   & 75/106   & 317/381  & 567/732  \\
Unique ($m = 2$) (obs/all)   & 24/108   & 204/405   & 296/741   \\
\end{tabular}
\end{ruledtabular}
\end{table}

\subsection{\label{sec:sral4_band_structure}%
Electronic Band Structure and Phonons}

\begin{figure}
\includegraphics[width=160mm]{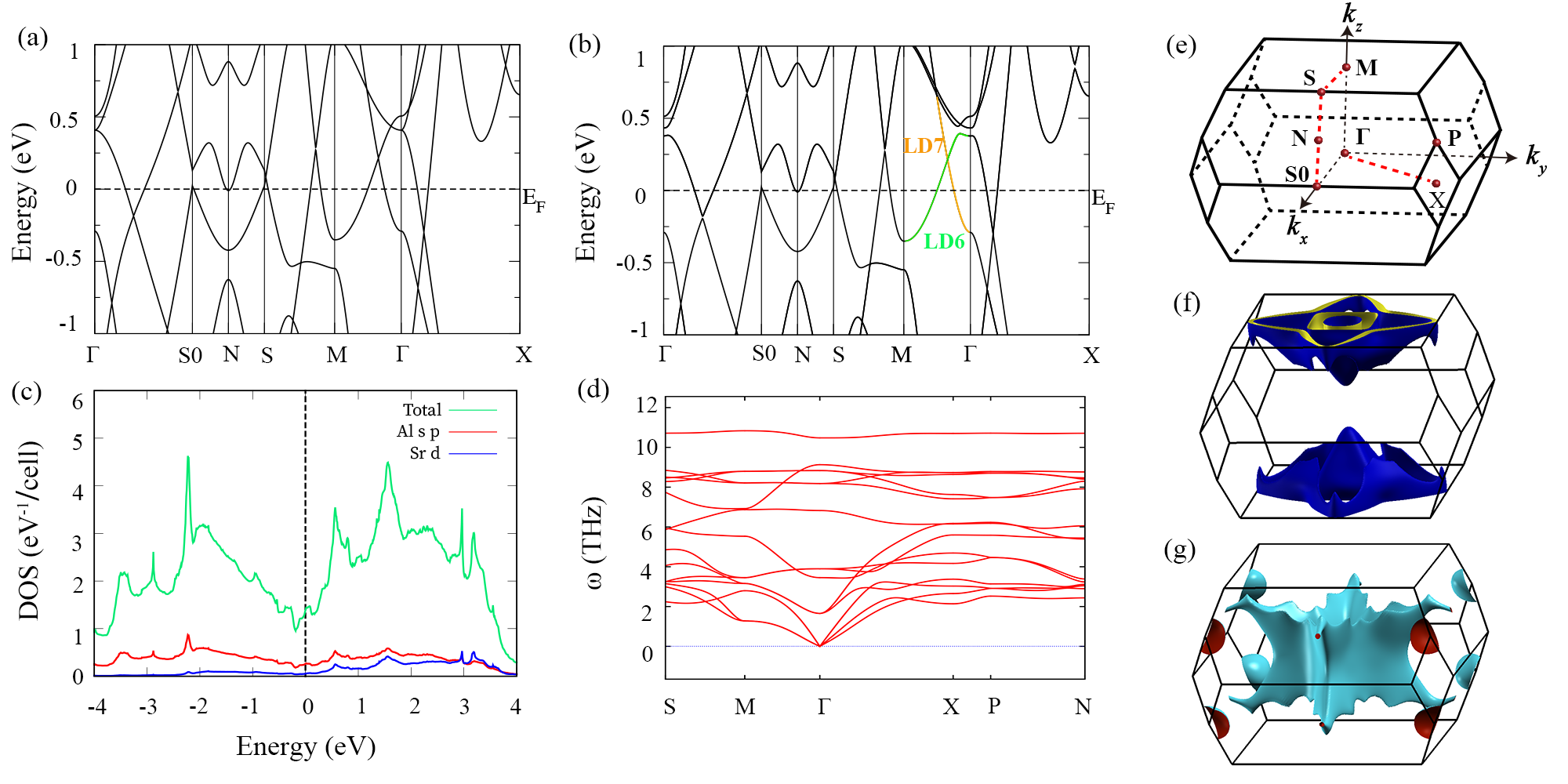}
\caption{\label{fig:sral4_dft}%
The bulk band structures of the tetragonal crystal
structure $I4/mmm$ of SrAl$_4$.
(a) Electronic band structure along high-symmetry
directions, as obtained without SOC.
(b) Electronic band structure calculated with SOC.
The dashed lines indicate the Fermi energy, $E_F$.
All band crossings are gapped out by SOC,
except the crossing of two bands along the M-$\Gamma$
direction, labeled LD6 (green) and LD7 (orange).
(c) The density-of-states (DOS) of the primitive
unit cell with SOC.
The green line gives the total DOS; the red line
gives the contributions of $2s$ and $2p$ orbitals
of Al; and the blue line stands for contributions
of the $3d$ orbital of Sr $d$.
(d) Phonon dispersion relations obtained by DFPT
calculations for a $4\times 4\times 2$ supercell
of the conventional unit cell, containing 320 atoms.
(e) First Brillouin zone of the primitive unit cell.
(f) Hole pockets, and
(g) electron pockets of the Fermi surface.}
\end{figure}

We have calculated the electronic structure
and topological properties of SrAl$_4$ for its
$I4/mmm$ crystal structure.
Dispersion relations along high-symmetry directions
within the Brillouin zone are given
in Fig. \ref{fig:sral4_dft}(a,b).
In the absence of SOC, several band crossings exist
between the highest valence band and the lowest
conduction band near
the Fermi level [Fig. \ref{fig:sral4_dft}(a)].
When the SOC is considered, most band crossings are gapped,
except one crossing along the line
M-$\Gamma$ [Fig. \ref{fig:sral4_dft}(b)].
Unfortunately, it appears at around $E = 0.2$ eV
above the Fermi level, which is not easy to find experimentally.

Using the software IRVSP \cite{Irvsp}, we found that these
two bands belong to the irreducible representations
of point group $4mm$, marked LD6 (green line) and
LD7(orange line), respectively
[Fig. \ref{fig:sral4_dft}(b)].
This indicates that the band crossing is a
topologically protected Dirac point.
It follows that SrAl$_4$ is a Dirac nodal line
semimetal without SOC and becomes a Dirac semimetal
when the SOC is considered,
like EuAl$_4$ \cite{nakamura2015a, ramakrishnan2022a}.

The total and atom-projected density of states (DOS)
near $E_F$ are given in Fig. \ref{fig:sral4_dft}(c).
They reveal that Al states (red line) dominate
the DOS near the Fermi level.
This observation parallels the situation observed
for EuAl$_4$ \cite{ramakrishnan2022a}.
Consequently, it is plausible to infer that the
CDW is associated with the Al atoms rather than the
Sr atoms.

Additional information on the mechanism of CDW
formation in SrAl$_4$ might be obtained from the
phonon dispersion relations and the Fermi surface.
Figure \ref{fig:sral4_dft}(d) shows the phonon
dispersion relations, as computed with aid of a
$4\times 4\times 2$ supercell of the conventional
basic cell, containing 320 atoms.
Density functional perturbative theory (DFPT)
\cite{DFPT1991,DFPT1997}
was used with a $4\times 4\times 3$ $k$-mesh.
No imaginary frequencies or soft modes can be
identified, indicating that electron-phonon
coupling (EPC) is too weak to induce a CDW transition.
This result is further checked by the application
of alternative computational methods.
The phonon dispersion relations were computed
with a highly optimised primitive cell
by the finite displacement method.
Small imaginary frequencies were obtained on
the line M-$\Gamma$,
when a $3\times 3\times 3$ $k$-mesh was used
for the computation
(see Fig. S18(a) in the Supplementary Material
\cite{sral4suppmat2023a}).
These imaginary frequencies are eliminated,
when the number of $k$-points is increased
toward to $5\times 5\times 5$
(Fig. S18(b) in \cite{sral4suppmat2023a}).
The software Quantum Espresso
\cite{Giannozzi_2009,Giannozzi_2017}
leads to the same conclusion
[compare Figs. S18(b),(c) and Fig. \ref{fig:sral4_dft}(d)].

The Fermi surface of SrAl$_4$ comprises hole pockets
centered on the point M [Fig. \ref{fig:sral4_dft}(f)],
as well as electron pockets surrounding $\Gamma$ and
those centered on point P [Fig. \ref{fig:sral4_dft}(g)].
To further check whether Fermi surface nesting (FSN)
could responsible for this CDW,
the bare charge susceptibility \cite{FSN} was calculated,
employing a $k$-mesh of $200\times 200 \times 200$,
and considering four conduction bands and four valence
bands near the Fermi level,
\begin{equation}
\begin{aligned}
\lim _{\omega \rightarrow 0}
\chi_0^{\prime\prime}(\boldsymbol{q}, \omega) / \omega &=
\sum_{n,n', \boldsymbol{k}}
\delta\left(\varepsilon_{n, \boldsymbol{k}}-\varepsilon_F\right)
\delta\left(\varepsilon_{n', \boldsymbol{k}+\boldsymbol{q}}-\varepsilon_F\right) \\
\chi_0^{\prime}(\boldsymbol{q}) &=
\sum_{n,n', \boldsymbol{k}}
\frac{f\left(\varepsilon_{n, \boldsymbol{k}}\right)
-f\left(\varepsilon_{n', \boldsymbol{k}+\boldsymbol{q}}\right)}
{\varepsilon_{n, \boldsymbol{k}}-\varepsilon_{n', \boldsymbol{k}+\boldsymbol{q}}},
\end{aligned}
\end{equation}
where $f\left(\varepsilon_{n, \boldsymbol{k}}\right)$
is the Fermi-Dirac distribution function.
The real and imaginary components of the bare charge
susceptibility of the SrAl$_4$ on the $k_{z}$ = $0.11$ $c^{*}$
plane are depicted as a function of wave vector $\mathbf{q}$
in Fig. S19(a,c) in the Supplemental Material \cite{sral4suppmat2023a}.
Both components exhibit peak values at the center.
We further conducted additional calculations of the bare charge
susceptibility along the $k_{z}$ line with $k_{x}$ = $k_{y}$ =$0$,
and present the outcomes in Fig. S19(b,d).
The real component shows a peak near the experimentally observed
$\mathbf{q}$ vector, whereas the imaginary part does not.
This discrepancy indicates that Fermi surface nesting (FSN)
is insufficient to induce the charge density wave (CDW) in
SrAl$_4$.
These conclusions are consistent with the analysis previously
done for SrAl$_4$ and EuAl$_4$
\cite{nakamura2015a,nakamura2016a,ramakrishnan2022a},
concluding that a clear mechanism of the CDW
cannot be determined directly.

\subsection{\label{sec:sral4_specific_heat}%
Specific heat}

\begin{figure}
\includegraphics[width=80mm]{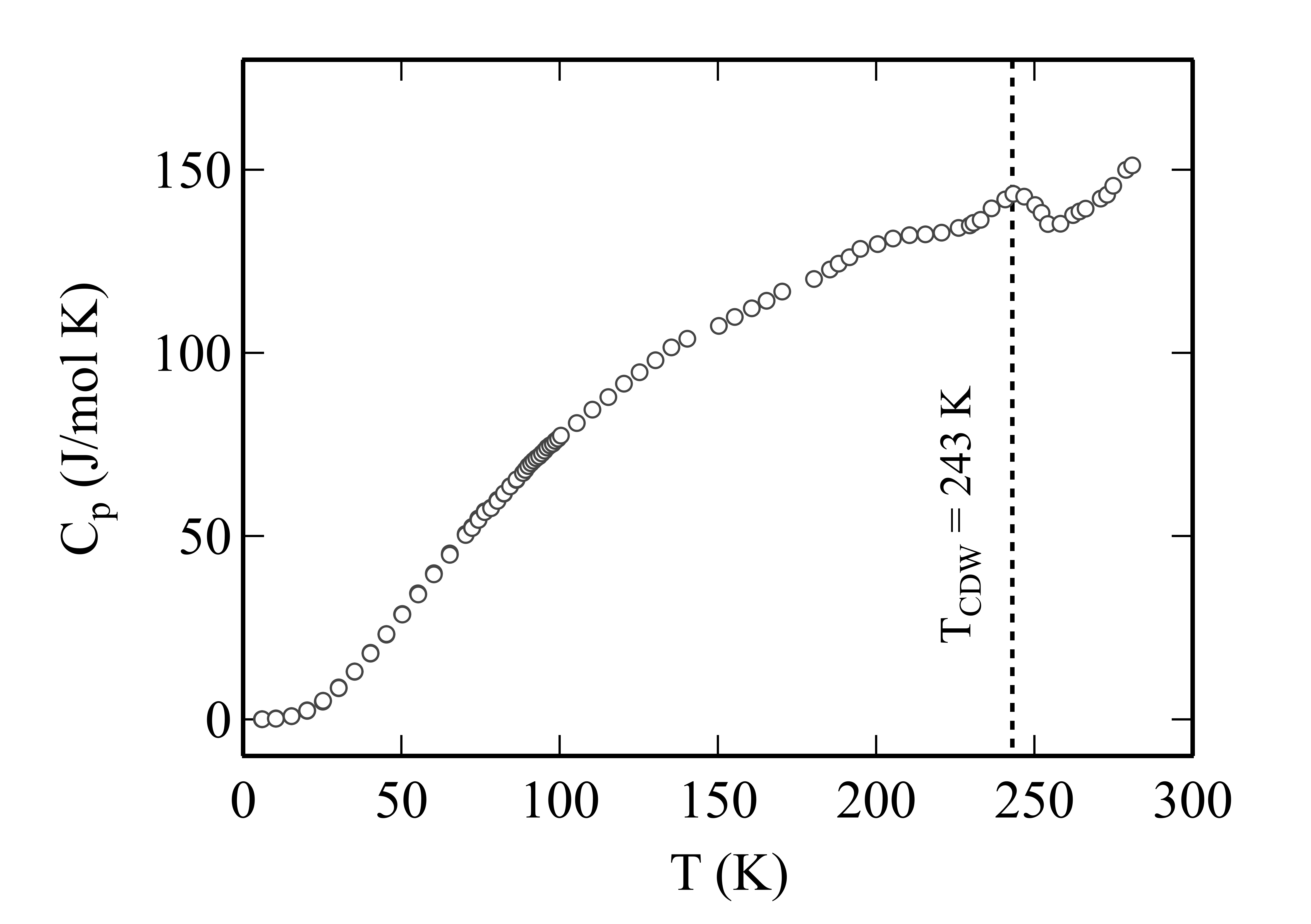}
\caption{\label{fig:sral4_heat_capacity}%
Temperature dependence of the specific heat $C_p$
of SrAl$_4$
for 3--280 K, as obtained during heating of the sample.
A broad maximum is observed
at $T_{CDW} = 243$ K (indicated by a vertical dashed line).
No anomaly could be detected near the second
structure transition at $T_{S}= 87$ K.}
\end{figure}

The temperature dependence of the specific heat ($C_p$)
shows a broad maximum at $T_{CDW}$
of magnitude $\Delta C_p\approx 12$ J\ mol$^{-1}$\ K$^{-1}$
(Fig. \ref{fig:sral4_heat_capacity}),
which is more pronounced than observed for EuAl$_4$
\cite{ramakrishnan2022a}.
The measured $C_p(T)$ data do not show any
anomaly at the temperature $T_{S}$ of the
structural phase transition
(Fig. \ref{fig:sral4_heat_capacity}).
This might be related to the relaxation method
of measurement, as employed in the PPMS instrument
\cite{suzukih2010a}.
Alternatively, the absence of a transition
at $T_S$ might be the result of the
presence of lattice defects in as-grown
material,
as previously described for CuV$_2$S$_4$
\cite{ramakrishnan2019a}.

\section{\label{sec:sral4_conclusions}%
Conclusions}

From SXRD experiments and the physical properties
measurements we confirm that SrAl$_4$ undergoes
two phase transitions.
At $T_{CDW} = 243$ K a CDW transition
takes place, at which the crystal symmetry is
lowered from centrosymmetric tetragonal to
noncentrosymmetric orthorhombic.
The CDW involves a transverse modulation, while
a small longitudinal component exists as
second harmonic modulation, \textit{i.e.} as
secondary modulation.
The variation of interatomic distances along the
incommensurate coordinate demonstrate that
Al1 atoms govern the CDW modulation.
For the structural transition at $T_{S} = 87$ K,
SXRD unambiguously shows that the low-temperature
phase is $\mathbf{b}$-axis monoclinic, and that
the CDW is virtually unaffected by this transition.

Replacement of non-magnetic Sr by magnetic Eu
has a minimal role in the crystal structure,
but it does affect the value of $T_{CDW}$, which
can be explained by different atomic radii of
Sr and Eu.
This idea has been expanded towards CaAl$_4$ and BaAl$_4$.
More generally, isostructural compounds with
the tetragonal BaAl$_4$ structure type exist for
$X$Al$_{4-x}$Ga$_{x}$ ($X$ = Ba, Eu, Sr, Ca; $0 < x < 4$).
We could show that only those compounds
undergo phase transitions for which $c/a$
falls within the narrow range
$2.51 < c/a < 2.54$ (Fig. \ref{fig:sral4_criteria}),
while CDW transitions may be found only in case
of the lack of chemical disorder, \textit{i.e.}
only in compounds with $x = 0,\;2,$ or 4.

Both SXRD and electronic band structure calculations
support the interpretation that the network of Al
atoms is the key contribution to CDW formation,
much like in EuAl$_4$.
The system possibly possesses a non-trivial band
topology and a complex Fermi surface, preventing
the mechanism of CDW formation to be simply
uncovered in SrAl$_4$.

We surmise that it may be possible to realise
multiple phases in EuAl$_4$, as the latter is
proposed to go towards monoclinic symmetry in
the AFM phases \cite{ramakrishnan2022a}.
For CaAl$_4$ there may exist an intermediary
orthorhombic phase between its tetragonal and
monoclinic phases.

\begin{acknowledgments}
Single crystals of SrAl$_4$ were grown by
Kerstin K\"{u}spert at the Laboratory of
Crystallography in Bayreuth,
and by Ruta Kulkarni at the Tata Institute
of Fundamental Research in Mumbai.
We thank Yilin Wang, Chao Cao, Simin Nie and Zhijun Wang
for useful discussions about theoretical
calculations and exploration of CDW mechanisms.
We thank C. Paulmann for
his assistance at Beamline P24.
We acknowledge DESY (Hamburg, Germany),
a member of the Helmholtz Association HGF, for
the provision of experimental facilities.
Parts of this research were carried out at
PETRA III, using beamline P24.
Beamtime was allocated for proposal I-20210805.
J. Chen is supported by China Postdoctoral
Science Foundation (No. 2022M722916).
This work was partly supported by
JSPS KAKENHI Grant Numbers JP21K03448 and JP23H04630.
The research at the University of Bayreuth has been
funded by the Deutsche Forschungsgemeinschaft
(DFG, German Research Foundation)--406658237.
\end{acknowledgments}

\bibliography{SrAl4}

\begin{thebibliography}{86}%
\makeatletter
\providecommand \@ifxundefined [1]{%
 \@ifx{#1\undefined}
}%
\providecommand \@ifnum [1]{%
 \ifnum #1\expandafter \@firstoftwo
 \else \expandafter \@secondoftwo
 \fi
}%
\providecommand \@ifx [1]{%
 \ifx #1\expandafter \@firstoftwo
 \else \expandafter \@secondoftwo
 \fi
}%
\providecommand \natexlab [1]{#1}%
\providecommand \enquote  [1]{``#1''}%
\providecommand \bibnamefont  [1]{#1}%
\providecommand \bibfnamefont [1]{#1}%
\providecommand \citenamefont [1]{#1}%
\providecommand \href@noop [0]{\@secondoftwo}%
\providecommand \href [0]{\begingroup \@sanitize@url \@href}%
\providecommand \@href[1]{\@@startlink{#1}\@@href}%
\providecommand \@@href[1]{\endgroup#1\@@endlink}%
\providecommand \@sanitize@url [0]{\catcode `\\12\catcode `\$12\catcode
  `\&12\catcode `\#12\catcode `\^12\catcode `\_12\catcode `\%12\relax}%
\providecommand \@@startlink[1]{}%
\providecommand \@@endlink[0]{}%
\providecommand \url  [0]{\begingroup\@sanitize@url \@url }%
\providecommand \@url [1]{\endgroup\@href {#1}{\urlprefix }}%
\providecommand \urlprefix  [0]{URL }%
\providecommand \Eprint [0]{\href }%
\providecommand \doibase [0]{https://doi.org/}%
\providecommand \selectlanguage [0]{\@gobble}%
\providecommand \bibinfo  [0]{\@secondoftwo}%
\providecommand \bibfield  [0]{\@secondoftwo}%
\providecommand \translation [1]{[#1]}%
\providecommand \BibitemOpen [0]{}%
\providecommand \bibitemStop [0]{}%
\providecommand \bibitemNoStop [0]{.\EOS\space}%
\providecommand \EOS [0]{\spacefactor3000\relax}%
\providecommand \BibitemShut  [1]{\csname bibitem#1\endcsname}%
\let\auto@bib@innerbib\@empty
\bibitem [{\citenamefont {Gr\"uner}(1988)}]{gruner1988a}%
  \BibitemOpen
  \bibfield  {author} {\bibinfo {author} {\bibfnamefont {G.}~\bibnamefont
  {Gr\"uner}},\ }\bibfield  {title} {\bibinfo {title} {The dynamics of
  charge-density waves},\ }\href@noop {} {\bibfield  {journal} {\bibinfo
  {journal} {Rev. Mod. Phys.}\ }\textbf {\bibinfo {volume} {60}},\ \bibinfo
  {pages} {1129} (\bibinfo {year} {1988})}\BibitemShut {NoStop}%
\bibitem [{\citenamefont {Gr\"{u}ner}(1994)}]{gruener1994a}%
  \BibitemOpen
  \bibfield  {author} {\bibinfo {author} {\bibfnamefont {G.}~\bibnamefont
  {Gr\"{u}ner}},\ }\href@noop {} {\emph {\bibinfo {title} {Charge Density Waves
  in Solids}}}\ (\bibinfo  {publisher} {Addison-Wesley},\ \bibinfo {address}
  {Reading, Massachusetts},\ \bibinfo {year} {1994})\BibitemShut {NoStop}%
\bibitem [{\citenamefont {Monceau}(2012)}]{monceau2012a}%
  \BibitemOpen
  \bibfield  {author} {\bibinfo {author} {\bibfnamefont {P.}~\bibnamefont
  {Monceau}},\ }\bibfield  {title} {\bibinfo {title} {Electronic crystals: an
  experimental overview},\ }\href@noop {} {\bibfield  {journal} {\bibinfo
  {journal} {Adv. Phys.}\ }\textbf {\bibinfo {volume} {61}},\ \bibinfo {pages}
  {325} (\bibinfo {year} {2012})}\BibitemShut {NoStop}%
\bibitem [{\citenamefont {Zhu}\ \emph {et~al.}(2017)\citenamefont {Zhu},
  \citenamefont {Guo}, \citenamefont {Zhang},\ and\ \citenamefont
  {Plummer}}]{zhu2017a}%
  \BibitemOpen
  \bibfield  {author} {\bibinfo {author} {\bibfnamefont {X.}~\bibnamefont
  {Zhu}}, \bibinfo {author} {\bibfnamefont {J.}~\bibnamefont {Guo}}, \bibinfo
  {author} {\bibfnamefont {J.}~\bibnamefont {Zhang}},\ and\ \bibinfo {author}
  {\bibfnamefont {E.~W.}\ \bibnamefont {Plummer}},\ }\bibfield  {title}
  {\bibinfo {title} {Misconceptions associated with the origin of charge
  density waves},\ }\href@noop {} {\bibfield  {journal} {\bibinfo  {journal}
  {Adv. Phys.: X}\ }\textbf {\bibinfo {volume} {2}},\ \bibinfo {pages} {622}
  (\bibinfo {year} {2017})}\BibitemShut {NoStop}%
\bibitem [{\citenamefont {Fleming}\ \emph {et~al.}(1981)\citenamefont
  {Fleming}, \citenamefont {DiSalvo}, \citenamefont {Cava},\ and\ \citenamefont
  {Waszczak}}]{flemingrm1981a}%
  \BibitemOpen
  \bibfield  {author} {\bibinfo {author} {\bibfnamefont {R.~M.}\ \bibnamefont
  {Fleming}}, \bibinfo {author} {\bibfnamefont {F.~J.}\ \bibnamefont
  {DiSalvo}}, \bibinfo {author} {\bibfnamefont {R.~J.}\ \bibnamefont {Cava}},\
  and\ \bibinfo {author} {\bibfnamefont {J.~V.}\ \bibnamefont {Waszczak}},\
  }\bibfield  {title} {\bibinfo {title} {Observation of charge-density waves in
  the cubic spinel structure {CuV$_2$S$_4$}},\ }\href
  {https://doi.org/10.1103/PhysRevB.24.2850} {\bibfield  {journal} {\bibinfo
  {journal} {Phys. Rev. B}\ }\textbf {\bibinfo {volume} {24}},\ \bibinfo
  {pages} {2850} (\bibinfo {year} {1981})}\BibitemShut {NoStop}%
\bibitem [{\citenamefont {Kawaguchi}\ \emph {et~al.}(2012)\citenamefont
  {Kawaguchi}, \citenamefont {Kubota}, \citenamefont {Tsuji}, \citenamefont
  {Kim}, \citenamefont {Kato}, \citenamefont {Takata},\ and\ \citenamefont
  {Ishibashi}}]{kawaguchi2012a}%
  \BibitemOpen
  \bibfield  {author} {\bibinfo {author} {\bibfnamefont {S.}~\bibnamefont
  {Kawaguchi}}, \bibinfo {author} {\bibfnamefont {Y.}~\bibnamefont {Kubota}},
  \bibinfo {author} {\bibfnamefont {N.}~\bibnamefont {Tsuji}}, \bibinfo
  {author} {\bibfnamefont {J.}~\bibnamefont {Kim}}, \bibinfo {author}
  {\bibfnamefont {K.}~\bibnamefont {Kato}}, \bibinfo {author} {\bibfnamefont
  {M.}~\bibnamefont {Takata}},\ and\ \bibinfo {author} {\bibfnamefont
  {H.}~\bibnamefont {Ishibashi}},\ }\bibfield  {title} {\bibinfo {title}
  {Structural analysis of spinel compound {CuV$_2$S$_4$} with incommensurate
  charge-density wave},\ }\href
  {https://doi.org/10.1088/1742-6596/391/1/012095} {\bibfield  {journal}
  {\bibinfo  {journal} {J. Phys. Conf. Ser.}\ }\textbf {\bibinfo {volume}
  {391}},\ \bibinfo {pages} {012095} (\bibinfo {year} {2012})}\BibitemShut
  {NoStop}%
\bibitem [{\citenamefont {Okada}\ \emph {et~al.}(2004)\citenamefont {Okada},
  \citenamefont {Koyama},\ and\ \citenamefont {Watanabe}}]{okadah2004a}%
  \BibitemOpen
  \bibfield  {author} {\bibinfo {author} {\bibfnamefont {H.}~\bibnamefont
  {Okada}}, \bibinfo {author} {\bibfnamefont {K.}~\bibnamefont {Koyama}},\ and\
  \bibinfo {author} {\bibfnamefont {K.}~\bibnamefont {Watanabe}},\ }\bibfield
  {title} {\bibinfo {title} {Two-step structural modulations and {F}ermi liquid
  state in spinel compound {CuV$_2$S$_4$}},\ }\href
  {https://doi.org/10.1143/jpsj.73.3227} {\bibfield  {journal} {\bibinfo
  {journal} {J. Phys. Soc. Jpn}\ }\textbf {\bibinfo {volume} {73}},\ \bibinfo
  {pages} {3227} (\bibinfo {year} {2004})}\BibitemShut {NoStop}%
\bibitem [{\citenamefont {Ramakrishnan}\ \emph {et~al.}(2019)\citenamefont
  {Ramakrishnan}, \citenamefont {Sch\"onleber}, \citenamefont {H\"ubschle},
  \citenamefont {Eisele}, \citenamefont {Schaller}, \citenamefont {Rekis},
  \citenamefont {Bui}, \citenamefont {Feulner}, \citenamefont {van Smaalen},
  \citenamefont {Bag}, \citenamefont {Ramakrishnan}, \citenamefont {Tolkiehn},\
  and\ \citenamefont {Paulmann}}]{ramakrishnan2019a}%
  \BibitemOpen
  \bibfield  {author} {\bibinfo {author} {\bibfnamefont {S.}~\bibnamefont
  {Ramakrishnan}}, \bibinfo {author} {\bibfnamefont {A.}~\bibnamefont
  {Sch\"onleber}}, \bibinfo {author} {\bibfnamefont {C.~B.}\ \bibnamefont
  {H\"ubschle}}, \bibinfo {author} {\bibfnamefont {C.}~\bibnamefont {Eisele}},
  \bibinfo {author} {\bibfnamefont {A.~M.}\ \bibnamefont {Schaller}}, \bibinfo
  {author} {\bibfnamefont {T.}~\bibnamefont {Rekis}}, \bibinfo {author}
  {\bibfnamefont {N.~H.~A.}\ \bibnamefont {Bui}}, \bibinfo {author}
  {\bibfnamefont {F.}~\bibnamefont {Feulner}}, \bibinfo {author} {\bibfnamefont
  {S.}~\bibnamefont {van Smaalen}}, \bibinfo {author} {\bibfnamefont
  {B.}~\bibnamefont {Bag}}, \bibinfo {author} {\bibfnamefont {S.}~\bibnamefont
  {Ramakrishnan}}, \bibinfo {author} {\bibfnamefont {M.}~\bibnamefont
  {Tolkiehn}},\ and\ \bibinfo {author} {\bibfnamefont {C.}~\bibnamefont
  {Paulmann}},\ }\bibfield  {title} {\bibinfo {title} {Charge density wave and
  lock-in transitions of {CuV$_{2}$S$_{4}$}},\ }\href
  {https://doi.org/10.1103/PhysRevB.99.195140} {\bibfield  {journal} {\bibinfo
  {journal} {Phys. Rev. B}\ }\textbf {\bibinfo {volume} {99}},\ \bibinfo
  {pages} {195140} (\bibinfo {year} {2019})}\BibitemShut {NoStop}%
\bibitem [{\citenamefont {Slebarski}\ and\ \citenamefont
  {Goraus}(2013)}]{slebarski2013a}%
  \BibitemOpen
  \bibfield  {author} {\bibinfo {author} {\bibfnamefont {A.}~\bibnamefont
  {Slebarski}}\ and\ \bibinfo {author} {\bibfnamefont {J.}~\bibnamefont
  {Goraus}},\ }\bibfield  {title} {\bibinfo {title} {Electronic structure and
  crystallographic properties of skutterudite-related {Ce$_{3}M_{4}$Sn$_{13}$}
  and {La$_{3}M_{4}$Sn$_{13}$} ({$M$} = {Co,} {Ru,} and {Rh})},\ }\href@noop {}
  {\bibfield  {journal} {\bibinfo  {journal} {Phys. Rev. B}\ }\textbf {\bibinfo
  {volume} {88}},\ \bibinfo {pages} {155122} (\bibinfo {year}
  {2013})}\BibitemShut {NoStop}%
\bibitem [{\citenamefont {Otomo}\ \emph {et~al.}(2016)\citenamefont {Otomo},
  \citenamefont {Iwasa}, \citenamefont {Suyama}, \citenamefont {Tomiyasu},
  \citenamefont {Sagayama}, \citenamefont {Sagayama}, \citenamefont {Nakao},
  \citenamefont {Kumai},\ and\ \citenamefont {Murakami}}]{otomo2016a}%
  \BibitemOpen
  \bibfield  {author} {\bibinfo {author} {\bibfnamefont {Y.}~\bibnamefont
  {Otomo}}, \bibinfo {author} {\bibfnamefont {K.}~\bibnamefont {Iwasa}},
  \bibinfo {author} {\bibfnamefont {K.}~\bibnamefont {Suyama}}, \bibinfo
  {author} {\bibfnamefont {K.}~\bibnamefont {Tomiyasu}}, \bibinfo {author}
  {\bibfnamefont {H.}~\bibnamefont {Sagayama}}, \bibinfo {author}
  {\bibfnamefont {R.}~\bibnamefont {Sagayama}}, \bibinfo {author}
  {\bibfnamefont {H.}~\bibnamefont {Nakao}}, \bibinfo {author} {\bibfnamefont
  {R.}~\bibnamefont {Kumai}},\ and\ \bibinfo {author} {\bibfnamefont
  {Y.}~\bibnamefont {Murakami}},\ }\bibfield  {title} {\bibinfo {title} {Chiral
  crystal-structure transformation of {$R_{3}$Co$_{4}$Sn$_{13}$} ({$R$} = {La}
  and {Ce})},\ }\href@noop {} {\bibfield  {journal} {\bibinfo  {journal} {Phys.
  Rev. B}\ }\textbf {\bibinfo {volume} {94}},\ \bibinfo {pages} {075109}
  (\bibinfo {year} {2016})}\BibitemShut {NoStop}%
\bibitem [{\citenamefont {Welsch}\ \emph {et~al.}(2019)\citenamefont {Welsch},
  \citenamefont {Ramakrishnan}, \citenamefont {Eisele}, \citenamefont {van
  Well}, \citenamefont {Sch\"onleber}, \citenamefont {van Smaalen},
  \citenamefont {Matteppanavar}, \citenamefont {Thamizhavel}, \citenamefont
  {Tolkiehn}, \citenamefont {Paulmann},\ and\ \citenamefont
  {Ramakrishnan}}]{welsch2019a}%
  \BibitemOpen
  \bibfield  {author} {\bibinfo {author} {\bibfnamefont {J.}~\bibnamefont
  {Welsch}}, \bibinfo {author} {\bibfnamefont {S.}~\bibnamefont
  {Ramakrishnan}}, \bibinfo {author} {\bibfnamefont {C.}~\bibnamefont
  {Eisele}}, \bibinfo {author} {\bibfnamefont {N.}~\bibnamefont {van Well}},
  \bibinfo {author} {\bibfnamefont {A.}~\bibnamefont {Sch\"onleber}}, \bibinfo
  {author} {\bibfnamefont {S.}~\bibnamefont {van Smaalen}}, \bibinfo {author}
  {\bibfnamefont {S.}~\bibnamefont {Matteppanavar}}, \bibinfo {author}
  {\bibfnamefont {A.}~\bibnamefont {Thamizhavel}}, \bibinfo {author}
  {\bibfnamefont {M.}~\bibnamefont {Tolkiehn}}, \bibinfo {author}
  {\bibfnamefont {C.}~\bibnamefont {Paulmann}},\ and\ \bibinfo {author}
  {\bibfnamefont {S.}~\bibnamefont {Ramakrishnan}},\ }\bibfield  {title}
  {\bibinfo {title} {Second-order charge-density-wave transition in single
  crystals of {La$_{3}$Co$_{4}$Sn$_{13}$}},\ }\href
  {https://doi.org/10.1103/PhysRevMaterials.3.125003} {\bibfield  {journal}
  {\bibinfo  {journal} {Phys. Rev. Materials}\ }\textbf {\bibinfo {volume}
  {3}},\ \bibinfo {pages} {125003} (\bibinfo {year} {2019})}\BibitemShut
  {NoStop}%
\bibitem [{\citenamefont {Ramakrishnan}\ and\ \citenamefont {van
  Smaalen}(2017)}]{ramakrishnan2017a}%
  \BibitemOpen
  \bibfield  {author} {\bibinfo {author} {\bibfnamefont {S.}~\bibnamefont
  {Ramakrishnan}}\ and\ \bibinfo {author} {\bibfnamefont {S.}~\bibnamefont {van
  Smaalen}},\ }\bibfield  {title} {\bibinfo {title} {Unusual ground states in
  {$R_5T_4X_{10}$} ({$R$} = rare earth; {$T$} = {Rh,} {Ir}; and {$X$} = {Si,}
  {Ge,} {Sn}): a review},\ }\href
  {https://doi.org/https://doi.org/10.1088/1361-6633/aa7d5f} {\bibfield
  {journal} {\bibinfo  {journal} {Rep. Prog. Phys.}\ }\textbf {\bibinfo
  {volume} {80}},\ \bibinfo {pages} {116501} (\bibinfo {year}
  {2017})}\BibitemShut {NoStop}%
\bibitem [{\citenamefont {Bugaris}\ \emph {et~al.}(2017)\citenamefont
  {Bugaris}, \citenamefont {Malliakas}, \citenamefont {Han}, \citenamefont
  {Calta}, \citenamefont {Sturza}, \citenamefont {Krogstad}, \citenamefont
  {Osborn}, \citenamefont {Rosenkranz}, \citenamefont {Ruff}, \citenamefont
  {Trimarchi}, \citenamefont {Bud’ko}, \citenamefont {Balasubramanian},
  \citenamefont {Chung},\ and\ \citenamefont {Kanatzidis}}]{bugaris2017a}%
  \BibitemOpen
  \bibfield  {author} {\bibinfo {author} {\bibfnamefont {D.~E.}\ \bibnamefont
  {Bugaris}}, \bibinfo {author} {\bibfnamefont {C.~D.}\ \bibnamefont
  {Malliakas}}, \bibinfo {author} {\bibfnamefont {F.}~\bibnamefont {Han}},
  \bibinfo {author} {\bibfnamefont {N.~P.}\ \bibnamefont {Calta}}, \bibinfo
  {author} {\bibfnamefont {M.}~\bibnamefont {Sturza}}, \bibinfo {author}
  {\bibfnamefont {M.~J.}\ \bibnamefont {Krogstad}}, \bibinfo {author}
  {\bibfnamefont {R.}~\bibnamefont {Osborn}}, \bibinfo {author} {\bibfnamefont
  {S.}~\bibnamefont {Rosenkranz}}, \bibinfo {author} {\bibfnamefont {J.~P.~C.}\
  \bibnamefont {Ruff}}, \bibinfo {author} {\bibfnamefont {G.}~\bibnamefont
  {Trimarchi}}, \bibinfo {author} {\bibfnamefont {S.~L.}\ \bibnamefont
  {Bud’ko}}, \bibinfo {author} {\bibfnamefont {M.}~\bibnamefont
  {Balasubramanian}}, \bibinfo {author} {\bibfnamefont {D.~Y.}\ \bibnamefont
  {Chung}},\ and\ \bibinfo {author} {\bibfnamefont {M.~G.}\ \bibnamefont
  {Kanatzidis}},\ }\bibfield  {title} {\bibinfo {title} {Charge density wave in
  the new polymorphs of {RE$_2$Ru$_3$Ge$_5$} ({RE} = {Pr,} {Sm,} {Dy})},\
  }\href@noop {} {\bibfield  {journal} {\bibinfo  {journal} {J. Amer. Chem.
  Soc.}\ }\textbf {\bibinfo {volume} {139}},\ \bibinfo {pages} {4130} (\bibinfo
  {year} {2017})}\BibitemShut {NoStop}%
\bibitem [{\citenamefont {Kuo}\ \emph {et~al.}(2020)\citenamefont {Kuo},
  \citenamefont {Hsu}, \citenamefont {Tseng}, \citenamefont {Chen},
  \citenamefont {Lin}, \citenamefont {Liu}, \citenamefont {Kuo},\ and\
  \citenamefont {Lue}}]{kuo2020a}%
  \BibitemOpen
  \bibfield  {author} {\bibinfo {author} {\bibfnamefont {C.~N.}\ \bibnamefont
  {Kuo}}, \bibinfo {author} {\bibfnamefont {C.~J.}\ \bibnamefont {Hsu}},
  \bibinfo {author} {\bibfnamefont {C.~W.}\ \bibnamefont {Tseng}}, \bibinfo
  {author} {\bibfnamefont {W.~T.}\ \bibnamefont {Chen}}, \bibinfo {author}
  {\bibfnamefont {S.~Y.}\ \bibnamefont {Lin}}, \bibinfo {author} {\bibfnamefont
  {W.~Z.}\ \bibnamefont {Liu}}, \bibinfo {author} {\bibfnamefont {Y.~K.}\
  \bibnamefont {Kuo}},\ and\ \bibinfo {author} {\bibfnamefont {C.~S.}\
  \bibnamefont {Lue}},\ }\bibfield  {title} {\bibinfo {title} {Charge density
  wave like behavior with magnetic ordering in orthorhombic
  {Sm$_{2}$Ru$_{3}$Ge$_{5}$}},\ }\href
  {https://doi.org/10.1103/PhysRevB.101.155140} {\bibfield  {journal} {\bibinfo
   {journal} {Phys. Rev. B}\ }\textbf {\bibinfo {volume} {101}},\ \bibinfo
  {pages} {155140} (\bibinfo {year} {2020})}\BibitemShut {NoStop}%
\bibitem [{\citenamefont {Kolincio}\ \emph {et~al.}(2020)\citenamefont
  {Kolincio}, \citenamefont {Roman},\ and\ \citenamefont
  {Klimczuk}}]{kolincio2020a}%
  \BibitemOpen
  \bibfield  {author} {\bibinfo {author} {\bibfnamefont {K.~K.}\ \bibnamefont
  {Kolincio}}, \bibinfo {author} {\bibfnamefont {M.}~\bibnamefont {Roman}},\
  and\ \bibinfo {author} {\bibfnamefont {T.}~\bibnamefont {Klimczuk}},\
  }\bibfield  {title} {\bibinfo {title} {Enhanced mobility and large linear
  nonsaturating magnetoresistance in the magnetically ordered states of
  {TmNiC$_{2}$}},\ }\href {https://doi.org/10.1103/PhysRevLett.125.176601}
  {\bibfield  {journal} {\bibinfo  {journal} {Phys. Rev. Lett.}\ }\textbf
  {\bibinfo {volume} {125}},\ \bibinfo {pages} {176601} (\bibinfo {year}
  {2020})}\BibitemShut {NoStop}%
\bibitem [{\citenamefont {Roman}\ \emph {et~al.}(2023)\citenamefont {Roman},
  \citenamefont {Fritthum}, \citenamefont {St\"oger}, \citenamefont {Adroja},\
  and\ \citenamefont {Michor}}]{roman2023a}%
  \BibitemOpen
  \bibfield  {author} {\bibinfo {author} {\bibfnamefont {M.}~\bibnamefont
  {Roman}}, \bibinfo {author} {\bibfnamefont {M.}~\bibnamefont {Fritthum}},
  \bibinfo {author} {\bibfnamefont {B.}~\bibnamefont {St\"oger}}, \bibinfo
  {author} {\bibfnamefont {D.~T.}\ \bibnamefont {Adroja}},\ and\ \bibinfo
  {author} {\bibfnamefont {H.}~\bibnamefont {Michor}},\ }\bibfield  {title}
  {\bibinfo {title} {Charge density wave and crystalline electric field effects
  in {TmNiC$_2$}},\ }\href@noop {} {\bibfield  {journal} {\bibinfo  {journal}
  {Phys. Rev. B}\ }\textbf {\bibinfo {volume} {107}},\ \bibinfo {pages}
  {125137} (\bibinfo {year} {2023})}\BibitemShut {NoStop}%
\bibitem [{\citenamefont {Ramakrishnan}\ \emph {et~al.}(2021)\citenamefont
  {Ramakrishnan}, \citenamefont {Sch\"onleber}, \citenamefont {Bao},
  \citenamefont {Rekis}, \citenamefont {Kotla}, \citenamefont {Schaller},
  \citenamefont {van Smaalen}, \citenamefont {Noohinejad}, \citenamefont
  {Tolkiehn}, \citenamefont {Paulmann}, \citenamefont {Sangeetha},
  \citenamefont {Pal}, \citenamefont {Thamizhavel},\ and\ \citenamefont
  {Ramakrishnan}}]{ramakrishnan2021a}%
  \BibitemOpen
  \bibfield  {author} {\bibinfo {author} {\bibfnamefont {S.}~\bibnamefont
  {Ramakrishnan}}, \bibinfo {author} {\bibfnamefont {A.}~\bibnamefont
  {Sch\"onleber}}, \bibinfo {author} {\bibfnamefont {J.-K.}\ \bibnamefont
  {Bao}}, \bibinfo {author} {\bibfnamefont {T.}~\bibnamefont {Rekis}}, \bibinfo
  {author} {\bibfnamefont {S.~R.}\ \bibnamefont {Kotla}}, \bibinfo {author}
  {\bibfnamefont {A.~M.}\ \bibnamefont {Schaller}}, \bibinfo {author}
  {\bibfnamefont {S.}~\bibnamefont {van Smaalen}}, \bibinfo {author}
  {\bibfnamefont {L.}~\bibnamefont {Noohinejad}}, \bibinfo {author}
  {\bibfnamefont {M.}~\bibnamefont {Tolkiehn}}, \bibinfo {author}
  {\bibfnamefont {C.}~\bibnamefont {Paulmann}}, \bibinfo {author}
  {\bibfnamefont {N.~S.}\ \bibnamefont {Sangeetha}}, \bibinfo {author}
  {\bibfnamefont {D.}~\bibnamefont {Pal}}, \bibinfo {author} {\bibfnamefont
  {A.}~\bibnamefont {Thamizhavel}},\ and\ \bibinfo {author} {\bibfnamefont
  {S.}~\bibnamefont {Ramakrishnan}},\ }\bibfield  {title} {\bibinfo {title}
  {Modulated crystal structure of the atypical charge density wave state of
  single-crystal {Lu$_{2}$Ir$_{3}$Si$_{5}$}},\ }\href@noop {} {\bibfield
  {journal} {\bibinfo  {journal} {Phys. Rev. B}\ }\textbf {\bibinfo {volume}
  {104}},\ \bibinfo {pages} {054116} (\bibinfo {year} {2021})}\BibitemShut
  {NoStop}%
\bibitem [{\citenamefont {Ramakrishnan}\ \emph {et~al.}(2020)\citenamefont
  {Ramakrishnan}, \citenamefont {Sch\"onleber}, \citenamefont {Rekis},
  \citenamefont {van Well}, \citenamefont {Noohinejad}, \citenamefont {van
  Smaalen}, \citenamefont {Tolkiehn}, \citenamefont {Paulmann}, \citenamefont
  {Bag}, \citenamefont {Thamizhavel}, \citenamefont {Pal},\ and\ \citenamefont
  {Ramakrishnan}}]{ramakrishnan2020a}%
  \BibitemOpen
  \bibfield  {author} {\bibinfo {author} {\bibfnamefont {S.}~\bibnamefont
  {Ramakrishnan}}, \bibinfo {author} {\bibfnamefont {A.}~\bibnamefont
  {Sch\"onleber}}, \bibinfo {author} {\bibfnamefont {T.}~\bibnamefont {Rekis}},
  \bibinfo {author} {\bibfnamefont {N.}~\bibnamefont {van Well}}, \bibinfo
  {author} {\bibfnamefont {L.}~\bibnamefont {Noohinejad}}, \bibinfo {author}
  {\bibfnamefont {S.}~\bibnamefont {van Smaalen}}, \bibinfo {author}
  {\bibfnamefont {M.}~\bibnamefont {Tolkiehn}}, \bibinfo {author}
  {\bibfnamefont {C.}~\bibnamefont {Paulmann}}, \bibinfo {author}
  {\bibfnamefont {B.}~\bibnamefont {Bag}}, \bibinfo {author} {\bibfnamefont
  {A.}~\bibnamefont {Thamizhavel}}, \bibinfo {author} {\bibfnamefont
  {D.}~\bibnamefont {Pal}},\ and\ \bibinfo {author} {\bibfnamefont
  {S.}~\bibnamefont {Ramakrishnan}},\ }\bibfield  {title} {\bibinfo {title}
  {Unusual charge density wave transition and absence of magnetic ordering in
  {Er$_{2}$Ir$_{3}$Si$_{5}$}},\ }\href
  {https://doi.org/10.1103/PhysRevB.101.060101} {\bibfield  {journal} {\bibinfo
   {journal} {Phys. Rev. B}\ }\textbf {\bibinfo {volume} {101}},\ \bibinfo
  {pages} {060101(R)} (\bibinfo {year} {2020})}\BibitemShut {NoStop}%
\bibitem [{\citenamefont {Ramakrishnan}\ \emph {et~al.}(2023)\citenamefont
  {Ramakrishnan}, \citenamefont {Bao}, \citenamefont {Eisele}, \citenamefont
  {Patra}, \citenamefont {Nohara}, \citenamefont {Bag}, \citenamefont
  {Noohinejad}, \citenamefont {Tolkiehn}, \citenamefont {Paulmann},
  \citenamefont {Schaller}, \citenamefont {Rekis}, \citenamefont {Kotla},
  \citenamefont {Schonleber}, \citenamefont {Thamizhavel}, \citenamefont
  {Singh}, \citenamefont {Ramakrishnan},\ and\ \citenamefont {van
  Smaalen}}]{ramakrishnan2023a}%
  \BibitemOpen
  \bibfield  {author} {\bibinfo {author} {\bibfnamefont {S.}~\bibnamefont
  {Ramakrishnan}}, \bibinfo {author} {\bibfnamefont {J.}~\bibnamefont {Bao}},
  \bibinfo {author} {\bibfnamefont {C.}~\bibnamefont {Eisele}}, \bibinfo
  {author} {\bibfnamefont {B.}~\bibnamefont {Patra}}, \bibinfo {author}
  {\bibfnamefont {M.}~\bibnamefont {Nohara}}, \bibinfo {author} {\bibfnamefont
  {B.}~\bibnamefont {Bag}}, \bibinfo {author} {\bibfnamefont {L.}~\bibnamefont
  {Noohinejad}}, \bibinfo {author} {\bibfnamefont {M.}~\bibnamefont
  {Tolkiehn}}, \bibinfo {author} {\bibfnamefont {C.}~\bibnamefont {Paulmann}},
  \bibinfo {author} {\bibfnamefont {A.~M.}\ \bibnamefont {Schaller}}, \bibinfo
  {author} {\bibfnamefont {T.}~\bibnamefont {Rekis}}, \bibinfo {author}
  {\bibfnamefont {S.~R.}\ \bibnamefont {Kotla}}, \bibinfo {author}
  {\bibfnamefont {A.}~\bibnamefont {Schonleber}}, \bibinfo {author}
  {\bibfnamefont {A.}~\bibnamefont {Thamizhavel}}, \bibinfo {author}
  {\bibfnamefont {B.}~\bibnamefont {Singh}}, \bibinfo {author} {\bibfnamefont
  {S.}~\bibnamefont {Ramakrishnan}},\ and\ \bibinfo {author} {\bibfnamefont
  {S.}~\bibnamefont {van Smaalen}},\ }\bibfield  {title} {\bibinfo {title}
  {Coupling between charge density wave ordering and magnetism in
  {Ho$_2$Ir$_3$Si$_5$}},\ }\href@noop {} {\bibfield  {journal} {\bibinfo
  {journal} {Chem. Mater.}\ }\textbf {\bibinfo {volume} {35}},\ \bibinfo
  {pages} {1980} (\bibinfo {year} {2023})}\BibitemShut {NoStop}%
\bibitem [{\citenamefont {Zeng}\ \emph {et~al.}(2022)\citenamefont {Zeng},
  \citenamefont {Hu}, \citenamefont {Wang}, \citenamefont {Sun}, \citenamefont
  {Yang}, \citenamefont {Boubeche}, \citenamefont {Luo}, \citenamefont {He},
  \citenamefont {Cheng}, \citenamefont {Yao},\ and\ \citenamefont
  {Luo}}]{zeng2022a}%
  \BibitemOpen
  \bibfield  {author} {\bibinfo {author} {\bibfnamefont {L.}~\bibnamefont
  {Zeng}}, \bibinfo {author} {\bibfnamefont {X.}~\bibnamefont {Hu}}, \bibinfo
  {author} {\bibfnamefont {N.}~\bibnamefont {Wang}}, \bibinfo {author}
  {\bibfnamefont {J.}~\bibnamefont {Sun}}, \bibinfo {author} {\bibfnamefont
  {P.}~\bibnamefont {Yang}}, \bibinfo {author} {\bibfnamefont {M.}~\bibnamefont
  {Boubeche}}, \bibinfo {author} {\bibfnamefont {S.}~\bibnamefont {Luo}},
  \bibinfo {author} {\bibfnamefont {Y.}~\bibnamefont {He}}, \bibinfo {author}
  {\bibfnamefont {J.}~\bibnamefont {Cheng}}, \bibinfo {author} {\bibfnamefont
  {D.-X.}\ \bibnamefont {Yao}},\ and\ \bibinfo {author} {\bibfnamefont
  {H.}~\bibnamefont {Luo}},\ }\bibfield  {title} {\bibinfo {title} {Interplay
  between charge-density-wave, superconductivity, and ferromagnetism in
  {CuIr$_{2–x}$Cr$_x$Te$_4$} chalcogenides},\ }\href@noop {} {\bibfield
  {journal} {\bibinfo  {journal} {J. Phys. Chem. Lett.}\ }\textbf {\bibinfo
  {volume} {13}},\ \bibinfo {pages} {2442} (\bibinfo {year}
  {2022})}\BibitemShut {NoStop}%
\bibitem [{\citenamefont {Guguchia}\ \emph {et~al.}(2023)\citenamefont
  {Guguchia}, \citenamefont {Mielke}, \citenamefont {Das}, \citenamefont
  {Gupta}, \citenamefont {Yin}, \citenamefont {Liu}, \citenamefont {Yin},
  \citenamefont {Christensen}, \citenamefont {Tu}, \citenamefont {Gong},
  \citenamefont {Shumiya}, \citenamefont {Hossain}, \citenamefont
  {Gamsakhurdashvili}, \citenamefont {Elender}, \citenamefont {Dai},
  \citenamefont {Amato}, \citenamefont {Shi}, \citenamefont {Lei},
  \citenamefont {Fernandes}, \citenamefont {Hasan}, \citenamefont {Luetkens},\
  and\ \citenamefont {Khasanov}}]{guguchia2023a}%
  \BibitemOpen
  \bibfield  {author} {\bibinfo {author} {\bibfnamefont {Z.}~\bibnamefont
  {Guguchia}}, \bibinfo {author} {\bibfnamefont {C.}~\bibnamefont {Mielke}},
  \bibinfo {author} {\bibfnamefont {D.}~\bibnamefont {Das}}, \bibinfo {author}
  {\bibfnamefont {R.}~\bibnamefont {Gupta}}, \bibinfo {author} {\bibfnamefont
  {J.-X.}\ \bibnamefont {Yin}}, \bibinfo {author} {\bibfnamefont
  {H.}~\bibnamefont {Liu}}, \bibinfo {author} {\bibfnamefont {Q.}~\bibnamefont
  {Yin}}, \bibinfo {author} {\bibfnamefont {M.~H.}\ \bibnamefont
  {Christensen}}, \bibinfo {author} {\bibfnamefont {Z.}~\bibnamefont {Tu}},
  \bibinfo {author} {\bibfnamefont {C.}~\bibnamefont {Gong}}, \bibinfo {author}
  {\bibfnamefont {N.}~\bibnamefont {Shumiya}}, \bibinfo {author} {\bibfnamefont
  {M.~S.}\ \bibnamefont {Hossain}}, \bibinfo {author} {\bibfnamefont
  {T.}~\bibnamefont {Gamsakhurdashvili}}, \bibinfo {author} {\bibfnamefont
  {M.}~\bibnamefont {Elender}}, \bibinfo {author} {\bibfnamefont
  {P.}~\bibnamefont {Dai}}, \bibinfo {author} {\bibfnamefont {A.}~\bibnamefont
  {Amato}}, \bibinfo {author} {\bibfnamefont {Y.}~\bibnamefont {Shi}}, \bibinfo
  {author} {\bibfnamefont {H.~C.}\ \bibnamefont {Lei}}, \bibinfo {author}
  {\bibfnamefont {R.~M.}\ \bibnamefont {Fernandes}}, \bibinfo {author}
  {\bibfnamefont {M.~Z.}\ \bibnamefont {Hasan}}, \bibinfo {author}
  {\bibfnamefont {H.}~\bibnamefont {Luetkens}},\ and\ \bibinfo {author}
  {\bibfnamefont {R.}~\bibnamefont {Khasanov}},\ }\bibfield  {title} {\bibinfo
  {title} {Tunable unconventional kagome superconductivity in charge ordered
  {RbV$_3$Sb$_5$} and {KV$_3$Sb$_5$}},\ }\href@noop {} {\bibfield  {journal}
  {\bibinfo  {journal} {Nature Commun.}\ }\textbf {\bibinfo {volume} {14}},\
  \bibinfo {pages} {153} (\bibinfo {year} {2023})}\BibitemShut {NoStop}%
\bibitem [{\citenamefont {Kautzsch}\ \emph {et~al.}(2023)\citenamefont
  {Kautzsch}, \citenamefont {Ortiz}, \citenamefont {Mallayya}, \citenamefont
  {Plumb}, \citenamefont {Pokharel}, \citenamefont {Ruff}, \citenamefont
  {Islam}, \citenamefont {Kim}, \citenamefont {Seshadri},\ and\ \citenamefont
  {Wilson}}]{kautzsch2023a}%
  \BibitemOpen
  \bibfield  {author} {\bibinfo {author} {\bibfnamefont {L.}~\bibnamefont
  {Kautzsch}}, \bibinfo {author} {\bibfnamefont {B.~R.}\ \bibnamefont {Ortiz}},
  \bibinfo {author} {\bibfnamefont {K.}~\bibnamefont {Mallayya}}, \bibinfo
  {author} {\bibfnamefont {J.}~\bibnamefont {Plumb}}, \bibinfo {author}
  {\bibfnamefont {G.}~\bibnamefont {Pokharel}}, \bibinfo {author}
  {\bibfnamefont {J.~P.~C.}\ \bibnamefont {Ruff}}, \bibinfo {author}
  {\bibfnamefont {Z.}~\bibnamefont {Islam}}, \bibinfo {author} {\bibfnamefont
  {E.-A.}\ \bibnamefont {Kim}}, \bibinfo {author} {\bibfnamefont
  {R.}~\bibnamefont {Seshadri}},\ and\ \bibinfo {author} {\bibfnamefont
  {S.~D.}\ \bibnamefont {Wilson}},\ }\bibfield  {title} {\bibinfo {title}
  {Structural evolution of the kagome superconductors {AV$_3$Sb$_5$} ({A} =
  {K,} {Rb,} and {Cs}) through charge density wave order},\ }\href@noop {}
  {\bibfield  {journal} {\bibinfo  {journal} {Phys. Rev. Mater.}\ }\textbf
  {\bibinfo {volume} {7}},\ \bibinfo {pages} {024806} (\bibinfo {year}
  {2023})}\BibitemShut {NoStop}%
\bibitem [{\citenamefont {Xiao}\ \emph {et~al.}(2023)\citenamefont {Xiao},
  \citenamefont {Lin}, \citenamefont {Li}, \citenamefont {Zheng}, \citenamefont
  {Francoual}, \citenamefont {Plueckthun}, \citenamefont {Xia}, \citenamefont
  {Qiu}, \citenamefont {Zhang}, \citenamefont {Guo}, \citenamefont {Feng},\
  and\ \citenamefont {Peng}}]{xiao2023a}%
  \BibitemOpen
  \bibfield  {author} {\bibinfo {author} {\bibfnamefont {Q.}~\bibnamefont
  {Xiao}}, \bibinfo {author} {\bibfnamefont {Y.}~\bibnamefont {Lin}}, \bibinfo
  {author} {\bibfnamefont {Q.}~\bibnamefont {Li}}, \bibinfo {author}
  {\bibfnamefont {X.}~\bibnamefont {Zheng}}, \bibinfo {author} {\bibfnamefont
  {S.}~\bibnamefont {Francoual}}, \bibinfo {author} {\bibfnamefont
  {C.}~\bibnamefont {Plueckthun}}, \bibinfo {author} {\bibfnamefont
  {W.}~\bibnamefont {Xia}}, \bibinfo {author} {\bibfnamefont {Q.}~\bibnamefont
  {Qiu}}, \bibinfo {author} {\bibfnamefont {S.}~\bibnamefont {Zhang}}, \bibinfo
  {author} {\bibfnamefont {Y.}~\bibnamefont {Guo}}, \bibinfo {author}
  {\bibfnamefont {J.}~\bibnamefont {Feng}},\ and\ \bibinfo {author}
  {\bibfnamefont {Y.}~\bibnamefont {Peng}},\ }\bibfield  {title} {\bibinfo
  {title} {Coexistence of multiple stacking charge density waves in kagome
  superconductor {CsV$_3$Sb$_5$}},\ }\href@noop {} {\bibfield  {journal}
  {\bibinfo  {journal} {Phys. Rev. Res.}\ }\textbf {\bibinfo {volume} {5}},\
  \bibinfo {pages} {L012032} (\bibinfo {year} {2023})}\BibitemShut {NoStop}%
\bibitem [{\citenamefont {Zhou}\ \emph
  {et~al.}(2023{\natexlab{a}})\citenamefont {Zhou}, \citenamefont {Li},
  \citenamefont {Fan}, \citenamefont {Hao}, \citenamefont {Xiang},
  \citenamefont {Liu}, \citenamefont {Dai}, \citenamefont {Wang}, \citenamefont
  {Yao},\ and\ \citenamefont {Wen}}]{zhou2023a}%
  \BibitemOpen
  \bibfield  {author} {\bibinfo {author} {\bibfnamefont {X.}~\bibnamefont
  {Zhou}}, \bibinfo {author} {\bibfnamefont {Y.}~\bibnamefont {Li}}, \bibinfo
  {author} {\bibfnamefont {X.}~\bibnamefont {Fan}}, \bibinfo {author}
  {\bibfnamefont {J.}~\bibnamefont {Hao}}, \bibinfo {author} {\bibfnamefont
  {Y.}~\bibnamefont {Xiang}}, \bibinfo {author} {\bibfnamefont
  {Z.}~\bibnamefont {Liu}}, \bibinfo {author} {\bibfnamefont {Y.}~\bibnamefont
  {Dai}}, \bibinfo {author} {\bibfnamefont {Z.}~\bibnamefont {Wang}}, \bibinfo
  {author} {\bibfnamefont {Y.}~\bibnamefont {Yao}},\ and\ \bibinfo {author}
  {\bibfnamefont {H.-H.}\ \bibnamefont {Wen}},\ }\bibfield  {title} {\bibinfo
  {title} {Electronic correlations and evolution of the charge density wave in
  the kagome metals {AV$_3$Sb$_5$} ({A} = {K,} {Rb,} and {Cs})},\ }\href@noop
  {} {\bibfield  {journal} {\bibinfo  {journal} {Phys. Rev. B}\ }\textbf
  {\bibinfo {volume} {107}},\ \bibinfo {pages} {165123} (\bibinfo {year}
  {2023}{\natexlab{a}})}\BibitemShut {NoStop}%
\bibitem [{\citenamefont {Wang}\ \emph
  {et~al.}(2023{\natexlab{a}})\citenamefont {Wang}, \citenamefont {Wu},
  \citenamefont {Li}, \citenamefont {Jiang},\ and\ \citenamefont
  {Hu}}]{wang2023b}%
  \BibitemOpen
  \bibfield  {author} {\bibinfo {author} {\bibfnamefont {Y.}~\bibnamefont
  {Wang}}, \bibinfo {author} {\bibfnamefont {T.}~\bibnamefont {Wu}}, \bibinfo
  {author} {\bibfnamefont {Z.}~\bibnamefont {Li}}, \bibinfo {author}
  {\bibfnamefont {K.}~\bibnamefont {Jiang}},\ and\ \bibinfo {author}
  {\bibfnamefont {J.}~\bibnamefont {Hu}},\ }\bibfield  {title} {\bibinfo
  {title} {Structure of the kagome superconductor {CsV$_3$Sb$_5$} in the charge
  density wave state},\ }\href@noop {} {\bibfield  {journal} {\bibinfo
  {journal} {Phys. Rev. B}\ }\textbf {\bibinfo {volume} {107}},\ \bibinfo
  {pages} {184106} (\bibinfo {year} {2023}{\natexlab{a}})}\BibitemShut
  {NoStop}%
\bibitem [{\citenamefont {Yang}\ \emph {et~al.}(2023)\citenamefont {Yang},
  \citenamefont {Xia}, \citenamefont {Mi}, \citenamefont {Zhang}, \citenamefont
  {Gan}, \citenamefont {Wang}, \citenamefont {Chai}, \citenamefont {Zhou},
  \citenamefont {Yang}, \citenamefont {Guo},\ and\ \citenamefont
  {He}}]{yang2023a}%
  \BibitemOpen
  \bibfield  {author} {\bibinfo {author} {\bibfnamefont {K.}~\bibnamefont
  {Yang}}, \bibinfo {author} {\bibfnamefont {W.}~\bibnamefont {Xia}}, \bibinfo
  {author} {\bibfnamefont {X.}~\bibnamefont {Mi}}, \bibinfo {author}
  {\bibfnamefont {L.}~\bibnamefont {Zhang}}, \bibinfo {author} {\bibfnamefont
  {Y.}~\bibnamefont {Gan}}, \bibinfo {author} {\bibfnamefont {A.}~\bibnamefont
  {Wang}}, \bibinfo {author} {\bibfnamefont {Y.}~\bibnamefont {Chai}}, \bibinfo
  {author} {\bibfnamefont {X.}~\bibnamefont {Zhou}}, \bibinfo {author}
  {\bibfnamefont {X.}~\bibnamefont {Yang}}, \bibinfo {author} {\bibfnamefont
  {Y.}~\bibnamefont {Guo}},\ and\ \bibinfo {author} {\bibfnamefont
  {M.}~\bibnamefont {He}},\ }\bibfield  {title} {\bibinfo {title} {Charge
  fluctuations above $t_{CDW}$ revealed by glasslike thermal transport in
  kagome metals {AV$_3$Sb$_5$} ({A} = {K,} {Rb,} and {Cs})},\ }\href@noop {}
  {\bibfield  {journal} {\bibinfo  {journal} {Phys. Rev. B}\ }\textbf {\bibinfo
  {volume} {107}},\ \bibinfo {pages} {184506} (\bibinfo {year}
  {2023})}\BibitemShut {NoStop}%
\bibitem [{\citenamefont {Teng}\ \emph {et~al.}(2022)\citenamefont {Teng},
  \citenamefont {Chen}, \citenamefont {Ye}, \citenamefont {Rosenberg},
  \citenamefont {Liu}, \citenamefont {Yin}, \citenamefont {Jiang},
  \citenamefont {Oh}, \citenamefont {Hasan}, \citenamefont {Neubauer},
  \citenamefont {Gao}, \citenamefont {Xie}, \citenamefont {Hashimoto},
  \citenamefont {Lu}, \citenamefont {Jozwiak}, \citenamefont {Bostwick},
  \citenamefont {Rotenberg}, \citenamefont {Birgeneau}, \citenamefont {Chu},
  \citenamefont {Yi},\ and\ \citenamefont {Dai}}]{teng2022a}%
  \BibitemOpen
  \bibfield  {author} {\bibinfo {author} {\bibfnamefont {X.}~\bibnamefont
  {Teng}}, \bibinfo {author} {\bibfnamefont {L.}~\bibnamefont {Chen}}, \bibinfo
  {author} {\bibfnamefont {F.}~\bibnamefont {Ye}}, \bibinfo {author}
  {\bibfnamefont {E.}~\bibnamefont {Rosenberg}}, \bibinfo {author}
  {\bibfnamefont {Z.}~\bibnamefont {Liu}}, \bibinfo {author} {\bibfnamefont
  {J.-X.}\ \bibnamefont {Yin}}, \bibinfo {author} {\bibfnamefont {Y.-X.}\
  \bibnamefont {Jiang}}, \bibinfo {author} {\bibfnamefont {J.~S.}\ \bibnamefont
  {Oh}}, \bibinfo {author} {\bibfnamefont {M.~Z.}\ \bibnamefont {Hasan}},
  \bibinfo {author} {\bibfnamefont {K.~J.}\ \bibnamefont {Neubauer}}, \bibinfo
  {author} {\bibfnamefont {B.}~\bibnamefont {Gao}}, \bibinfo {author}
  {\bibfnamefont {Y.}~\bibnamefont {Xie}}, \bibinfo {author} {\bibfnamefont
  {M.}~\bibnamefont {Hashimoto}}, \bibinfo {author} {\bibfnamefont
  {D.}~\bibnamefont {Lu}}, \bibinfo {author} {\bibfnamefont {C.}~\bibnamefont
  {Jozwiak}}, \bibinfo {author} {\bibfnamefont {A.}~\bibnamefont {Bostwick}},
  \bibinfo {author} {\bibfnamefont {E.}~\bibnamefont {Rotenberg}}, \bibinfo
  {author} {\bibfnamefont {R.~J.}\ \bibnamefont {Birgeneau}}, \bibinfo {author}
  {\bibfnamefont {J.-H.}\ \bibnamefont {Chu}}, \bibinfo {author} {\bibfnamefont
  {M.}~\bibnamefont {Yi}},\ and\ \bibinfo {author} {\bibfnamefont
  {P.}~\bibnamefont {Dai}},\ }\bibfield  {title} {\bibinfo {title} {Discovery
  of charge density wave in a kagome lattice antiferromagnet},\ }\href@noop {}
  {\bibfield  {journal} {\bibinfo  {journal} {Nature}\ }\textbf {\bibinfo
  {volume} {609}},\ \bibinfo {pages} {490} (\bibinfo {year}
  {2022})}\BibitemShut {NoStop}%
\bibitem [{\citenamefont {Teng}\ \emph {et~al.}(2023)\citenamefont {Teng},
  \citenamefont {Oh}, \citenamefont {Tan}, \citenamefont {Chen}, \citenamefont
  {Huang}, \citenamefont {Gao}, \citenamefont {Yin}, \citenamefont {Chu},
  \citenamefont {Hashimoto}, \citenamefont {Lu}, \citenamefont {Jozwiak},
  \citenamefont {Bostwick}, \citenamefont {Rotenberg}, \citenamefont
  {Granroth}, \citenamefont {Yan}, \citenamefont {Birgeneau}, \citenamefont
  {Dai},\ and\ \citenamefont {Yi}}]{teng2023a}%
  \BibitemOpen
  \bibfield  {author} {\bibinfo {author} {\bibfnamefont {X.}~\bibnamefont
  {Teng}}, \bibinfo {author} {\bibfnamefont {J.~S.}\ \bibnamefont {Oh}},
  \bibinfo {author} {\bibfnamefont {H.}~\bibnamefont {Tan}}, \bibinfo {author}
  {\bibfnamefont {L.}~\bibnamefont {Chen}}, \bibinfo {author} {\bibfnamefont
  {J.}~\bibnamefont {Huang}}, \bibinfo {author} {\bibfnamefont
  {B.}~\bibnamefont {Gao}}, \bibinfo {author} {\bibfnamefont {J.-X.}\
  \bibnamefont {Yin}}, \bibinfo {author} {\bibfnamefont {J.-H.}\ \bibnamefont
  {Chu}}, \bibinfo {author} {\bibfnamefont {M.}~\bibnamefont {Hashimoto}},
  \bibinfo {author} {\bibfnamefont {D.}~\bibnamefont {Lu}}, \bibinfo {author}
  {\bibfnamefont {C.}~\bibnamefont {Jozwiak}}, \bibinfo {author} {\bibfnamefont
  {A.}~\bibnamefont {Bostwick}}, \bibinfo {author} {\bibfnamefont
  {E.}~\bibnamefont {Rotenberg}}, \bibinfo {author} {\bibfnamefont {G.~E.}\
  \bibnamefont {Granroth}}, \bibinfo {author} {\bibfnamefont {B.}~\bibnamefont
  {Yan}}, \bibinfo {author} {\bibfnamefont {R.~J.}\ \bibnamefont {Birgeneau}},
  \bibinfo {author} {\bibfnamefont {P.}~\bibnamefont {Dai}},\ and\ \bibinfo
  {author} {\bibfnamefont {M.}~\bibnamefont {Yi}},\ }\bibfield  {title}
  {\bibinfo {title} {Magnetism and charge density wave order in kagome
  {FeGe}},\ }\href@noop {} {\bibfield  {journal} {\bibinfo  {journal} {Nat.
  Phys.}\ }\textbf {\bibinfo {volume} {19}} (\bibinfo {year}
  {2023})}\BibitemShut {NoStop}%
\bibitem [{\citenamefont {Zhou}\ \emph
  {et~al.}(2023{\natexlab{b}})\citenamefont {Zhou}, \citenamefont {Yan},
  \citenamefont {Fan}, \citenamefont {Wang},\ and\ \citenamefont
  {Wan}}]{zhou2023b}%
  \BibitemOpen
  \bibfield  {author} {\bibinfo {author} {\bibfnamefont {H.}~\bibnamefont
  {Zhou}}, \bibinfo {author} {\bibfnamefont {S.}~\bibnamefont {Yan}}, \bibinfo
  {author} {\bibfnamefont {D.}~\bibnamefont {Fan}}, \bibinfo {author}
  {\bibfnamefont {D.}~\bibnamefont {Wang}},\ and\ \bibinfo {author}
  {\bibfnamefont {X.}~\bibnamefont {Wan}},\ }\bibfield  {title} {\bibinfo
  {title} {Magnetic interactions and possible structural distortion in kagome
  {FeGe} from first-principles calculations and symmetry analysis},\
  }\href@noop {} {\bibfield  {journal} {\bibinfo  {journal} {Phys. Rev. B}\
  }\textbf {\bibinfo {volume} {108}},\ \bibinfo {pages} {035138} (\bibinfo
  {year} {2023}{\natexlab{b}})}\BibitemShut {NoStop}%
\bibitem [{\citenamefont {Haussermann}\ \emph {et~al.}(2002)\citenamefont
  {Haussermann}, \citenamefont {Amerioun}, \citenamefont {Eriksson},
  \citenamefont {Lee},\ and\ \citenamefont {Miller}}]{haussermann2002a}%
  \BibitemOpen
  \bibfield  {author} {\bibinfo {author} {\bibfnamefont {U.}~\bibnamefont
  {Haussermann}}, \bibinfo {author} {\bibfnamefont {S.}~\bibnamefont
  {Amerioun}}, \bibinfo {author} {\bibfnamefont {L.}~\bibnamefont {Eriksson}},
  \bibinfo {author} {\bibfnamefont {C.-S.}\ \bibnamefont {Lee}},\ and\ \bibinfo
  {author} {\bibfnamefont {G.~J.}\ \bibnamefont {Miller}},\ }\bibfield  {title}
  {\bibinfo {title} {The $s$--$p$ bonded representatives of the prominent
  {BaAl$_4$} structure type: A case study on structural stability of polar
  intermetallic network structures},\ }\href@noop {} {\bibfield  {journal}
  {\bibinfo  {journal} {J. Amer. Chem. Soc.}\ }\textbf {\bibinfo {volume}
  {124}},\ \bibinfo {pages} {4371} (\bibinfo {year} {2002})}\BibitemShut
  {NoStop}%
\bibitem [{\citenamefont {Stavinoha}\ \emph {et~al.}(2018)\citenamefont
  {Stavinoha}, \citenamefont {Cooley}, \citenamefont {Minasian}, \citenamefont
  {McQueen}, \citenamefont {Kauzlarich}, \citenamefont {Huang},\ and\
  \citenamefont {Morosan}}]{stavinoha2018a}%
  \BibitemOpen
  \bibfield  {author} {\bibinfo {author} {\bibfnamefont {M.}~\bibnamefont
  {Stavinoha}}, \bibinfo {author} {\bibfnamefont {J.~A.}\ \bibnamefont
  {Cooley}}, \bibinfo {author} {\bibfnamefont {S.~G.}\ \bibnamefont
  {Minasian}}, \bibinfo {author} {\bibfnamefont {T.~M.}\ \bibnamefont
  {McQueen}}, \bibinfo {author} {\bibfnamefont {S.~M.}\ \bibnamefont
  {Kauzlarich}}, \bibinfo {author} {\bibfnamefont {C.-L.}\ \bibnamefont
  {Huang}},\ and\ \bibinfo {author} {\bibfnamefont {E.}~\bibnamefont
  {Morosan}},\ }\bibfield  {title} {\bibinfo {title} {Charge density wave
  behavior and order-disorder in the antiferromagnetic metallic series
  {Eu(Ga$_{1-x}$Al$_x$)$_4$}},\ }\href@noop {} {\bibfield  {journal} {\bibinfo
  {journal} {Phys. Rev. B}\ }\textbf {\bibinfo {volume} {97}},\ \bibinfo
  {pages} {195146} (\bibinfo {year} {2018})}\BibitemShut {NoStop}%
\bibitem [{\citenamefont {Wang}\ \emph {et~al.}(2021)\citenamefont {Wang},
  \citenamefont {Mori}, \citenamefont {Wang}, \citenamefont {Wang},
  \citenamefont {Ma}, \citenamefont {Latzke}, \citenamefont {Graf},
  \citenamefont {Denlinger}, \citenamefont {Campbell}, \citenamefont
  {Bernevig}, \citenamefont {Lanzara},\ and\ \citenamefont
  {Paglione}}]{wang2021a}%
  \BibitemOpen
  \bibfield  {author} {\bibinfo {author} {\bibfnamefont {K.}~\bibnamefont
  {Wang}}, \bibinfo {author} {\bibfnamefont {R.}~\bibnamefont {Mori}}, \bibinfo
  {author} {\bibfnamefont {Z.}~\bibnamefont {Wang}}, \bibinfo {author}
  {\bibfnamefont {L.}~\bibnamefont {Wang}}, \bibinfo {author} {\bibfnamefont
  {J.~H.~S.}\ \bibnamefont {Ma}}, \bibinfo {author} {\bibfnamefont {D.~W.}\
  \bibnamefont {Latzke}}, \bibinfo {author} {\bibfnamefont {D.~E.}\
  \bibnamefont {Graf}}, \bibinfo {author} {\bibfnamefont {J.~D.}\ \bibnamefont
  {Denlinger}}, \bibinfo {author} {\bibfnamefont {D.}~\bibnamefont {Campbell}},
  \bibinfo {author} {\bibfnamefont {B.~A.}\ \bibnamefont {Bernevig}}, \bibinfo
  {author} {\bibfnamefont {A.}~\bibnamefont {Lanzara}},\ and\ \bibinfo {author}
  {\bibfnamefont {J.}~\bibnamefont {Paglione}},\ }\bibfield  {title} {\bibinfo
  {title} {Crystalline symmetry-protected non-trivial topology in prototype
  compound {BaAl$_4$}},\ }\href@noop {} {\bibfield  {journal} {\bibinfo
  {journal} {NPJ Quant. Mater.}\ }\textbf {\bibinfo {volume} {6}},\ \bibinfo
  {pages} {28} (\bibinfo {year} {2021})}\BibitemShut {NoStop}%
\bibitem [{\citenamefont {Shang}\ \emph {et~al.}(2021)\citenamefont {Shang},
  \citenamefont {Xu}, \citenamefont {Gawryluk}, \citenamefont {Ma},
  \citenamefont {Shiroka}, \citenamefont {Shi},\ and\ \citenamefont
  {Pomjakushina}}]{shang2021a}%
  \BibitemOpen
  \bibfield  {author} {\bibinfo {author} {\bibfnamefont {T.}~\bibnamefont
  {Shang}}, \bibinfo {author} {\bibfnamefont {Y.}~\bibnamefont {Xu}}, \bibinfo
  {author} {\bibfnamefont {D.~J.}\ \bibnamefont {Gawryluk}}, \bibinfo {author}
  {\bibfnamefont {J.~Z.}\ \bibnamefont {Ma}}, \bibinfo {author} {\bibfnamefont
  {T.}~\bibnamefont {Shiroka}}, \bibinfo {author} {\bibfnamefont
  {M.}~\bibnamefont {Shi}},\ and\ \bibinfo {author} {\bibfnamefont
  {E.}~\bibnamefont {Pomjakushina}},\ }\bibfield  {title} {\bibinfo {title}
  {Anomalous hall resistivity and possible topological hall effect in the
  {EuAl$_4$} antiferromagnet},\ }\href@noop {} {\bibfield  {journal} {\bibinfo
  {journal} {Phys. Rev. B}\ }\textbf {\bibinfo {volume} {103}},\ \bibinfo
  {pages} {L020405} (\bibinfo {year} {2021})}\BibitemShut {NoStop}%
\bibitem [{\citenamefont {Mori}\ \emph {et~al.}(2022)\citenamefont {Mori},
  \citenamefont {Wang}, \citenamefont {Morimoto}, \citenamefont {Ciocys},
  \citenamefont {Denlinger}, \citenamefont {Paglione},\ and\ \citenamefont
  {Lanzara}}]{mori2022a}%
  \BibitemOpen
  \bibfield  {author} {\bibinfo {author} {\bibfnamefont {R.}~\bibnamefont
  {Mori}}, \bibinfo {author} {\bibfnamefont {K.}~\bibnamefont {Wang}}, \bibinfo
  {author} {\bibfnamefont {T.}~\bibnamefont {Morimoto}}, \bibinfo {author}
  {\bibfnamefont {S.}~\bibnamefont {Ciocys}}, \bibinfo {author} {\bibfnamefont
  {J.~D.}\ \bibnamefont {Denlinger}}, \bibinfo {author} {\bibfnamefont
  {J.}~\bibnamefont {Paglione}},\ and\ \bibinfo {author} {\bibfnamefont
  {A.}~\bibnamefont {Lanzara}},\ }\bibfield  {title} {\bibinfo {title}
  {Observation of a flat and extended surface state in a topological
  semimetal},\ }\href@noop {} {\bibfield  {journal} {\bibinfo  {journal}
  {Materials}\ }\textbf {\bibinfo {volume} {15}} (\bibinfo {year}
  {2022})}\BibitemShut {NoStop}%
\bibitem [{\citenamefont {Gen}\ \emph {et~al.}(2023)\citenamefont {Gen},
  \citenamefont {Takagi}, \citenamefont {Watanabe}, \citenamefont {Kitou},
  \citenamefont {Sagayama}, \citenamefont {Matsuyama}, \citenamefont {Kohama},
  \citenamefont {Ikeda}, \citenamefont {Onuki}, \citenamefont {Kurumaji},
  \citenamefont {Arima},\ and\ \citenamefont {Seki}}]{gen2023a}%
  \BibitemOpen
  \bibfield  {author} {\bibinfo {author} {\bibfnamefont {M.}~\bibnamefont
  {Gen}}, \bibinfo {author} {\bibfnamefont {R.}~\bibnamefont {Takagi}},
  \bibinfo {author} {\bibfnamefont {Y.}~\bibnamefont {Watanabe}}, \bibinfo
  {author} {\bibfnamefont {S.}~\bibnamefont {Kitou}}, \bibinfo {author}
  {\bibfnamefont {H.}~\bibnamefont {Sagayama}}, \bibinfo {author}
  {\bibfnamefont {N.}~\bibnamefont {Matsuyama}}, \bibinfo {author}
  {\bibfnamefont {Y.}~\bibnamefont {Kohama}}, \bibinfo {author} {\bibfnamefont
  {A.}~\bibnamefont {Ikeda}}, \bibinfo {author} {\bibfnamefont
  {Y.}~\bibnamefont {Onuki}}, \bibinfo {author} {\bibfnamefont
  {T.}~\bibnamefont {Kurumaji}}, \bibinfo {author} {\bibfnamefont {T.-h.}\
  \bibnamefont {Arima}},\ and\ \bibinfo {author} {\bibfnamefont
  {S.}~\bibnamefont {Seki}},\ }\bibfield  {title} {\bibinfo {title} {Rhombic
  skyrmion lattice coupled with orthorhombic structural distortion in
  {EuAl$_4$}},\ }\href {https://doi.org/10.1103/PhysRevB.107.L020410}
  {\bibfield  {journal} {\bibinfo  {journal} {Phys. Rev. B}\ }\textbf {\bibinfo
  {volume} {107}},\ \bibinfo {pages} {L020410} (\bibinfo {year}
  {2023})}\BibitemShut {NoStop}%
\bibitem [{\citenamefont {Wang}\ \emph
  {et~al.}(2023{\natexlab{b}})\citenamefont {Wang}, \citenamefont {Nepal},\
  and\ \citenamefont {Canfield}}]{wang2023a}%
  \BibitemOpen
  \bibfield  {author} {\bibinfo {author} {\bibfnamefont {L.-L.}\ \bibnamefont
  {Wang}}, \bibinfo {author} {\bibfnamefont {N.~K.}\ \bibnamefont {Nepal}},\
  and\ \bibinfo {author} {\bibfnamefont {P.~C.}\ \bibnamefont {Canfield}},\
  }\href@noop {} {\bibinfo {title} {Origin of charge density wave in
  topological semimetals {SrAl$_4$} and {EuAl$_4$}}} (\bibinfo {year}
  {2023}{\natexlab{b}}),\ \Eprint {https://arxiv.org/abs/2306.15068}
  {arXiv:2306.15068 [cond-mat.mtrl-sci]} \BibitemShut {NoStop}%
\bibitem [{\citenamefont {Nakamura}\ \emph {et~al.}(2015)\citenamefont
  {Nakamura}, \citenamefont {Uejo}, \citenamefont {Honda}, \citenamefont
  {Takeuchi}, \citenamefont {Harima}, \citenamefont {Yamamoto}, \citenamefont
  {Haga}, \citenamefont {Matsubayashi}, \citenamefont {Uwatoko}, \citenamefont
  {Hedo}, \citenamefont {Nakama},\ and\ \citenamefont
  {Ōnuki}}]{nakamura2015a}%
  \BibitemOpen
  \bibfield  {author} {\bibinfo {author} {\bibfnamefont {A.}~\bibnamefont
  {Nakamura}}, \bibinfo {author} {\bibfnamefont {T.}~\bibnamefont {Uejo}},
  \bibinfo {author} {\bibfnamefont {F.}~\bibnamefont {Honda}}, \bibinfo
  {author} {\bibfnamefont {T.}~\bibnamefont {Takeuchi}}, \bibinfo {author}
  {\bibfnamefont {H.}~\bibnamefont {Harima}}, \bibinfo {author} {\bibfnamefont
  {E.}~\bibnamefont {Yamamoto}}, \bibinfo {author} {\bibfnamefont
  {Y.}~\bibnamefont {Haga}}, \bibinfo {author} {\bibfnamefont {K.}~\bibnamefont
  {Matsubayashi}}, \bibinfo {author} {\bibfnamefont {Y.}~\bibnamefont
  {Uwatoko}}, \bibinfo {author} {\bibfnamefont {M.}~\bibnamefont {Hedo}},
  \bibinfo {author} {\bibfnamefont {T.}~\bibnamefont {Nakama}},\ and\ \bibinfo
  {author} {\bibfnamefont {Y.}~\bibnamefont {Ōnuki}},\ }\bibfield  {title}
  {\bibinfo {title} {Transport and magnetic properties of {EuAl$_4$} and
  {EuGa$_4$}},\ }\href@noop {} {\bibfield  {journal} {\bibinfo  {journal} {J.
  Phys. Soc. Jpn}\ }\textbf {\bibinfo {volume} {84}},\ \bibinfo {pages}
  {124711} (\bibinfo {year} {2015})}\BibitemShut {NoStop}%
\bibitem [{\citenamefont {Shimomura}\ \emph {et~al.}(2019)\citenamefont
  {Shimomura}, \citenamefont {Murao}, \citenamefont {Tsutsui}, \citenamefont
  {Nakao}, \citenamefont {Nakamura}, \citenamefont {Hedo}, \citenamefont
  {Nakama},\ and\ \citenamefont {Ōnuki}}]{shimomura2019a}%
  \BibitemOpen
  \bibfield  {author} {\bibinfo {author} {\bibfnamefont {S.}~\bibnamefont
  {Shimomura}}, \bibinfo {author} {\bibfnamefont {H.}~\bibnamefont {Murao}},
  \bibinfo {author} {\bibfnamefont {S.}~\bibnamefont {Tsutsui}}, \bibinfo
  {author} {\bibfnamefont {H.}~\bibnamefont {Nakao}}, \bibinfo {author}
  {\bibfnamefont {A.}~\bibnamefont {Nakamura}}, \bibinfo {author}
  {\bibfnamefont {M.}~\bibnamefont {Hedo}}, \bibinfo {author} {\bibfnamefont
  {T.}~\bibnamefont {Nakama}},\ and\ \bibinfo {author} {\bibfnamefont
  {Y.}~\bibnamefont {Ōnuki}},\ }\bibfield  {title} {\bibinfo {title} {Lattice
  modulation and structural phase transition in the antiferromagnet
  {EuAl$_4$}},\ }\href@noop {} {\bibfield  {journal} {\bibinfo  {journal} {J.
  Phys. Soc. Jpn}\ }\textbf {\bibinfo {volume} {88}},\ \bibinfo {pages}
  {014602} (\bibinfo {year} {2019})}\BibitemShut {NoStop}%
\bibitem [{\citenamefont {Kaneko}\ \emph {et~al.}(2021)\citenamefont {Kaneko},
  \citenamefont {Kawasaki}, \citenamefont {Nakamura}, \citenamefont {Munakata},
  \citenamefont {Nakao}, \citenamefont {Hanashima}, \citenamefont {Kiyanagi},
  \citenamefont {Ohhara}, \citenamefont {Hedo}, \citenamefont {Nakama},\ and\
  \citenamefont {Ōnuki}}]{kaneko2021a}%
  \BibitemOpen
  \bibfield  {author} {\bibinfo {author} {\bibfnamefont {K.}~\bibnamefont
  {Kaneko}}, \bibinfo {author} {\bibfnamefont {T.}~\bibnamefont {Kawasaki}},
  \bibinfo {author} {\bibfnamefont {A.}~\bibnamefont {Nakamura}}, \bibinfo
  {author} {\bibfnamefont {K.}~\bibnamefont {Munakata}}, \bibinfo {author}
  {\bibfnamefont {A.}~\bibnamefont {Nakao}}, \bibinfo {author} {\bibfnamefont
  {T.}~\bibnamefont {Hanashima}}, \bibinfo {author} {\bibfnamefont
  {R.}~\bibnamefont {Kiyanagi}}, \bibinfo {author} {\bibfnamefont
  {T.}~\bibnamefont {Ohhara}}, \bibinfo {author} {\bibfnamefont
  {M.}~\bibnamefont {Hedo}}, \bibinfo {author} {\bibfnamefont {T.}~\bibnamefont
  {Nakama}},\ and\ \bibinfo {author} {\bibfnamefont {Y.}~\bibnamefont
  {Ōnuki}},\ }\bibfield  {title} {\bibinfo {title} {Charge-density-wave order
  and multiple magnetic transitions in divalent europium compound {EuAl$_4$}},\
  }\href@noop {} {\bibfield  {journal} {\bibinfo  {journal} {J. Phys. Soc.
  Jpn}\ }\textbf {\bibinfo {volume} {90}},\ \bibinfo {pages} {064704} (\bibinfo
  {year} {2021})}\BibitemShut {NoStop}%
\bibitem [{\citenamefont {Ramakrishnan}\ \emph {et~al.}(2022)\citenamefont
  {Ramakrishnan}, \citenamefont {Kotla}, \citenamefont {Rekis}, \citenamefont
  {Bao}, \citenamefont {Eisele}, \citenamefont {Noohinejad}, \citenamefont
  {Tolkiehn}, \citenamefont {Paulmann}, \citenamefont {Singh}, \citenamefont
  {Verma}, \citenamefont {Bag}, \citenamefont {Kulkarni}, \citenamefont
  {Thamizhavel}, \citenamefont {Singh}, \citenamefont {Ramakrishnan},\ and\
  \citenamefont {van Smaalen}}]{ramakrishnan2022a}%
  \BibitemOpen
  \bibfield  {author} {\bibinfo {author} {\bibfnamefont {S.}~\bibnamefont
  {Ramakrishnan}}, \bibinfo {author} {\bibfnamefont {S.~R.}\ \bibnamefont
  {Kotla}}, \bibinfo {author} {\bibfnamefont {T.}~\bibnamefont {Rekis}},
  \bibinfo {author} {\bibfnamefont {J.-K.}\ \bibnamefont {Bao}}, \bibinfo
  {author} {\bibfnamefont {C.}~\bibnamefont {Eisele}}, \bibinfo {author}
  {\bibfnamefont {L.}~\bibnamefont {Noohinejad}}, \bibinfo {author}
  {\bibfnamefont {M.}~\bibnamefont {Tolkiehn}}, \bibinfo {author}
  {\bibfnamefont {C.}~\bibnamefont {Paulmann}}, \bibinfo {author}
  {\bibfnamefont {B.}~\bibnamefont {Singh}}, \bibinfo {author} {\bibfnamefont
  {R.}~\bibnamefont {Verma}}, \bibinfo {author} {\bibfnamefont
  {B.}~\bibnamefont {Bag}}, \bibinfo {author} {\bibfnamefont {R.}~\bibnamefont
  {Kulkarni}}, \bibinfo {author} {\bibfnamefont {A.}~\bibnamefont
  {Thamizhavel}}, \bibinfo {author} {\bibfnamefont {B.}~\bibnamefont {Singh}},
  \bibinfo {author} {\bibfnamefont {S.}~\bibnamefont {Ramakrishnan}},\ and\
  \bibinfo {author} {\bibfnamefont {S.}~\bibnamefont {van Smaalen}},\
  }\bibfield  {title} {\bibinfo {title} {Orthorhombic charge density wave on
  the tetragonal lattice of {EuAl$_{4}$}},\ }\href
  {https://doi.org/10.1107/S2052252522003888} {\bibfield  {journal} {\bibinfo
  {journal} {IUCrJ}\ }\textbf {\bibinfo {volume} {9}},\ \bibinfo {pages} {378}
  (\bibinfo {year} {2022})}\BibitemShut {NoStop}%
\bibitem [{\citenamefont {Meier}\ \emph {et~al.}(2022)\citenamefont {Meier},
  \citenamefont {Torres}, \citenamefont {Hermann}, \citenamefont {Zhao},
  \citenamefont {Lavina}, \citenamefont {Sales},\ and\ \citenamefont
  {May}}]{meier2022a}%
  \BibitemOpen
  \bibfield  {author} {\bibinfo {author} {\bibfnamefont {W.~R.}\ \bibnamefont
  {Meier}}, \bibinfo {author} {\bibfnamefont {J.~R.}\ \bibnamefont {Torres}},
  \bibinfo {author} {\bibfnamefont {R.~P.}\ \bibnamefont {Hermann}}, \bibinfo
  {author} {\bibfnamefont {J.}~\bibnamefont {Zhao}}, \bibinfo {author}
  {\bibfnamefont {B.}~\bibnamefont {Lavina}}, \bibinfo {author} {\bibfnamefont
  {B.~C.}\ \bibnamefont {Sales}},\ and\ \bibinfo {author} {\bibfnamefont
  {A.~F.}\ \bibnamefont {May}},\ }\bibfield  {title} {\bibinfo {title}
  {Thermodynamic insights into the intricate magnetic phase diagram of
  {EuAl$_4$}},\ }\href {https://doi.org/10.1103/PhysRevB.106.094421} {\bibfield
   {journal} {\bibinfo  {journal} {Phys. Rev. B}\ }\textbf {\bibinfo {volume}
  {106}},\ \bibinfo {pages} {094421} (\bibinfo {year} {2022})}\BibitemShut
  {NoStop}%
\bibitem [{\citenamefont {Takagi}\ \emph {et~al.}(2022)\citenamefont {Takagi},
  \citenamefont {Matsuyama}, \citenamefont {Ukleev}, \citenamefont {Yu},
  \citenamefont {White}, \citenamefont {Francoual}, \citenamefont {Mardegan},
  \citenamefont {Hayami}, \citenamefont {Saito}, \citenamefont {Kaneko},
  \citenamefont {Ohishi}, \citenamefont {Ōnuki}, \citenamefont {Arima},
  \citenamefont {Tokura}, \citenamefont {Nakajima},\ and\ \citenamefont
  {Seki}}]{takagir2022a}%
  \BibitemOpen
  \bibfield  {author} {\bibinfo {author} {\bibfnamefont {R.}~\bibnamefont
  {Takagi}}, \bibinfo {author} {\bibfnamefont {N.}~\bibnamefont {Matsuyama}},
  \bibinfo {author} {\bibfnamefont {V.}~\bibnamefont {Ukleev}}, \bibinfo
  {author} {\bibfnamefont {L.}~\bibnamefont {Yu}}, \bibinfo {author}
  {\bibfnamefont {J.~S.}\ \bibnamefont {White}}, \bibinfo {author}
  {\bibfnamefont {S.}~\bibnamefont {Francoual}}, \bibinfo {author}
  {\bibfnamefont {J.~R.~L.}\ \bibnamefont {Mardegan}}, \bibinfo {author}
  {\bibfnamefont {S.}~\bibnamefont {Hayami}}, \bibinfo {author} {\bibfnamefont
  {H.}~\bibnamefont {Saito}}, \bibinfo {author} {\bibfnamefont
  {K.}~\bibnamefont {Kaneko}}, \bibinfo {author} {\bibfnamefont
  {K.}~\bibnamefont {Ohishi}}, \bibinfo {author} {\bibfnamefont
  {Y.}~\bibnamefont {Ōnuki}}, \bibinfo {author} {\bibfnamefont {T.-h.}\
  \bibnamefont {Arima}}, \bibinfo {author} {\bibfnamefont {Y.}~\bibnamefont
  {Tokura}}, \bibinfo {author} {\bibfnamefont {T.}~\bibnamefont {Nakajima}},\
  and\ \bibinfo {author} {\bibfnamefont {S.}~\bibnamefont {Seki}},\ }\bibfield
  {title} {\bibinfo {title} {Square and rhombic lattices of magnetic skyrmions
  in a centrosymmetric binary compound},\ }\href
  {https://doi.org/10.1038/s41467-022-29131-9} {\bibfield  {journal} {\bibinfo
  {journal} {Nature Commun.}\ }\textbf {\bibinfo {volume} {13}},\ \bibinfo
  {pages} {1472} (\bibinfo {year} {2022})}\BibitemShut {NoStop}%
\bibitem [{\citenamefont {Kaneko}\ \emph {et~al.}(2019)\citenamefont {Kaneko},
  \citenamefont {Frontzek}, \citenamefont {Matsuda}, \citenamefont {Nakao},
  \citenamefont {Munakata}, \citenamefont {Ohhara}, \citenamefont {Kakihana},
  \citenamefont {Haga}, \citenamefont {Hedo}, \citenamefont {Nakama},\ and\
  \citenamefont {Onuki}}]{kaneko2019a}%
  \BibitemOpen
  \bibfield  {author} {\bibinfo {author} {\bibfnamefont {K.}~\bibnamefont
  {Kaneko}}, \bibinfo {author} {\bibfnamefont {M.~D.}\ \bibnamefont
  {Frontzek}}, \bibinfo {author} {\bibfnamefont {M.}~\bibnamefont {Matsuda}},
  \bibinfo {author} {\bibfnamefont {A.}~\bibnamefont {Nakao}}, \bibinfo
  {author} {\bibfnamefont {K.}~\bibnamefont {Munakata}}, \bibinfo {author}
  {\bibfnamefont {T.}~\bibnamefont {Ohhara}}, \bibinfo {author} {\bibfnamefont
  {M.}~\bibnamefont {Kakihana}}, \bibinfo {author} {\bibfnamefont
  {Y.}~\bibnamefont {Haga}}, \bibinfo {author} {\bibfnamefont {M.}~\bibnamefont
  {Hedo}}, \bibinfo {author} {\bibfnamefont {T.}~\bibnamefont {Nakama}},\ and\
  \bibinfo {author} {\bibfnamefont {Y.}~\bibnamefont {Onuki}},\ }\bibfield
  {title} {\bibinfo {title} {Unique helical magnetic order and field-induced
  phase in trillium lattice antiferromagnet {EuPtSi}},\ }\href@noop {}
  {\bibfield  {journal} {\bibinfo  {journal} {J. Phys. Soc. Jpn}\ }\textbf
  {\bibinfo {volume} {88}},\ \bibinfo {pages} {013702} (\bibinfo {year}
  {2019})}\BibitemShut {NoStop}%
\bibitem [{\citenamefont {Onuki}\ \emph {et~al.}(2020)\citenamefont {Onuki},
  \citenamefont {Hedo},\ and\ \citenamefont {Honda}}]{onuki2020a}%
  \BibitemOpen
  \bibfield  {author} {\bibinfo {author} {\bibfnamefont {Y.}~\bibnamefont
  {Onuki}}, \bibinfo {author} {\bibfnamefont {M.}~\bibnamefont {Hedo}},\ and\
  \bibinfo {author} {\bibfnamefont {F.}~\bibnamefont {Honda}},\ }\bibfield
  {title} {\bibinfo {title} {Unique electronic states of {Eu}-based
  compounds},\ }\href@noop {} {\bibfield  {journal} {\bibinfo  {journal} {J.
  Phys. Soc. Jpn}\ }\textbf {\bibinfo {volume} {89}},\ \bibinfo {pages}
  {102001} (\bibinfo {year} {2020})}\BibitemShut {NoStop}%
\bibitem [{\citenamefont {Zhu}\ \emph {et~al.}(2022)\citenamefont {Zhu},
  \citenamefont {Zhang}, \citenamefont {Gawryluk}, \citenamefont {Zhen},
  \citenamefont {Yu}, \citenamefont {Ju}, \citenamefont {Xie}, \citenamefont
  {Jiang}, \citenamefont {Cheng}, \citenamefont {Xu}, \citenamefont {Shi},
  \citenamefont {Pomjakushina}, \citenamefont {Zhan}, \citenamefont {Shiroka},\
  and\ \citenamefont {Shang}}]{zhu2022a}%
  \BibitemOpen
  \bibfield  {author} {\bibinfo {author} {\bibfnamefont {X.~Y.}\ \bibnamefont
  {Zhu}}, \bibinfo {author} {\bibfnamefont {H.}~\bibnamefont {Zhang}}, \bibinfo
  {author} {\bibfnamefont {D.~J.}\ \bibnamefont {Gawryluk}}, \bibinfo {author}
  {\bibfnamefont {Z.~X.}\ \bibnamefont {Zhen}}, \bibinfo {author}
  {\bibfnamefont {B.~C.}\ \bibnamefont {Yu}}, \bibinfo {author} {\bibfnamefont
  {S.~L.}\ \bibnamefont {Ju}}, \bibinfo {author} {\bibfnamefont
  {W.}~\bibnamefont {Xie}}, \bibinfo {author} {\bibfnamefont {D.~M.}\
  \bibnamefont {Jiang}}, \bibinfo {author} {\bibfnamefont {W.~J.}\ \bibnamefont
  {Cheng}}, \bibinfo {author} {\bibfnamefont {Y.}~\bibnamefont {Xu}}, \bibinfo
  {author} {\bibfnamefont {M.}~\bibnamefont {Shi}}, \bibinfo {author}
  {\bibfnamefont {E.}~\bibnamefont {Pomjakushina}}, \bibinfo {author}
  {\bibfnamefont {Q.~F.}\ \bibnamefont {Zhan}}, \bibinfo {author}
  {\bibfnamefont {T.}~\bibnamefont {Shiroka}},\ and\ \bibinfo {author}
  {\bibfnamefont {T.}~\bibnamefont {Shang}},\ }\bibfield  {title} {\bibinfo
  {title} {Spin order and fluctuations in the {EuAl$_4$} and {EuGa$_4$}
  topological antiferromagnets: A ${\mu}\mathrm{SR}$ study},\ }\href@noop {}
  {\bibfield  {journal} {\bibinfo  {journal} {Phys. Rev. B}\ }\textbf {\bibinfo
  {volume} {105}},\ \bibinfo {pages} {014423} (\bibinfo {year}
  {2022})}\BibitemShut {NoStop}%
\bibitem [{\citenamefont {Ni}\ \emph {et~al.}(2024)\citenamefont {Ni},
  \citenamefont {Meier}, \citenamefont {Miao}, \citenamefont {May},
  \citenamefont {Sales}, \citenamefont {Zuo},\ and\ \citenamefont
  {Chi}}]{nih2024a}%
  \BibitemOpen
  \bibfield  {author} {\bibinfo {author} {\bibfnamefont {H.}~\bibnamefont
  {Ni}}, \bibinfo {author} {\bibfnamefont {W.~R.}\ \bibnamefont {Meier}},
  \bibinfo {author} {\bibfnamefont {H.}~\bibnamefont {Miao}}, \bibinfo {author}
  {\bibfnamefont {A.~J.}\ \bibnamefont {May}}, \bibinfo {author} {\bibfnamefont
  {B.~C.}\ \bibnamefont {Sales}}, \bibinfo {author} {\bibfnamefont {J.-m.}\
  \bibnamefont {Zuo}},\ and\ \bibinfo {author} {\bibfnamefont {M.}~\bibnamefont
  {Chi}},\ }\bibfield  {title} {\bibinfo {title} {Real-space visualization of
  charge density wave induced local inversion-symmetry breaking in a skyrmion
  magnet},\ }\href@noop {} {\bibfield  {journal} {\bibinfo  {journal} {arXiv}\
  }\textbf {\bibinfo {volume} {arXiv:2311.17682}} (\bibinfo {year}
  {2024})}\BibitemShut {NoStop}%
\bibitem [{\citenamefont {Yang}\ \emph {et~al.}(2024)\citenamefont {Yang},
  \citenamefont {Le}, \citenamefont {Zhu}, \citenamefont {Wang}, \citenamefont
  {Shang}, \citenamefont {Dai}, \citenamefont {Hu},\ and\ \citenamefont
  {Dressel}}]{yangr2024a}%
  \BibitemOpen
  \bibfield  {author} {\bibinfo {author} {\bibfnamefont {R.}~\bibnamefont
  {Yang}}, \bibinfo {author} {\bibfnamefont {C.~C.}\ \bibnamefont {Le}},
  \bibinfo {author} {\bibfnamefont {P.}~\bibnamefont {Zhu}}, \bibinfo {author}
  {\bibfnamefont {Z.~W.}\ \bibnamefont {Wang}}, \bibinfo {author}
  {\bibfnamefont {T.}~\bibnamefont {Shang}}, \bibinfo {author} {\bibfnamefont
  {Y.~M.}\ \bibnamefont {Dai}}, \bibinfo {author} {\bibfnamefont {J.~P.}\
  \bibnamefont {Hu}},\ and\ \bibinfo {author} {\bibfnamefont {M.}~\bibnamefont
  {Dressel}},\ }\bibfield  {title} {\bibinfo {title} {Charge density wave
  transition in the magnetic topological semimetal {EuAl$_{4}$}},\ }\href
  {https://doi.org/10.1103/PhysRevB.109.L041113} {\bibfield  {journal}
  {\bibinfo  {journal} {Phys. Rev. B}\ }\textbf {\bibinfo {volume} {109}},\
  \bibinfo {pages} {L041113} (\bibinfo {year} {2024})}\BibitemShut {NoStop}%
\bibitem [{\citenamefont {Korshunov}\ \emph {et~al.}(2024)\citenamefont
  {Korshunov}, \citenamefont {Sukhanov}, \citenamefont {Gebel}, \citenamefont
  {Pavlovskii}, \citenamefont {Andriushin}, \citenamefont {Gao}, \citenamefont
  {Moya}, \citenamefont {Morosan},\ and\ \citenamefont
  {Rahn}}]{korshunovan2024a}%
  \BibitemOpen
  \bibfield  {author} {\bibinfo {author} {\bibfnamefont {A.~N.}\ \bibnamefont
  {Korshunov}}, \bibinfo {author} {\bibfnamefont {A.~S.}\ \bibnamefont
  {Sukhanov}}, \bibinfo {author} {\bibfnamefont {S.}~\bibnamefont {Gebel}},
  \bibinfo {author} {\bibfnamefont {M.~S.}\ \bibnamefont {Pavlovskii}},
  \bibinfo {author} {\bibfnamefont {N.~D.}\ \bibnamefont {Andriushin}},
  \bibinfo {author} {\bibfnamefont {Y.}~\bibnamefont {Gao}}, \bibinfo {author}
  {\bibfnamefont {J.~M.}\ \bibnamefont {Moya}}, \bibinfo {author}
  {\bibfnamefont {E.}~\bibnamefont {Morosan}},\ and\ \bibinfo {author}
  {\bibfnamefont {M.~C.}\ \bibnamefont {Rahn}},\ }\bibfield  {title} {\bibinfo
  {title} {Phonon softening and atomic modulations in {EuAl$_4$}},\ }\bibfield
  {journal} {\bibinfo  {journal} {arXiv}\ }\textbf {\bibinfo {volume}
  {arXiv:2402.15397v1}},\ \href {https://doi.org/10.48550/arXiv.2402.15397}
  {10.48550/arXiv.2402.15397} (\bibinfo {year} {2024})\BibitemShut {NoStop}%
\bibitem [{\citenamefont {Nakamura}\ \emph {et~al.}(2016)\citenamefont
  {Nakamura}, \citenamefont {Uejo}, \citenamefont {Harima}, \citenamefont
  {Araki}, \citenamefont {Kobayashi}, \citenamefont {Nakashima}, \citenamefont
  {Amako}, \citenamefont {Hedo}, \citenamefont {Nakama},\ and\ \citenamefont
  {Onuki}}]{nakamura2016a}%
  \BibitemOpen
  \bibfield  {author} {\bibinfo {author} {\bibfnamefont {A.}~\bibnamefont
  {Nakamura}}, \bibinfo {author} {\bibfnamefont {T.}~\bibnamefont {Uejo}},
  \bibinfo {author} {\bibfnamefont {H.}~\bibnamefont {Harima}}, \bibinfo
  {author} {\bibfnamefont {S.}~\bibnamefont {Araki}}, \bibinfo {author}
  {\bibfnamefont {T.~C.}\ \bibnamefont {Kobayashi}}, \bibinfo {author}
  {\bibfnamefont {M.}~\bibnamefont {Nakashima}}, \bibinfo {author}
  {\bibfnamefont {Y.}~\bibnamefont {Amako}}, \bibinfo {author} {\bibfnamefont
  {M.}~\bibnamefont {Hedo}}, \bibinfo {author} {\bibfnamefont {T.}~\bibnamefont
  {Nakama}},\ and\ \bibinfo {author} {\bibfnamefont {Y.}~\bibnamefont
  {Onuki}},\ }\bibfield  {title} {\bibinfo {title} {Characteristic fermi
  surfaces and charge density wave in {SrAl$_4$} and related compounds with the
  {BaAl$_4$}-type tetragonal structure},\ }\href
  {https://doi.org/https://doi.org/10.1016/j.jallcom.2015.08.193} {\bibfield
  {journal} {\bibinfo  {journal} {J. Alloys Compounds}\ }\textbf {\bibinfo
  {volume} {654}},\ \bibinfo {pages} {290} (\bibinfo {year}
  {2016})}\BibitemShut {NoStop}%
\bibitem [{\citenamefont {Niki}\ \emph {et~al.}(2020)\citenamefont {Niki},
  \citenamefont {Kuroshima}, \citenamefont {Higa}, \citenamefont {Morishima},
  \citenamefont {Yogi}, \citenamefont {Nakamura}, \citenamefont {Niki},
  \citenamefont {Maehira}, \citenamefont {Hedo}, \citenamefont {Nakama},\ and\
  \citenamefont {Ōnuki}}]{niki2020a}%
  \BibitemOpen
  \bibfield  {author} {\bibinfo {author} {\bibfnamefont {H.}~\bibnamefont
  {Niki}}, \bibinfo {author} {\bibfnamefont {H.}~\bibnamefont {Kuroshima}},
  \bibinfo {author} {\bibfnamefont {N.}~\bibnamefont {Higa}}, \bibinfo {author}
  {\bibfnamefont {M.}~\bibnamefont {Morishima}}, \bibinfo {author}
  {\bibfnamefont {M.}~\bibnamefont {Yogi}}, \bibinfo {author} {\bibfnamefont
  {A.}~\bibnamefont {Nakamura}}, \bibinfo {author} {\bibfnamefont
  {K.}~\bibnamefont {Niki}}, \bibinfo {author} {\bibfnamefont {T.}~\bibnamefont
  {Maehira}}, \bibinfo {author} {\bibfnamefont {M.}~\bibnamefont {Hedo}},
  \bibinfo {author} {\bibfnamefont {T.}~\bibnamefont {Nakama}},\ and\ \bibinfo
  {author} {\bibfnamefont {Y.}~\bibnamefont {Ōnuki}},\ }\bibfield  {title}
  {\bibinfo {title} {{NMR} study of characteristic {CDW} transition in
  {SrAl$_4$}},\ }\href@noop {} {\bibfield  {journal} {\bibinfo  {journal}
  {Proc. Int'l Conf. Strongly Correlated Electron Systems (SCES2019)}\ }
  (\bibinfo {year} {2020})}\BibitemShut {NoStop}%
\bibitem [{\citenamefont {Miller}\ \emph {et~al.}(1993)\citenamefont {Miller},
  \citenamefont {Li},\ and\ \citenamefont {Franzen}}]{miller1993a}%
  \BibitemOpen
  \bibfield  {author} {\bibinfo {author} {\bibfnamefont {G.~J.}\ \bibnamefont
  {Miller}}, \bibinfo {author} {\bibfnamefont {F.}~\bibnamefont {Li}},\ and\
  \bibinfo {author} {\bibfnamefont {H.~F.}\ \bibnamefont {Franzen}},\
  }\bibfield  {title} {\bibinfo {title} {The structural phase transition in
  calcium-aluminum compound ({CaAl$_4$}): a concerted application of landau
  theory and energy band theory},\ }\href@noop {} {\bibfield  {journal}
  {\bibinfo  {journal} {J. Amer. Chem. Soc.}\ }\textbf {\bibinfo {volume}
  {115}},\ \bibinfo {pages} {3739} (\bibinfo {year} {1993})}\BibitemShut
  {NoStop}%
\bibitem [{\citenamefont {Kobata}\ \emph {et~al.}(2016)\citenamefont {Kobata},
  \citenamefont {Fujimor}, \citenamefont {Takeda}, \citenamefont {Okane},
  \citenamefont {Saitoh}, \citenamefont {Kobayashi}, \citenamefont {Yamagami},
  \citenamefont {Nakamura}, \citenamefont {Hedo}, \citenamefont {Nakama},\ and\
  \citenamefont {Onuki}}]{kobata2016a}%
  \BibitemOpen
  \bibfield  {author} {\bibinfo {author} {\bibfnamefont {M.}~\bibnamefont
  {Kobata}}, \bibinfo {author} {\bibfnamefont {S.-i.}\ \bibnamefont {Fujimor}},
  \bibinfo {author} {\bibfnamefont {Y.}~\bibnamefont {Takeda}}, \bibinfo
  {author} {\bibfnamefont {T.}~\bibnamefont {Okane}}, \bibinfo {author}
  {\bibfnamefont {Y.}~\bibnamefont {Saitoh}}, \bibinfo {author} {\bibfnamefont
  {K.}~\bibnamefont {Kobayashi}}, \bibinfo {author} {\bibfnamefont
  {H.}~\bibnamefont {Yamagami}}, \bibinfo {author} {\bibfnamefont
  {A.}~\bibnamefont {Nakamura}}, \bibinfo {author} {\bibfnamefont
  {M.}~\bibnamefont {Hedo}}, \bibinfo {author} {\bibfnamefont {T.}~\bibnamefont
  {Nakama}},\ and\ \bibinfo {author} {\bibfnamefont {Y.}~\bibnamefont
  {Onuki}},\ }\bibfield  {title} {\bibinfo {title} {Electronic structure of
  {EuAl$_4$} studied by photoelectron spectroscopy},\ }\href@noop {} {\bibfield
   {journal} {\bibinfo  {journal} {J. Phys. Soc. Jpn}\ }\textbf {\bibinfo
  {volume} {85}},\ \bibinfo {pages} {094703} (\bibinfo {year}
  {2016})}\BibitemShut {NoStop}%
\bibitem [{\citenamefont {Wang}\ \emph
  {et~al.}(2023{\natexlab{c}})\citenamefont {Wang}, \citenamefont {Nepal},\
  and\ \citenamefont {Canfield}}]{wangll2023a}%
  \BibitemOpen
  \bibfield  {author} {\bibinfo {author} {\bibfnamefont {L.-L.}\ \bibnamefont
  {Wang}}, \bibinfo {author} {\bibfnamefont {N.~K.}\ \bibnamefont {Nepal}},\
  and\ \bibinfo {author} {\bibfnamefont {P.~C.}\ \bibnamefont {Canfield}},\
  }\bibfield  {title} {\bibinfo {title} {Origin of charge density wave in
  topological semimetals {SrAl$_4$} and {EuAl$_4$}},\ }\bibfield  {journal}
  {\bibinfo  {journal} {arXiv}\ }\textbf {\bibinfo {volume}
  {arXiv:2306.15068}},\ \href {https://doi.org/10.48550/arXiv.2306.15068}
  {10.48550/arXiv.2306.15068} (\bibinfo {year}
  {2023}{\natexlab{c}})\BibitemShut {NoStop}%
\bibitem [{sra()}]{sral4suppmat2023a}%
  \BibitemOpen
  \href@noop {} {}\bibinfo {note} {See Supplemental Material at [URL will be
  inserted by publisher] for details on the diffraction experiments and values
  of the structural parameters, including the citation
  \cite{parsons2003a}.}\BibitemShut {Stop}%
\bibitem [{\citenamefont {Jiliang}\ and\ \citenamefont
  {Svilen}(2013)}]{zhang2013a}%
  \BibitemOpen
  \bibfield  {author} {\bibinfo {author} {\bibfnamefont {Z.}~\bibnamefont
  {Jiliang}}\ and\ \bibinfo {author} {\bibfnamefont {B.}~\bibnamefont
  {Svilen}},\ }\bibfield  {title} {\bibinfo {title} {Synthesis, structural
  characterization and properties of {SrAl$_{4-x}$Ge$_x$},
  {BaAl$_{4-x}$Ge$_x$,} and {EuAl$_{4-x}$Ge$_x$} $(x \approx 0.3-0.4)$ -- rare
  examples of electron-rich phases with the {BaAl$_4$} structure type},\
  }\href@noop {} {\bibfield  {journal} {\bibinfo  {journal} {J. Sol. State
  Chem.}\ }\textbf {\bibinfo {volume} {205}},\ \bibinfo {pages} {21} (\bibinfo
  {year} {2013})}\BibitemShut {NoStop}%
\bibitem [{\citenamefont {Zevalkink}\ \emph {et~al.}(2017)\citenamefont
  {Zevalkink}, \citenamefont {Bobnar}, \citenamefont {Schwarz},\ and\
  \citenamefont {Grin}}]{zevalkink2017a}%
  \BibitemOpen
  \bibfield  {author} {\bibinfo {author} {\bibfnamefont {A.}~\bibnamefont
  {Zevalkink}}, \bibinfo {author} {\bibfnamefont {M.}~\bibnamefont {Bobnar}},
  \bibinfo {author} {\bibfnamefont {U.}~\bibnamefont {Schwarz}},\ and\ \bibinfo
  {author} {\bibfnamefont {Y.}~\bibnamefont {Grin}},\ }\bibfield  {title}
  {\bibinfo {title} {Making and breaking bonds in superconducting
  {SrAl$_{4-x}$Si$_x$} $(0 \leq x \leq 2)$},\ }\href@noop {} {\bibfield
  {journal} {\bibinfo  {journal} {Chem. Mater.}\ }\textbf {\bibinfo {volume}
  {29}},\ \bibinfo {pages} {1236} (\bibinfo {year} {2017})}\BibitemShut
  {NoStop}%
\bibitem [{\citenamefont {Paulmann}(2022)}]{paulmann2020a}%
  \BibitemOpen
  \bibfield  {author} {\bibinfo {author} {\bibfnamefont {C.}~\bibnamefont
  {Paulmann}},\ }\href@noop {} {\bibinfo {title} {{P24ToolsCP}, v1.12, software
  tools for postprocessing area detector, x-ray diffraction data}} (\bibinfo
  {year} {2022})\BibitemShut {NoStop}%
\bibitem [{\citenamefont {Schreurs}\ \emph {et~al.}(2010)\citenamefont
  {Schreurs}, \citenamefont {Xian},\ and\ \citenamefont
  {Kroon-Batenburg}}]{schreursamm2010a}%
  \BibitemOpen
  \bibfield  {author} {\bibinfo {author} {\bibfnamefont {A.~M.~M.}\
  \bibnamefont {Schreurs}}, \bibinfo {author} {\bibfnamefont {X.}~\bibnamefont
  {Xian}},\ and\ \bibinfo {author} {\bibfnamefont {L.~M.~J.}\ \bibnamefont
  {Kroon-Batenburg}},\ }\bibfield  {title} {\bibinfo {title} {{EVAL15:} a
  diffraction data integration method based on ab initio predicted profiles},\
  }\href {https://doi.org/10.1107/S0021889809043234} {\bibfield  {journal}
  {\bibinfo  {journal} {J. Appl. Crystallogr.}\ }\textbf {\bibinfo {volume}
  {43}},\ \bibinfo {pages} {70} (\bibinfo {year} {2010})}\BibitemShut {NoStop}%
\bibitem [{\citenamefont {Sheldrick}(2008)}]{sheldrick2008}%
  \BibitemOpen
  \bibfield  {author} {\bibinfo {author} {\bibfnamefont {G.~M.}\ \bibnamefont
  {Sheldrick}},\ }\href@noop {} {\emph {\bibinfo {title} {{SADABS,} {V}ersion
  2008/1}}}\ (\bibinfo  {publisher} {G\"{o}ttingen: University of
  G\"{o}ttingen},\ \bibinfo {year} {2008})\BibitemShut {NoStop}%
\bibitem [{\citenamefont {Petricek}\ \emph {et~al.}(2014)\citenamefont
  {Petricek}, \citenamefont {Dusek},\ and\ \citenamefont
  {Palatinus}}]{petricekv2014a}%
  \BibitemOpen
  \bibfield  {author} {\bibinfo {author} {\bibfnamefont {V.}~\bibnamefont
  {Petricek}}, \bibinfo {author} {\bibfnamefont {M.}~\bibnamefont {Dusek}},\
  and\ \bibinfo {author} {\bibfnamefont {L.}~\bibnamefont {Palatinus}},\
  }\bibfield  {title} {\bibinfo {title} {Crystallographic computing system
  {JANA2006:} general features},\ }\href
  {https://doi.org/10.1515/zkri-2014-1737} {\bibfield  {journal} {\bibinfo
  {journal} {Z. Kristallogr.}\ }\textbf {\bibinfo {volume} {229}},\ \bibinfo
  {pages} {345} (\bibinfo {year} {2014})}\BibitemShut {NoStop}%
\bibitem [{\citenamefont {Petricek}\ \emph {et~al.}(2016)\citenamefont
  {Petricek}, \citenamefont {Eigner}, \citenamefont {Dusek},\ and\
  \citenamefont {Cejchan}}]{petricekv2016a}%
  \BibitemOpen
  \bibfield  {author} {\bibinfo {author} {\bibfnamefont {V.}~\bibnamefont
  {Petricek}}, \bibinfo {author} {\bibfnamefont {V.}~\bibnamefont {Eigner}},
  \bibinfo {author} {\bibfnamefont {M.}~\bibnamefont {Dusek}},\ and\ \bibinfo
  {author} {\bibfnamefont {A.}~\bibnamefont {Cejchan}},\ }\bibfield  {title}
  {\bibinfo {title} {Discontinuous modulation functions and their application
  for analysis of modulated structures with the computing system {JANA2006}},\
  }\href@noop {} {\bibfield  {journal} {\bibinfo  {journal} {Z. Kristallogr.}\
  }\textbf {\bibinfo {volume} {231}},\ \bibinfo {pages} {301} (\bibinfo {year}
  {2016})}\BibitemShut {NoStop}%
\bibitem [{\citenamefont {Stokes}\ \emph {et~al.}(2011)\citenamefont {Stokes},
  \citenamefont {Campbell},\ and\ \citenamefont {van Smaalen}}]{stokesht2011a}%
  \BibitemOpen
  \bibfield  {author} {\bibinfo {author} {\bibfnamefont {H.~T.}\ \bibnamefont
  {Stokes}}, \bibinfo {author} {\bibfnamefont {B.~J.}\ \bibnamefont
  {Campbell}},\ and\ \bibinfo {author} {\bibfnamefont {S.}~\bibnamefont {van
  Smaalen}},\ }\bibfield  {title} {\bibinfo {title} {Generation of
  {$(3+d)$}-dimensional superspace groups for describing the symmetry of
  modulated crystalline structures},\ }\href
  {https://doi.org/10.1107/S0108767310042297} {\bibfield  {journal} {\bibinfo
  {journal} {Acta Crystallogr. A}\ }\textbf {\bibinfo {volume} {67}},\ \bibinfo
  {pages} {45} (\bibinfo {year} {2011})}\BibitemShut {NoStop}%
\bibitem [{\citenamefont {Kresse}\ and\ \citenamefont
  {Furthm\"uller}(1996)}]{kresse1996efficient}%
  \BibitemOpen
  \bibfield  {author} {\bibinfo {author} {\bibfnamefont {G.}~\bibnamefont
  {Kresse}}\ and\ \bibinfo {author} {\bibfnamefont {J.}~\bibnamefont
  {Furthm\"uller}},\ }\bibfield  {title} {\bibinfo {title} {Efficient iterative
  schemes for ab initio total-energy calculations using a plane-wave basis
  set},\ }\href@noop {} {\bibfield  {journal} {\bibinfo  {journal} {Phys. Rev.
  B}\ }\textbf {\bibinfo {volume} {54}},\ \bibinfo {pages} {11169} (\bibinfo
  {year} {1996})}\BibitemShut {NoStop}%
\bibitem [{\citenamefont {Bl\"ochl}(1994)}]{blochl1994projector}%
  \BibitemOpen
  \bibfield  {author} {\bibinfo {author} {\bibfnamefont {P.~E.}\ \bibnamefont
  {Bl\"ochl}},\ }\bibfield  {title} {\bibinfo {title} {Projector augmented-wave
  method},\ }\href {https://doi.org/10.1103/PhysRevB.50.17953} {\bibfield
  {journal} {\bibinfo  {journal} {Phys. Rev. B}\ }\textbf {\bibinfo {volume}
  {50}},\ \bibinfo {pages} {17953} (\bibinfo {year} {1994})}\BibitemShut
  {NoStop}%
\bibitem [{\citenamefont {Kresse}\ and\ \citenamefont
  {Joubert}(1999)}]{PAW1999}%
  \BibitemOpen
  \bibfield  {author} {\bibinfo {author} {\bibfnamefont {G.}~\bibnamefont
  {Kresse}}\ and\ \bibinfo {author} {\bibfnamefont {D.}~\bibnamefont
  {Joubert}},\ }\bibfield  {title} {\bibinfo {title} {From ultrasoft
  pseudopotentials to the projector augmented-wave method},\ }\href
  {https://doi.org/10.1103/PhysRevB.59.1758} {\bibfield  {journal} {\bibinfo
  {journal} {Phys. Rev. B}\ }\textbf {\bibinfo {volume} {59}},\ \bibinfo
  {pages} {1758} (\bibinfo {year} {1999})}\BibitemShut {NoStop}%
\bibitem [{\citenamefont {Perdew}\ \emph {et~al.}(1996)\citenamefont {Perdew},
  \citenamefont {Burke},\ and\ \citenamefont
  {Ernzerhof}}]{perdew1996generalized}%
  \BibitemOpen
  \bibfield  {author} {\bibinfo {author} {\bibfnamefont {J.~P.}\ \bibnamefont
  {Perdew}}, \bibinfo {author} {\bibfnamefont {K.}~\bibnamefont {Burke}},\ and\
  \bibinfo {author} {\bibfnamefont {M.}~\bibnamefont {Ernzerhof}},\ }\bibfield
  {title} {\bibinfo {title} {Generalized gradient approximation made simple},\
  }\href@noop {} {\bibfield  {journal} {\bibinfo  {journal} {Phys. Rev. Lett.}\
  }\textbf {\bibinfo {volume} {77}},\ \bibinfo {pages} {3865} (\bibinfo {year}
  {1996})}\BibitemShut {NoStop}%
\bibitem [{\citenamefont {Wu}\ \emph {et~al.}(2018)\citenamefont {Wu},
  \citenamefont {Zhang}, \citenamefont {Song}, \citenamefont {Troyer},\ and\
  \citenamefont {Soluyanov}}]{WU2018}%
  \BibitemOpen
  \bibfield  {author} {\bibinfo {author} {\bibfnamefont {Q.}~\bibnamefont
  {Wu}}, \bibinfo {author} {\bibfnamefont {S.}~\bibnamefont {Zhang}}, \bibinfo
  {author} {\bibfnamefont {H.-F.}\ \bibnamefont {Song}}, \bibinfo {author}
  {\bibfnamefont {M.}~\bibnamefont {Troyer}},\ and\ \bibinfo {author}
  {\bibfnamefont {A.~A.}\ \bibnamefont {Soluyanov}},\ }\bibfield  {title}
  {\bibinfo {title} {Wanniertools: An open-source software package for novel
  topological materials},\ }\href
  {https://doi.org/https://doi.org/10.1016/j.cpc.2017.09.033} {\bibfield
  {journal} {\bibinfo  {journal} {Computer Physics Communications}\ }\textbf
  {\bibinfo {volume} {224}},\ \bibinfo {pages} {405 } (\bibinfo {year}
  {2018})}\BibitemShut {NoStop}%
\bibitem [{\citenamefont {Mostofi}\ \emph {et~al.}(2008)\citenamefont
  {Mostofi}, \citenamefont {Yates}, \citenamefont {Lee}, \citenamefont {Souza},
  \citenamefont {Vanderbilt},\ and\ \citenamefont {Marzari}}]{wan2008}%
  \BibitemOpen
  \bibfield  {author} {\bibinfo {author} {\bibfnamefont {A.~A.}\ \bibnamefont
  {Mostofi}}, \bibinfo {author} {\bibfnamefont {J.~R.}\ \bibnamefont {Yates}},
  \bibinfo {author} {\bibfnamefont {Y.-S.}\ \bibnamefont {Lee}}, \bibinfo
  {author} {\bibfnamefont {I.}~\bibnamefont {Souza}}, \bibinfo {author}
  {\bibfnamefont {D.}~\bibnamefont {Vanderbilt}},\ and\ \bibinfo {author}
  {\bibfnamefont {N.}~\bibnamefont {Marzari}},\ }\bibfield  {title} {\bibinfo
  {title} {wannier90: A tool for obtaining maximally-localised wannier
  functions},\ }\href
  {https://doi.org/https://doi.org/10.1016/j.cpc.2007.11.016} {\bibfield
  {journal} {\bibinfo  {journal} {Computer Physics Communications}\ }\textbf
  {\bibinfo {volume} {178}},\ \bibinfo {pages} {685} (\bibinfo {year}
  {2008})}\BibitemShut {NoStop}%
\bibitem [{\citenamefont {Mostofi}\ \emph {et~al.}(2014)\citenamefont
  {Mostofi}, \citenamefont {Yates}, \citenamefont {Pizzi}, \citenamefont {Lee},
  \citenamefont {Souza}, \citenamefont {Vanderbilt},\ and\ \citenamefont
  {Marzari}}]{wan90}%
  \BibitemOpen
  \bibfield  {author} {\bibinfo {author} {\bibfnamefont {A.~A.}\ \bibnamefont
  {Mostofi}}, \bibinfo {author} {\bibfnamefont {J.~R.}\ \bibnamefont {Yates}},
  \bibinfo {author} {\bibfnamefont {G.}~\bibnamefont {Pizzi}}, \bibinfo
  {author} {\bibfnamefont {Y.-S.}\ \bibnamefont {Lee}}, \bibinfo {author}
  {\bibfnamefont {I.}~\bibnamefont {Souza}}, \bibinfo {author} {\bibfnamefont
  {D.}~\bibnamefont {Vanderbilt}},\ and\ \bibinfo {author} {\bibfnamefont
  {N.}~\bibnamefont {Marzari}},\ }\bibfield  {title} {\bibinfo {title} {An
  updated version of wannier90: A tool for obtaining maximally-localised
  wannier functions},\ }\href@noop {} {\bibfield  {journal} {\bibinfo
  {journal} {Computer Physics Communications}\ }\textbf {\bibinfo {volume}
  {185}},\ \bibinfo {pages} {2309} (\bibinfo {year} {2014})}\BibitemShut
  {NoStop}%
\bibitem [{\citenamefont {Marzari}\ \emph {et~al.}(2012)\citenamefont
  {Marzari}, \citenamefont {Mostofi}, \citenamefont {Yates}, \citenamefont
  {Souza},\ and\ \citenamefont {Vanderbilt}}]{MLWF2012}%
  \BibitemOpen
  \bibfield  {author} {\bibinfo {author} {\bibfnamefont {N.}~\bibnamefont
  {Marzari}}, \bibinfo {author} {\bibfnamefont {A.~A.}\ \bibnamefont
  {Mostofi}}, \bibinfo {author} {\bibfnamefont {J.~R.}\ \bibnamefont {Yates}},
  \bibinfo {author} {\bibfnamefont {I.}~\bibnamefont {Souza}},\ and\ \bibinfo
  {author} {\bibfnamefont {D.}~\bibnamefont {Vanderbilt}},\ }\bibfield  {title}
  {\bibinfo {title} {Maximally localized wannier functions: Theory and
  applications},\ }\href {https://doi.org/10.1103/RevModPhys.84.1419}
  {\bibfield  {journal} {\bibinfo  {journal} {Rev. Mod. Phys.}\ }\textbf
  {\bibinfo {volume} {84}},\ \bibinfo {pages} {1419} (\bibinfo {year}
  {2012})}\BibitemShut {NoStop}%
\bibitem [{\citenamefont {Giannozzi}\ \emph {et~al.}(1991)\citenamefont
  {Giannozzi}, \citenamefont {de~Gironcoli}, \citenamefont {Pavone},\ and\
  \citenamefont {Baroni}}]{DFPT1991}%
  \BibitemOpen
  \bibfield  {author} {\bibinfo {author} {\bibfnamefont {P.}~\bibnamefont
  {Giannozzi}}, \bibinfo {author} {\bibfnamefont {S.}~\bibnamefont
  {de~Gironcoli}}, \bibinfo {author} {\bibfnamefont {P.}~\bibnamefont
  {Pavone}},\ and\ \bibinfo {author} {\bibfnamefont {S.}~\bibnamefont
  {Baroni}},\ }\bibfield  {title} {\bibinfo {title} {Ab initio calculation of
  phonon dispersions in semiconductors},\ }\href
  {https://doi.org/10.1103/PhysRevB.43.7231} {\bibfield  {journal} {\bibinfo
  {journal} {Phys. Rev. B}\ }\textbf {\bibinfo {volume} {43}},\ \bibinfo
  {pages} {7231} (\bibinfo {year} {1991})}\BibitemShut {NoStop}%
\bibitem [{\citenamefont {Gonze}\ and\ \citenamefont {Lee}(1997)}]{DFPT1997}%
  \BibitemOpen
  \bibfield  {author} {\bibinfo {author} {\bibfnamefont {X.}~\bibnamefont
  {Gonze}}\ and\ \bibinfo {author} {\bibfnamefont {C.}~\bibnamefont {Lee}},\
  }\bibfield  {title} {\bibinfo {title} {Dynamical matrices, born effective
  charges, dielectric permittivity tensors, and interatomic force constants
  from density-functional perturbation theory},\ }\href
  {https://doi.org/10.1103/PhysRevB.55.10355} {\bibfield  {journal} {\bibinfo
  {journal} {Phys. Rev. B}\ }\textbf {\bibinfo {volume} {55}},\ \bibinfo
  {pages} {10355} (\bibinfo {year} {1997})}\BibitemShut {NoStop}%
\bibitem [{\citenamefont {Togo}\ and\ \citenamefont
  {Tanaka}(2015)}]{togo2015first}%
  \BibitemOpen
  \bibfield  {author} {\bibinfo {author} {\bibfnamefont {A.}~\bibnamefont
  {Togo}}\ and\ \bibinfo {author} {\bibfnamefont {I.}~\bibnamefont {Tanaka}},\
  }\bibfield  {title} {\bibinfo {title} {First principles phonon calculations
  in materials science},\ }\href@noop {} {\bibfield  {journal} {\bibinfo
  {journal} {Scripta Materialia}\ }\textbf {\bibinfo {volume} {108}},\ \bibinfo
  {pages} {1} (\bibinfo {year} {2015})}\BibitemShut {NoStop}%
\bibitem [{\citenamefont {Giannozzi}\ \emph {et~al.}(2009)\citenamefont
  {Giannozzi}, \citenamefont {Baroni}, \citenamefont {Bonini}, \citenamefont
  {Calandra}, \citenamefont {Car}, \citenamefont {Cavazzoni}, \citenamefont
  {Ceresoli}, \citenamefont {Chiarotti}, \citenamefont {Cococcioni},
  \citenamefont {Dabo}, \citenamefont {Corso}, \citenamefont {de~Gironcoli},
  \citenamefont {Fabris}, \citenamefont {Fratesi}, \citenamefont {Gebauer},
  \citenamefont {Gerstmann}, \citenamefont {Gougoussis}, \citenamefont
  {Kokalj}, \citenamefont {Lazzeri}, \citenamefont {Martin-Samos},
  \citenamefont {Marzari}, \citenamefont {Mauri}, \citenamefont {Mazzarello},
  \citenamefont {Paolini}, \citenamefont {Pasquarello}, \citenamefont
  {Paulatto}, \citenamefont {Sbraccia}, \citenamefont {Scandolo}, \citenamefont
  {Sclauzero}, \citenamefont {Seitsonen}, \citenamefont {Smogunov},
  \citenamefont {Umari},\ and\ \citenamefont {Wentzcovitch}}]{Giannozzi_2009}%
  \BibitemOpen
  \bibfield  {author} {\bibinfo {author} {\bibfnamefont {P.}~\bibnamefont
  {Giannozzi}}, \bibinfo {author} {\bibfnamefont {S.}~\bibnamefont {Baroni}},
  \bibinfo {author} {\bibfnamefont {N.}~\bibnamefont {Bonini}}, \bibinfo
  {author} {\bibfnamefont {M.}~\bibnamefont {Calandra}}, \bibinfo {author}
  {\bibfnamefont {R.}~\bibnamefont {Car}}, \bibinfo {author} {\bibfnamefont
  {C.}~\bibnamefont {Cavazzoni}}, \bibinfo {author} {\bibfnamefont
  {D.}~\bibnamefont {Ceresoli}}, \bibinfo {author} {\bibfnamefont {G.~L.}\
  \bibnamefont {Chiarotti}}, \bibinfo {author} {\bibfnamefont {M.}~\bibnamefont
  {Cococcioni}}, \bibinfo {author} {\bibfnamefont {I.}~\bibnamefont {Dabo}},
  \bibinfo {author} {\bibfnamefont {A.~D.}\ \bibnamefont {Corso}}, \bibinfo
  {author} {\bibfnamefont {S.}~\bibnamefont {de~Gironcoli}}, \bibinfo {author}
  {\bibfnamefont {S.}~\bibnamefont {Fabris}}, \bibinfo {author} {\bibfnamefont
  {G.}~\bibnamefont {Fratesi}}, \bibinfo {author} {\bibfnamefont
  {R.}~\bibnamefont {Gebauer}}, \bibinfo {author} {\bibfnamefont
  {U.}~\bibnamefont {Gerstmann}}, \bibinfo {author} {\bibfnamefont
  {C.}~\bibnamefont {Gougoussis}}, \bibinfo {author} {\bibfnamefont
  {A.}~\bibnamefont {Kokalj}}, \bibinfo {author} {\bibfnamefont
  {M.}~\bibnamefont {Lazzeri}}, \bibinfo {author} {\bibfnamefont
  {L.}~\bibnamefont {Martin-Samos}}, \bibinfo {author} {\bibfnamefont
  {N.}~\bibnamefont {Marzari}}, \bibinfo {author} {\bibfnamefont
  {F.}~\bibnamefont {Mauri}}, \bibinfo {author} {\bibfnamefont
  {R.}~\bibnamefont {Mazzarello}}, \bibinfo {author} {\bibfnamefont
  {S.}~\bibnamefont {Paolini}}, \bibinfo {author} {\bibfnamefont
  {A.}~\bibnamefont {Pasquarello}}, \bibinfo {author} {\bibfnamefont
  {L.}~\bibnamefont {Paulatto}}, \bibinfo {author} {\bibfnamefont
  {C.}~\bibnamefont {Sbraccia}}, \bibinfo {author} {\bibfnamefont
  {S.}~\bibnamefont {Scandolo}}, \bibinfo {author} {\bibfnamefont
  {G.}~\bibnamefont {Sclauzero}}, \bibinfo {author} {\bibfnamefont {A.~P.}\
  \bibnamefont {Seitsonen}}, \bibinfo {author} {\bibfnamefont {A.}~\bibnamefont
  {Smogunov}}, \bibinfo {author} {\bibfnamefont {P.}~\bibnamefont {Umari}},\
  and\ \bibinfo {author} {\bibfnamefont {R.~M.}\ \bibnamefont {Wentzcovitch}},\
  }\bibfield  {title} {\bibinfo {title} {Quantum espresso: a modular and
  open-source software project for quantum simulations of materials},\ }\href
  {https://doi.org/10.1088/0953-8984/21/39/395502} {\bibfield  {journal}
  {\bibinfo  {journal} {J. Phys.: Condens. Matter}\ }\textbf {\bibinfo {volume}
  {21}},\ \bibinfo {pages} {395502} (\bibinfo {year} {2009})}\BibitemShut
  {NoStop}%
\bibitem [{\citenamefont {Giannozzi}\ \emph {et~al.}(2017)\citenamefont
  {Giannozzi}, \citenamefont {Andreussi}, \citenamefont {Brumme}, \citenamefont
  {Bunau}, \citenamefont {Nardelli}, \citenamefont {Calandra}, \citenamefont
  {Car}, \citenamefont {Cavazzoni}, \citenamefont {Ceresoli}, \citenamefont
  {Cococcioni}, \citenamefont {Colonna}, \citenamefont {Carnimeo},
  \citenamefont {Corso}, \citenamefont {de~Gironcoli}, \citenamefont {Delugas},
  \citenamefont {DiStasio}, \citenamefont {Ferretti}, \citenamefont {Floris},
  \citenamefont {Fratesi}, \citenamefont {Fugallo}, \citenamefont {Gebauer},
  \citenamefont {Gerstmann}, \citenamefont {Giustino}, \citenamefont {Gorni},
  \citenamefont {Jia}, \citenamefont {Kawamura}, \citenamefont {Ko},
  \citenamefont {Kokalj}, \citenamefont {Küçükbenli}, \citenamefont
  {Lazzeri}, \citenamefont {Marsili}, \citenamefont {Marzari}, \citenamefont
  {Mauri}, \citenamefont {Nguyen}, \citenamefont {Nguyen}, \citenamefont {de-la
  Roza}, \citenamefont {Paulatto}, \citenamefont {Poncé}, \citenamefont
  {Rocca}, \citenamefont {Sabatini}, \citenamefont {Santra}, \citenamefont
  {Schlipf}, \citenamefont {Seitsonen}, \citenamefont {Smogunov}, \citenamefont
  {Timrov}, \citenamefont {Thonhauser}, \citenamefont {Umari}, \citenamefont
  {Vast}, \citenamefont {Wu},\ and\ \citenamefont {Baroni}}]{Giannozzi_2017}%
  \BibitemOpen
  \bibfield  {author} {\bibinfo {author} {\bibfnamefont {P.}~\bibnamefont
  {Giannozzi}}, \bibinfo {author} {\bibfnamefont {O.}~\bibnamefont
  {Andreussi}}, \bibinfo {author} {\bibfnamefont {T.}~\bibnamefont {Brumme}},
  \bibinfo {author} {\bibfnamefont {O.}~\bibnamefont {Bunau}}, \bibinfo
  {author} {\bibfnamefont {M.~B.}\ \bibnamefont {Nardelli}}, \bibinfo {author}
  {\bibfnamefont {M.}~\bibnamefont {Calandra}}, \bibinfo {author}
  {\bibfnamefont {R.}~\bibnamefont {Car}}, \bibinfo {author} {\bibfnamefont
  {C.}~\bibnamefont {Cavazzoni}}, \bibinfo {author} {\bibfnamefont
  {D.}~\bibnamefont {Ceresoli}}, \bibinfo {author} {\bibfnamefont
  {M.}~\bibnamefont {Cococcioni}}, \bibinfo {author} {\bibfnamefont
  {N.}~\bibnamefont {Colonna}}, \bibinfo {author} {\bibfnamefont
  {I.}~\bibnamefont {Carnimeo}}, \bibinfo {author} {\bibfnamefont {A.~D.}\
  \bibnamefont {Corso}}, \bibinfo {author} {\bibfnamefont {S.}~\bibnamefont
  {de~Gironcoli}}, \bibinfo {author} {\bibfnamefont {P.}~\bibnamefont
  {Delugas}}, \bibinfo {author} {\bibfnamefont {R.~A.}\ \bibnamefont
  {DiStasio}}, \bibinfo {author} {\bibfnamefont {A.}~\bibnamefont {Ferretti}},
  \bibinfo {author} {\bibfnamefont {A.}~\bibnamefont {Floris}}, \bibinfo
  {author} {\bibfnamefont {G.}~\bibnamefont {Fratesi}}, \bibinfo {author}
  {\bibfnamefont {G.}~\bibnamefont {Fugallo}}, \bibinfo {author} {\bibfnamefont
  {R.}~\bibnamefont {Gebauer}}, \bibinfo {author} {\bibfnamefont
  {U.}~\bibnamefont {Gerstmann}}, \bibinfo {author} {\bibfnamefont
  {F.}~\bibnamefont {Giustino}}, \bibinfo {author} {\bibfnamefont
  {T.}~\bibnamefont {Gorni}}, \bibinfo {author} {\bibfnamefont
  {J.}~\bibnamefont {Jia}}, \bibinfo {author} {\bibfnamefont {M.}~\bibnamefont
  {Kawamura}}, \bibinfo {author} {\bibfnamefont {H.-Y.}\ \bibnamefont {Ko}},
  \bibinfo {author} {\bibfnamefont {A.}~\bibnamefont {Kokalj}}, \bibinfo
  {author} {\bibfnamefont {E.}~\bibnamefont {Küçükbenli}}, \bibinfo {author}
  {\bibfnamefont {M.}~\bibnamefont {Lazzeri}}, \bibinfo {author} {\bibfnamefont
  {M.}~\bibnamefont {Marsili}}, \bibinfo {author} {\bibfnamefont
  {N.}~\bibnamefont {Marzari}}, \bibinfo {author} {\bibfnamefont
  {F.}~\bibnamefont {Mauri}}, \bibinfo {author} {\bibfnamefont {N.~L.}\
  \bibnamefont {Nguyen}}, \bibinfo {author} {\bibfnamefont {H.-V.}\
  \bibnamefont {Nguyen}}, \bibinfo {author} {\bibfnamefont {A.~O.}\
  \bibnamefont {de-la Roza}}, \bibinfo {author} {\bibfnamefont
  {L.}~\bibnamefont {Paulatto}}, \bibinfo {author} {\bibfnamefont
  {S.}~\bibnamefont {Poncé}}, \bibinfo {author} {\bibfnamefont
  {D.}~\bibnamefont {Rocca}}, \bibinfo {author} {\bibfnamefont
  {R.}~\bibnamefont {Sabatini}}, \bibinfo {author} {\bibfnamefont
  {B.}~\bibnamefont {Santra}}, \bibinfo {author} {\bibfnamefont
  {M.}~\bibnamefont {Schlipf}}, \bibinfo {author} {\bibfnamefont {A.~P.}\
  \bibnamefont {Seitsonen}}, \bibinfo {author} {\bibfnamefont {A.}~\bibnamefont
  {Smogunov}}, \bibinfo {author} {\bibfnamefont {I.}~\bibnamefont {Timrov}},
  \bibinfo {author} {\bibfnamefont {T.}~\bibnamefont {Thonhauser}}, \bibinfo
  {author} {\bibfnamefont {P.}~\bibnamefont {Umari}}, \bibinfo {author}
  {\bibfnamefont {N.}~\bibnamefont {Vast}}, \bibinfo {author} {\bibfnamefont
  {X.}~\bibnamefont {Wu}},\ and\ \bibinfo {author} {\bibfnamefont
  {S.}~\bibnamefont {Baroni}},\ }\bibfield  {title} {\bibinfo {title} {Advanced
  capabilities for materials modelling with quantum espresso},\ }\href
  {https://doi.org/10.1088/1361-648X/aa8f79} {\bibfield  {journal} {\bibinfo
  {journal} {J. Phys.: Condens. Matter}\ }\textbf {\bibinfo {volume} {29}},\
  \bibinfo {pages} {465901} (\bibinfo {year} {2017})}\BibitemShut {NoStop}%
\bibitem [{\citenamefont {Moya}\ \emph {et~al.}(2022)\citenamefont {Moya},
  \citenamefont {Lei}, \citenamefont {Clements}, \citenamefont {Kengle},
  \citenamefont {Sun}, \citenamefont {Allen}, \citenamefont {Li}, \citenamefont
  {Peng}, \citenamefont {Husain}, \citenamefont {Mitrano}, \citenamefont
  {Krogstad}, \citenamefont {Osborn}, \citenamefont {Puthirath}, \citenamefont
  {Chi}, \citenamefont {Debeer-Schmitt}, \citenamefont {Gaudet}, \citenamefont
  {Abbamonte}, \citenamefont {Lynn},\ and\ \citenamefont
  {Morosan}}]{moyajm2022a}%
  \BibitemOpen
  \bibfield  {author} {\bibinfo {author} {\bibfnamefont {J.~M.}\ \bibnamefont
  {Moya}}, \bibinfo {author} {\bibfnamefont {S.}~\bibnamefont {Lei}}, \bibinfo
  {author} {\bibfnamefont {E.~M.}\ \bibnamefont {Clements}}, \bibinfo {author}
  {\bibfnamefont {C.~S.}\ \bibnamefont {Kengle}}, \bibinfo {author}
  {\bibfnamefont {S.}~\bibnamefont {Sun}}, \bibinfo {author} {\bibfnamefont
  {K.}~\bibnamefont {Allen}}, \bibinfo {author} {\bibfnamefont
  {Q.}~\bibnamefont {Li}}, \bibinfo {author} {\bibfnamefont {Y.~Y.}\
  \bibnamefont {Peng}}, \bibinfo {author} {\bibfnamefont {A.~A.}\ \bibnamefont
  {Husain}}, \bibinfo {author} {\bibfnamefont {M.}~\bibnamefont {Mitrano}},
  \bibinfo {author} {\bibfnamefont {M.~J.}\ \bibnamefont {Krogstad}}, \bibinfo
  {author} {\bibfnamefont {R.}~\bibnamefont {Osborn}}, \bibinfo {author}
  {\bibfnamefont {A.~B.}\ \bibnamefont {Puthirath}}, \bibinfo {author}
  {\bibfnamefont {S.}~\bibnamefont {Chi}}, \bibinfo {author} {\bibfnamefont
  {L.}~\bibnamefont {Debeer-Schmitt}}, \bibinfo {author} {\bibfnamefont
  {J.}~\bibnamefont {Gaudet}}, \bibinfo {author} {\bibfnamefont
  {P.}~\bibnamefont {Abbamonte}}, \bibinfo {author} {\bibfnamefont {J.~W.}\
  \bibnamefont {Lynn}},\ and\ \bibinfo {author} {\bibfnamefont
  {E.}~\bibnamefont {Morosan}},\ }\bibfield  {title} {\bibinfo {title}
  {Incommensurate magnetic orders and topological hall effect in the square-net
  centrosymmetric {EuGa$_{2}$Al$_{2}$} system},\ }\href
  {https://doi.org/10.1103/PhysRevMaterials.6.074201} {\bibfield  {journal}
  {\bibinfo  {journal} {Phys. Rev. Mater.}\ }\textbf {\bibinfo {volume} {6}},\
  \bibinfo {pages} {074201} (\bibinfo {year} {2022})}\BibitemShut {NoStop}%
\bibitem [{\citenamefont {Rigaku}(2019)}]{crysalis}%
  \BibitemOpen
  \bibfield  {author} {\bibinfo {author} {\bibnamefont {Rigaku}},\ }\href@noop
  {} {\bibinfo {title} {Crysalis pro version 171.40.53, rigaku oxford
  diffraction.}} (\bibinfo {year} {2019})\BibitemShut {NoStop}%
\bibitem [{\citenamefont {Parth{\'{e}}}\ \emph {et~al.}(1983)\citenamefont
  {Parth{\'{e}}}, \citenamefont {Chabot}, \citenamefont {Braun},\ and\
  \citenamefont {Engel}}]{parthe1983a}%
  \BibitemOpen
  \bibfield  {author} {\bibinfo {author} {\bibfnamefont {E.}~\bibnamefont
  {Parth{\'{e}}}}, \bibinfo {author} {\bibfnamefont {B.}~\bibnamefont
  {Chabot}}, \bibinfo {author} {\bibfnamefont {H.~F.}\ \bibnamefont {Braun}},\
  and\ \bibinfo {author} {\bibfnamefont {N.}~\bibnamefont {Engel}},\ }\bibfield
   {title} {\bibinfo {title} {{Ternary {BaAl$_{4}$}-type derivative
  structures}},\ }\href@noop {} {\bibfield  {journal} {\bibinfo  {journal}
  {Acta Crystallogr. B}\ }\textbf {\bibinfo {volume} {39}},\ \bibinfo {pages}
  {588} (\bibinfo {year} {1983})}\BibitemShut {NoStop}%
\bibitem [{\citenamefont {Momma}\ and\ \citenamefont
  {Izumi}(2008)}]{momma2008vesta}%
  \BibitemOpen
  \bibfield  {author} {\bibinfo {author} {\bibfnamefont {K.}~\bibnamefont
  {Momma}}\ and\ \bibinfo {author} {\bibfnamefont {F.}~\bibnamefont {Izumi}},\
  }\bibfield  {title} {\bibinfo {title} {{VESTA:} a three-dimensional
  visualization system for electronic and structural analysis},\ }\href@noop {}
  {\bibfield  {journal} {\bibinfo  {journal} {J. Appl. Crystallogr.}\ }\textbf
  {\bibinfo {volume} {41}},\ \bibinfo {pages} {653} (\bibinfo {year}
  {2008})}\BibitemShut {NoStop}%
\bibitem [{\citenamefont {van Smaalen}(2007)}]{vansmaalen2007a}%
  \BibitemOpen
  \bibfield  {author} {\bibinfo {author} {\bibfnamefont {S.}~\bibnamefont {van
  Smaalen}},\ }\href@noop {} {\emph {\bibinfo {title} {Incommensurate
  Crystallography}}}\ (\bibinfo  {publisher} {Oxford University Press},\
  \bibinfo {address} {Oxford},\ \bibinfo {year} {2007})\BibitemShut {NoStop}%
\bibitem [{\citenamefont {Liu}\ \emph {et~al.}(2010)\citenamefont {Liu},
  \citenamefont {Li}, \citenamefont {Wang}, \citenamefont {Yin}, \citenamefont
  {Shi},\ and\ \citenamefont {Xiong}}]{liuy2010a}%
  \BibitemOpen
  \bibfield  {author} {\bibinfo {author} {\bibfnamefont {Y.}~\bibnamefont
  {Liu}}, \bibinfo {author} {\bibfnamefont {C.}~\bibnamefont {Li}}, \bibinfo
  {author} {\bibfnamefont {J.}~\bibnamefont {Wang}}, \bibinfo {author}
  {\bibfnamefont {D.}~\bibnamefont {Yin}}, \bibinfo {author} {\bibfnamefont
  {J.}~\bibnamefont {Shi}},\ and\ \bibinfo {author} {\bibfnamefont
  {R.}~\bibnamefont {Xiong}},\ }\bibfield  {title} {\bibinfo {title} {Thermal
  transport properties and electronic structure of {W}-doped rubidium blue
  bronzes {Rb$_{0.3}$Mo$_{1-x}$W$_{x}$O$_3$} (x = 0, 0.001, 0.003, 0.005)},\
  }\href {https://doi.org/10.1016/j.physb.2010.04.006} {\bibfield  {journal}
  {\bibinfo  {journal} {Physica B}\ }\textbf {\bibinfo {volume} {405}},\
  \bibinfo {pages} {2857} (\bibinfo {year} {2010})}\BibitemShut {NoStop}%
\bibitem [{\citenamefont {Yan}\ \emph {et~al.}(2019)\citenamefont {Yan},
  \citenamefont {Zeng}, \citenamefont {Lin}, \citenamefont {Yin}, \citenamefont
  {He}, \citenamefont {Zhang}, \citenamefont {Huang}, \citenamefont {Shen},
  \citenamefont {Wang}, \citenamefont {Wang}, \citenamefont {Yao},\ and\
  \citenamefont {Luo}}]{yand2019a}%
  \BibitemOpen
  \bibfield  {author} {\bibinfo {author} {\bibfnamefont {D.}~\bibnamefont
  {Yan}}, \bibinfo {author} {\bibfnamefont {L.}~\bibnamefont {Zeng}}, \bibinfo
  {author} {\bibfnamefont {Y.}~\bibnamefont {Lin}}, \bibinfo {author}
  {\bibfnamefont {J.}~\bibnamefont {Yin}}, \bibinfo {author} {\bibfnamefont
  {Y.}~\bibnamefont {He}}, \bibinfo {author} {\bibfnamefont {X.}~\bibnamefont
  {Zhang}}, \bibinfo {author} {\bibfnamefont {M.}~\bibnamefont {Huang}},
  \bibinfo {author} {\bibfnamefont {B.}~\bibnamefont {Shen}}, \bibinfo {author}
  {\bibfnamefont {M.}~\bibnamefont {Wang}}, \bibinfo {author} {\bibfnamefont
  {Y.}~\bibnamefont {Wang}}, \bibinfo {author} {\bibfnamefont {D.}~\bibnamefont
  {Yao}},\ and\ \bibinfo {author} {\bibfnamefont {H.}~\bibnamefont {Luo}},\
  }\bibfield  {title} {\bibinfo {title} {Superconductivity in {Ru}-doped
  {CuIr$_2$Te$_4$} telluride chalcogenide},\ }\href
  {https://doi.org/10.1103/PhysRevB.100.174504} {\bibfield  {journal} {\bibinfo
   {journal} {Phys. Rev. B}\ }\textbf {\bibinfo {volume} {100}},\ \bibinfo
  {pages} {174504} (\bibinfo {year} {2019})}\BibitemShut {NoStop}%
\bibitem [{\citenamefont {Gao}\ \emph {et~al.}(2021)\citenamefont {Gao},
  \citenamefont {Wu}, \citenamefont {Persson},\ and\ \citenamefont
  {Wang}}]{Irvsp}%
  \BibitemOpen
  \bibfield  {author} {\bibinfo {author} {\bibfnamefont {J.}~\bibnamefont
  {Gao}}, \bibinfo {author} {\bibfnamefont {Q.}~\bibnamefont {Wu}}, \bibinfo
  {author} {\bibfnamefont {C.}~\bibnamefont {Persson}},\ and\ \bibinfo {author}
  {\bibfnamefont {Z.}~\bibnamefont {Wang}},\ }\bibfield  {title} {\bibinfo
  {title} {Irvsp: To obtain irreducible representations of electronic states in
  the {VASP}},\ }\href@noop {} {\bibfield  {journal} {\bibinfo  {journal}
  {Computer Physics Communications}\ }\textbf {\bibinfo {volume} {261}},\
  \bibinfo {pages} {107760} (\bibinfo {year} {2021})}\BibitemShut {NoStop}%
\bibitem [{\citenamefont {Johannes}\ and\ \citenamefont {Mazin}(2008)}]{FSN}%
  \BibitemOpen
  \bibfield  {author} {\bibinfo {author} {\bibfnamefont {M.~D.}\ \bibnamefont
  {Johannes}}\ and\ \bibinfo {author} {\bibfnamefont {I.~I.}\ \bibnamefont
  {Mazin}},\ }\bibfield  {title} {\bibinfo {title} {{F}ermi surface nesting and
  the origin of charge density waves in metals},\ }\href@noop {} {\bibfield
  {journal} {\bibinfo  {journal} {Phys. Rev. B}\ }\textbf {\bibinfo {volume}
  {77}},\ \bibinfo {pages} {165135} (\bibinfo {year} {2008})}\BibitemShut
  {NoStop}%
\bibitem [{\citenamefont {Suzuki}\ \emph {et~al.}(2010)\citenamefont {Suzuki},
  \citenamefont {Inaba},\ and\ \citenamefont {Meingast}}]{suzukih2010a}%
  \BibitemOpen
  \bibfield  {author} {\bibinfo {author} {\bibfnamefont {H.}~\bibnamefont
  {Suzuki}}, \bibinfo {author} {\bibfnamefont {A.}~\bibnamefont {Inaba}},\ and\
  \bibinfo {author} {\bibfnamefont {C.}~\bibnamefont {Meingast}},\ }\bibfield
  {title} {\bibinfo {title} {Accurate heat capacity data at phase transitions
  from relaxation calorimetry},\ }\href
  {https://doi.org/10.1016/j.cryogenics.2010.07.003} {\bibfield  {journal}
  {\bibinfo  {journal} {Cryogenics}\ }\textbf {\bibinfo {volume} {50}},\
  \bibinfo {pages} {693} (\bibinfo {year} {2010})}\BibitemShut {NoStop}%
\bibitem [{\citenamefont {Parsons}(2003)}]{parsons2003a}%
  \BibitemOpen
  \bibfield  {author} {\bibinfo {author} {\bibfnamefont {S.}~\bibnamefont
  {Parsons}},\ }\bibfield  {title} {\bibinfo {title} {Introduction to
  twinning},\ }\href@noop {} {\bibfield  {journal} {\bibinfo  {journal} {Acta
  Crystallogr. D}\ }\textbf {\bibinfo {volume} {59}},\ \bibinfo {pages} {1995}
  (\bibinfo {year} {2003})}\BibitemShut {NoStop}%
\end{thebibliography}%

\end{document}